\documentclass[a4paper,fleqn]{cas-sc}

\usepackage[numbers]{natbib}

\usepackage{float}
\usepackage{acronym}
\usepackage{subfiles}
\usepackage[linewidth=1pt]{mdframed}

\usepackage{graphicx}
\usepackage{multicol}
\usepackage{flafter}
\usepackage{setspace}
\usepackage{enumitem}   
\usepackage{tablefootnote}
\usepackage{url}
\usepackage{pgfplots} 
\pgfplotsset{compat=1.7}
\usepackage{tikz}
\usepackage{array}
\usepackage{longtable}
\newcolumntype{P}[1]{>{\centering\arraybackslash}p{#1}}
\usepackage{stfloats}
\usepackage{booktabs, multirow} 
\usepackage{soul}
\usepackage{changepage,threeparttable} 
\usepackage{booktabs}
\usepackage{layout}

\newcolumntype{M}[1]{>{\centering\arraybackslash}m{#1}}

\graphicspath{ {./Figures/} }

\def\tsc#1{\csdef{#1}{\textsc{\lowercase{#1}}\xspace}}
\tsc{WGM}
\tsc{QE}
\tsc{EP}
\tsc{PMS}
\tsc{BEC}
\tsc{DE}
\tsc{HPC}

\begin{document}
\let\WriteBookmarks\relax
\def\floatpagepagefraction{1}
\def\textpagefraction{.001}

\shorttitle{HPC Applications to Power System Optimization}

\shortauthors{Ahmed Al-Shafei et~al.}

\title [mode = title]{A Review of High-Performance Computing and Parallel Techniques Applied to Power Systems Optimization}                      

\author[1]{Ahmed Al-Shafei} 

\ead{ahmed.alshafei@ucalgary.ca}

\author[1]{Hamidreza Zareipour}
\ead{hzareipo@ucalgary.ca}

\author[2]{Yankai Cao}
\ead{yankai.cao@ubc.ca}

\cormark[1]

    

\cortext[cor1]{Corresponding author.}

\begin{abstract}
The accelerating technological landscape and drive towards net-zero emission made the power system grow in scale and complexity. Serial computational approaches for grid planning and operation struggle to execute necessary calculations within reasonable times. Resorting to high-performance and parallel computing approaches has become paramount. Moreover, the ambitious plans for the future grid and IoT integration make a shift towards utilizing Cloud computing inevitable. This article recounts the dawn of parallel computation and its appearance in power system studies, reviewing the most recent literature and research on exploiting the available computational resources and technologies today. The article starts with a brief introduction to the field. The relevant hardware and paradigms are then explained thoroughly in this article providing a base for the reader to understand the literature. Later, parallel power system studies are reviewed, reciting the study development from older papers up to the 21st century, emphasizing the most impactful work of the last decade. The studies included system stability studies, state estimation and power system operation, and market optimization. The reviews are split into \ac{CPU} based,\ac{GPU} based, and Cloud-based studies. Finally, the state-of-the-art is discussed, highlighting the issue of standardization and the future of computation in power system studies.    
\end{abstract}

\begin{highlights}
\item Bird's-eye view of high-performance computing adoption in power system optimization studies. 
\item Opportunities and challenges presented in current implementations and literature.
\item Framework suggestion to encourage collaboration and accelerate parallel computing adoption.
\end{highlights}


\makeatletter
\renewcommand\section{\@startsection{section}{1}{\z@}%
    {15pt \@plus 3\p@ \@minus 3\p@}%
    {6\p@}%
    {
     \sectionfont\raggedright\hst[13pt]}}
\renewcommand\subsection{\@startsection{subsection}{2}{\z@}%
    {10pt \@plus 3\p@ \@minus 2\p@}%
    {4.5\p@}%
    {
     \ssectionfont\raggedright }}
\renewcommand\subsubsection{\@startsection{subsubsection}{3}{\z@}%
    {10pt \@plus 1\p@ \@minus .3\p@}%
    {3\p@}%
    {
     \sssectionfont\raggedright}}
     
\makeatother

\maketitle

Word Count = 25543

\tableofcontents

\begin{mdframed}
\section*{Abbreviations}
\begin{multicols}{2}
\begin{acronym}
\small
\acro{MIMD}{Multiple Instruction Multiple Data}
\acro{SIMD}{Single Instruction Multiple Data}
\acro{CPU}{Central Processing Unit}
\acro{GPU}{Graphical Processing Unit}
\acro{RAM}{Random Access Memory}
\acro{ACOPF}{AC Optimal Power Flow}
\acro{DCOPF}{DC Optimal Power Flow}
\acro{ED}{Economic Dispatch}
\acro{OPF}{Optimal Power Flow}
\acro{PF}{Power Flow}
\acro{UC}{Unit Commitment}
\acro{SCUC}{Security Constrained Unit Commitment}
\acro{API}{Application Programming Interfaces}
\acro{FPGA}{Field Programmable Gate Arrays}
\acro{CUDA}{Compute Unified Device Architecture}
\acro{MPI}{Message Passing Interface}
\acro{HPC}{High Performance Computing}
\acro{VO}{Virtual Organization}
\acro{AWS}{Amazon Web Services}
\acro{PVM}{Parallel Virtual Machines}
\acro{PNNL}{Pacific Northwest National Laboratory}
\acro{OS}{Operating System}
\acro{NR}{Newton Raphson}
\acro{IPM}{Interior Point Method}
\acro{FDPF}{Fast-Decoupled Power Flow}
\acro{SOR}{Successive Over-relaxation}
\acro{GMRES}{Generalized Minimal Residual Method}
\acro{SLS}{Sparse Linear Systems}
\acro{SCOPF}{Security Constrained Optimal Power Flow}
\acro{SSA}{System Stability Analysis}
\acro{APP}{Auxiliary Problem Principle}
\acro{ADMM}{Alternating Direction Method of Multiplier}
\acro{MILP}{Mixed Integer Non-Linear Programming}
\acro{NLP}{Non-Linear Programming}
\acro{MINLP}{Mixed Integer Non-Linear Programming}
\acro{PMU}{Phasor Measurement Unit}
\acro{WLS}{Weighted Least Square} 
\acro{TSCOPF}{Transient Stability Constrained OPF}
\acro{CPLP}{Confidentiality-Preserving Linear Programming}
\acro{DSM}{Demand Side Management}
\acro{TSCUC}{Transient Stability Constrained UC}
\acro{ACPF}{AC Power Flow}
\acro{PPF}{Probabilistic Power Flow}
\acro{EMT}{Electromagnetic Transient}
\acro{MC}{Monte Carlo}
\acro{RE}{Renewable Energy}
\acro{DER}{Distributed Energy Sources}
\acro{DCPF}{DC Power Flow}
\acro{SLR}{Surrogate Lagrangian Relaxation}
\acro{ATC}{Analytical Target Cascading}
\acro{SSE}{System State Estimation}
\end{acronym}
\end{multicols}
\end{mdframed}

\newpage

\section{Introduction}

Power systems problems are growing not only in scale but also in complexity. The goals toward grid decarbonization, the changing grid topology, electricity market decentralization, and modernization mean innovations and new elements are continuously introduced to the inventory of factors considered in grid operation and planning. Moreover, With the accelerating technological landscape and policy changes, the number of potential future paths to Net-Zero increases, and finding the optimal transition plan becomes an inconceivable task. 

North America \cite{FACTSHEET_2021}, the EU \cite{CoalExitTimeline} and many other countries \cite{Mapped_2020} set target to completely retire coal plants earlier than 2035 and decarbonize the power system by 2050. In addition, the development of Carbon Capture and Storage Facilities is growing \cite{Carboncapture}. Renewable energy penetration targets are set, with evidence of fast-growing proliferation across the globe, including both transmission connected Variable Renewable Energy (VRE) \cite{Ritchie_Roser_Rosado_2020} and behind the meter distributed resources \cite{Renewableaccelerating}. The demand profile is changing with increased electrification of various industrial sectors \cite{ltdelectrif} and the transportation sector \cite{TrendsEV} building electrification, energy efficiency \cite{GONZALEZTORRES2022626} \cite{B2EIndustry} and the venture into a Sharing Economy \cite{ZHU2021125209}. 

The emerging IoT, facilitated by low latency, low-cost next-generation 5G communication networks, helps the roll-outs of advanced control technologies and Advanced Metering Infrastructure \cite{Frequentlysmart} \cite{BORENIUS2021107367}. This gives more options for contingency remedial operational actions to increase the grid reliability and cost-effectiveness, such as Transmission Switching \cite{9505699}, Demand Response \cite{8663395}, adding more micro-grids, and other Transmission-Distribution co-ordination mechanisms \cite{9215152}. Additionally, they allow lower investment in transmission lines and look for other future planning solutions such as flow management devices and FACTs \cite{9250454}, Distributed Variable Renewable Energy \cite{8931650} and Bulk Energy Storage \cite{GONZALEZROMERO2019154}. 

Moreover, the future grid faces non-parametric uncertainties in the form of new policies such different as carbon taxation, pricing and certifications schemes \cite{CarbonTaxBasics}, feed-in-tariffs \cite{9011131} and time of use tariffs \cite{8970543}, and other incentives. More uncertainties are introduced in smart grid visions and network topological and economic model conceptual transformations. These include the Internet of Energy \cite{SHAHZAD2020106739}, Power-to-X integration \cite{9248892}, the transactive grid through Peer-to-Peer energy trading facilitated by distributed ledgers or Blockchain Energy \cite{9535388}. Such disruptions create access to cheap renewable energy and potential revenue streams for prosumers and encourage load shifting and dynamic pricing. Many of these concepts already have pilot projects in various locations in the world \cite{PowerledgerEnergyProjects}. With all those mentioned above, cost-effective real-time operations, decision-making while maintaining reliability becomes extremely difficult, much less planning the network transition to sustain such a dynamic nature and stochastic future. 

Many current power system operational models are non-convex, mixed-integer, and non-linear programming problems \cite{Conejo2019} and incorporate stochastic framework accounting for weather, load, and contingency planning \cite{CooperWilliamW.2010}. Operators must solve such problems for day-ahead and real-time electricity markets and ensure meeting reliability standards. In NERC, for example, the reliability standards require transmission operators to perform a real-time reliability assessment every 30 minutes \cite{NERC2019}. The computational burden to solve these decision-making problems, even with our current grid topology and standards, forces the recourse to cutting-edge computational technology and high-performance computing strategies for online real-time applications and offline calculations to achieve tractability. 

Operators already use high-performance computation facilities or services. Areas where \ac{HPC} is utilized are Transmission and Generation Investment Planning, Grid Control, Cost Optimization, Losses Reduction, Power System Simulation and Analysis and Control Room Visualization, as seen in \cite{Ma2016}, and \cite{Alam2018}. However, for operational purposes where problems need to be solved on a short time horizon, system models are usually simplified, and heuristic methods are used, relying on the experience of operators such as in \cite{Chen2020}. As a consequence, these models tend to be conservative in fashion, reaching a reliable solution at the expense of reduced efficiency \cite{E.M.Constantinescu2011}. According to The Federal Energy Regulatory Commission, finding more accurate solution techniques for highly complex problems such as \ac{ACOPF} could potentially save billions of dollars \cite{FERC2013}. This motivates the search for methods to produce high-quality solutions in a reasonable time. Finding appropriate techniques, formulations, and proper parallel implementation on \ac{HPC}s for power system studies has been a research area of interest. Progress has been made to make solving complex, accurate power system models for real-time decisions favorable. 

The first work loosely related to parallelism on a high-level task in a power system might have been by Harvey H. Happ in 1969 \cite{Happ1969} where a hierarchical decentralized multi-computer configuration was suggested targeting \ac{UC} and \ac{ED}. Other similar work in multi-level multi-computer frameworks followed soon targeting Security and voltage control in the 70s \cite{Ewart1971,Narita1973,Wu1976} .P. M. Anderson from the Electrical Power Research Institute created a special report in 1977, collecting various studies and simulations performed at the time exploring the potential applications for power system analysis on parallel processing machines \cite{P.M.Anderson}. Also, several papers came out suggesting new hardware to accommodate power system-related calculations \cite{Podmore1982}. 

In 1992, C. Tylasky et al. made what might be the first comprehensive review about the state of the art of parallel processing for power systems studies \cite{Tylavsky1992}. It discussed challenges that are still relevant today, such as different machine architectures and transparency and portability of the codes used. A few parallel power system study reviews have been conducted throughout the development of computational hardware. Some had similar goals to this paper reviewing \ac{HPC} applications for power systems \cite{Bose1993} \cite{Falcao1996} and on distributed computing for online real-time applications \cite{Ramesh1996}. Computational paradigms changed exponentially, reducing those reviews to pieces of history. The latest relevant, comprehensive reviews in the topic were by R. C. Green et al. for general power system applications \cite{Ii2011}, and again focused on innovative smart grid applications \cite{Green2013}. These two handle a variety of topics in power systems. 

This work adds to existing reviews a fresh overview of state of the art. It distinguishes itself by providing the full context and history of parallel computation and \ac{HPC} in power system optimization and its development up to the latest work in the field. Moreover, it focuses on deterministic equation solving and mathematical optimization problems ignoring metaheuristics or machine learning solution methods. It also provides a thorough base and background for newcomers to the field of power system optimization in terms of both computational paradigms and applied algorithms. Finally, it brings to light the necessity of integrating \ac{HPC} in future studies amidst the energy transition and suggests a framework that encourages future collaboration to accelerate \ac{HPC} deployment.

This paper is organized as follows: Sections 2 \& 3  Identify the main \ac{HPC} components and parallel computation paradigms. The next six sections review parallel algorithms under both \ac{MIMD} and \ac{SIMD} paradigms split into their early development and state of the art for each study, starting with Section 4 \ac{PF}. Section 5 \ac{OPF}. Section 6  \ac{UC}. Section 7 Power System Stability Studies. Section 8 \ac{SSE}. Section 9 Reviews Unique Formulations \& Studies. Section 10 Reviews Gird and Cloud Computation application studies. In these studies, novel approaches and algorithms and modifications to existing complex models are made, parallelizing them to achieve tractability or reducing their processing time below that needed for real-time applications. Many of the studies showed promising outcomes and made a strong case for further opportunities in using complex system models on \ac{HPC} and Cloud for operational applications. Section 11 highlights and discusses the present challenges in the studies and re-projects the future of \ac{HPC} in power systems and energy markets and recommends a framework for future studies. Section 12 concludes the review.  

This review does not include machine learning or meta-heuristic parallel applications such as particle swarm and genetic algorithms. Furthermore, while it does include some studies related to co-optimization of Transmission and Distribution systems, the study excludes parallel analysis and simulations of distribution systems. It is also important to note that this review does not include Transmission and generation system planning problems or models related to grid transitioning because of the lack of parallel or \ac{HPC} application in the literature on such models, which is addressed in the discussion section. 

\section{Parallel Hardware}

Parallel computation involves breaking down a task into several ones and distributing the separate tasks to multiple workers to be performed simultaneously with some coordination mechanism and communication in place. Depending on the nature of the code or algorithm, parallel programs can be performed on a single processing unit or a single computer with multiple processors, given that the computer falls under a parallel system classification. As long as multiple workers execute tasks simultaneously, there is parallelism. A worker here is a loose term, and pieces of hardware such as a Multi-core processor can be considered as a worker containing workers or its own (the cores). This depends on the implemented parallel algorithm and how it is defined. 

\begin{figure}[ht]
    \centering
    \includegraphics[width=0.4\textwidth]{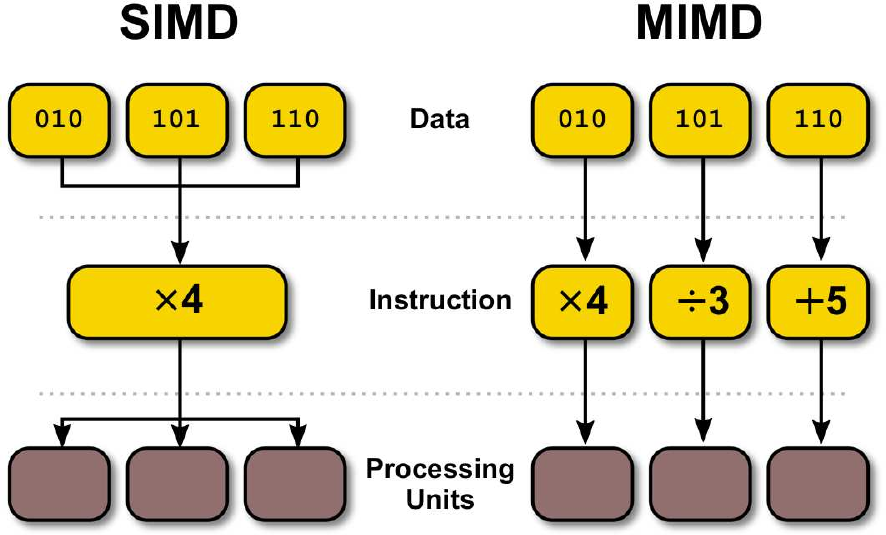}
    \caption{Data operation in \ac{MIMD} and \ac{SIMD} architecture}
   \label{fig1}
\end{figure}

Predominantly, parallel machines can be placed under two categories based on The Von Neumann architecture \cite{TAN2019727}, \ac{MIMD} and \ac{SIMD} machines. \ac{MIMD} includes any machines where multiple heterogeneous instructions or operations can be carried on the pool of available data. A multi-core \ac{CPU} is the main example, as each core has the functionality to perform full instructions. However, the obvious one is a multi-node computer cluster since it contains multiple \ac{CPU}s. \ac{SIMD} architecture, on the other hand, performs the same operation on multiple data points as the name implies. Fig. \ref{fig1} demonstrates the abstract idea behind the two. 

The versatility of \ac{MIMD} illustrated in Fig. \ref{fig1} is traded off with simplicity in \ac{SIMD}. Also, note that instructions are not limited to arithmetics and can take different forms (logical, transfer, etc.). The first-ever general-purpose programmable microprocessor was made in the 70s (intel 4004), and more powerful intel microprocessors were developed later \cite{ChipFame_2018}. These processors were too costly to scale and create \ac{MIMD} supercomputers. That changed, however, in the 90s as the drop in price-to-performance ratio of general-purpose microprocessors meant they superseded vector processors in supercomputers \cite{Stringer_2016}. Transputers and microprocessors designed specifically for aggregation and scalability started dominating parallel computing systems as well \cite{Transputers1990}. This shift from \ac{SIMD} to \ac{MIMD} can also be observed in power system studies as more \ac{MIMD} suitable algorithms were being used. More ``grain" started to be added to algorithms, where entire subproblems are sent to each processor, even though \ac{MIMD} machines at the time did not have a large number of processors. This fact is reflected in the earlier studies mentioned in this paper, which weren't vastly broken down. 

\subsection{CPUs} 

\begin{figure}[ht]
    \centering
    \includegraphics[width=0.4\textwidth]{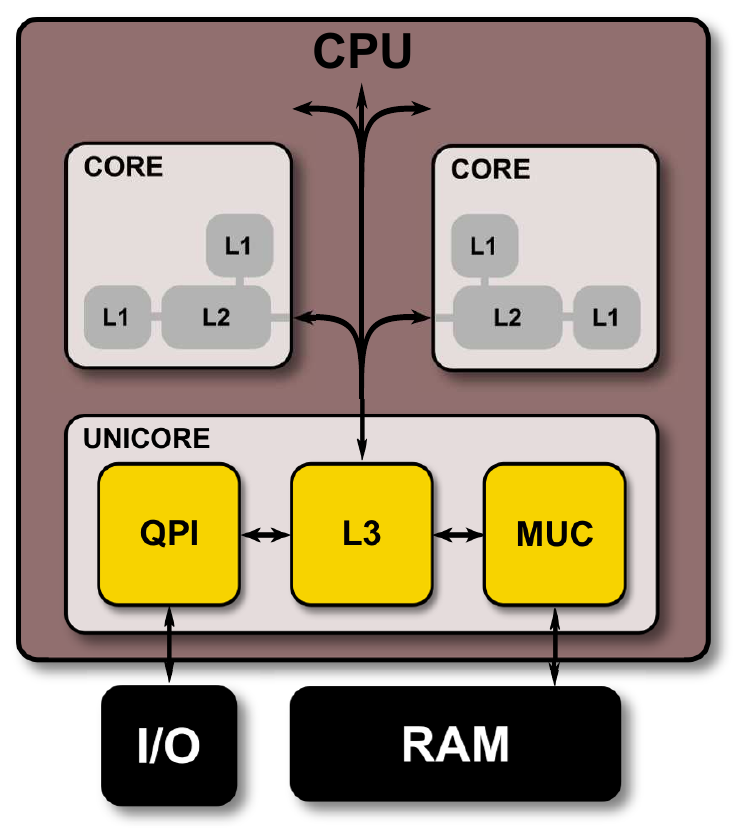}
    \caption{Architecture Illustration inspired by Intel Xeon series \cite{Productxeon} showing the interconnection of Quick Path Interconnect (QPI), Memory Control Unit (MCU), with caches and RAM}
   \label{fig2}
\end{figure}

\ac{CPU}s were initially optimized for scalar programming and executing complex logical tasks effectively. They kept developing as sequential task processors up until 2005 when their development hit a power wall \cite{Bose2011}. Since then, they have relied on multiple cores to increase their performance. \ac{CPU}s of the modern era function as miniature superscalar computers containing multiple cores capable of complex logical operations, enabling pipelining, task-level parallelism and multi-threading \cite{IntelProc}. They contain an extra \ac{SIMD} layer that supports data-level parallelism, vectorization, and vector processing from a typical eight 64-bit register. They can contain up to 32, 512-bit vector registers and units such as Fused-Multiply Add that combine addition and multiplication in one step \cite{vectorization2021} \cite{intel512}. Furthermore, \ac{CPU}s use a hierarchy of memory and caches that allow the execution of sequences of instructions on multiple data points very quickly; this hierarchy can be observed in the illustration in Fig. \ref{fig2}.

Fig. \ref{fig2} shows caches from high-speed, low-capacity registers that are closest to the \ac{CPU} cores to lower-speed, higher-capacity caches descending from L1 to L2 to L3 caches that save data for the processors. They give the \ac{CPU} a distinct functional advantage over \ac{GPU}s in that data is readily accessible, allowing the complex operations without fetching data from main memory. 

\ac{CPU}s employ a variety self optimizing techniques such as ``Speculation", ``Hyperthreading or ``Simultaneous Multi-threading", ``Auto vectorization" and ``task dependency detection" \cite{IntelProc}. The emphasis on those features and low latency make modern \ac{CPU}s very efficient in enhancing the performance of multiple general complex tasks and functions without much interference or specific instructions. Workstation \ac{CPU}s can have up to 16 processing cores, and server-level \ac{CPU}s can have up to 128 cores in certain products \cite{altra2021}. \\

Often, any software that utilizes Multi-threading uses a portable \ac{API} such as Cilk or OpenMP, which allows function and loop level parallelism to be achieved. However, the true value of scaling parallel algorithms today is achieved by using several \ac{CPU}s in multi-processing to solve massive decomposed problems, which is facilitated by \ac{API}s such as \ac{MPI}. Multi-threading is usually embedded at the process level in the compiled code. E.g., when the distributed subproblems of a decomposed optimization problem are solved on every \ac{CPU} using a general solver which applies multi-threading techniques. 

\subsection{GPUs}

\begin{figure}[ht]
    \centering
    \includegraphics[width=0.7\textwidth]{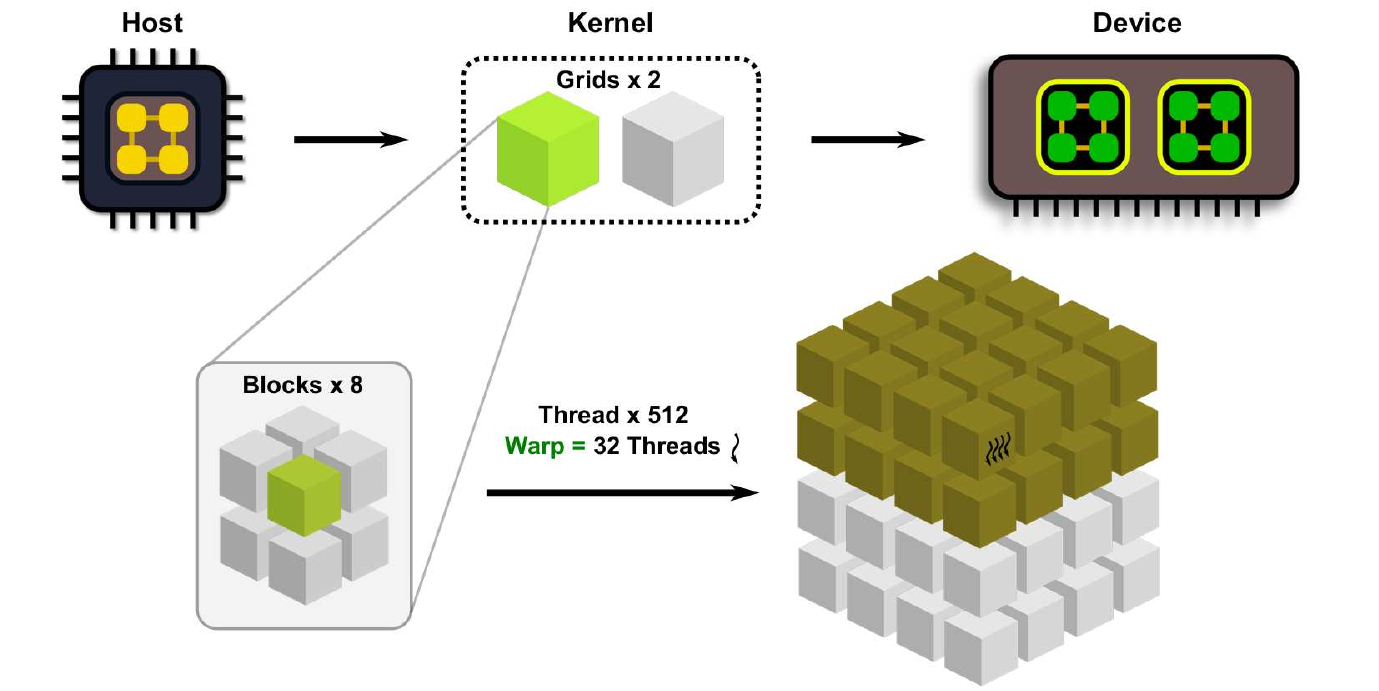}
    \caption{\ac{GPU} instruction cycle and CUDA abstractions}
   \label{fig3}
\end{figure}

\ac{GPU}s function very similarly to Vector Processing Units or Array processors, which used to dominate supercomputer design. They are additional components to a ``Host" machine that sends kernels which is essentially the \ac{CPU}. \ac{GPU}s were originally designed to render 3D graphics. They are especially good at vector calculations. The representation of 3D graphics has a ``grid" nature and requires the same process for a vast number of input data points. This execution has been extended to many applications in scientific computing and machine learning, solving massive symmetrical problems or performing symmetrical tasks. \ac{GPU}s have many streaming processors with a bank of registers, low latency shared memory, warp scheduler, and many cores and multi-processors. Each contains an instruction cache and access to global memory. It also contains ALUs capable of vectorizing basic mathematical operations, including trigonometric functions. Unlike \ac{CPU}s, achieving efficiency in \ac{GPU}s parallelism is a more tedious task due to the fine-grained \ac{SIMD} nature and rigid architecture. Fig. \ref{fig3} shows a simple breakdown of \ac{GPU} instruction cycle. 

The \ac{GPU} (Device) interfaces with the \ac{CPU} (Host) through PCI express bus from which it receives instruction ``Kernels". In each cycle, a Kernel function is sent and processed by vast amounts of \ac{GPU} threads with limited communication between them. Thus, the symmetry of the parallelized task is a requisite, and the number of parallel threads has to be of a specific multiple factor to avoid the sequential execution of tasks. Specifically, they need to be executed in multiples of 32 threads (a warp) and multiples of 2 streaming processors per block for the highest efficiencies. 

\ac{GPU}s can be programmed in C or C++. However, many \ac{API}s exist to program \ac{GPU}s such as OpenCL, HIP, C++ AMP, DirectCompute, and OpenACC. These \ac{API}s provide high-level functions, instructions, and hardware abstractions, making \ac{GPU} utilization more accessible. The most relevant interface is the \ac{CUDA} by NVIDIA since it dominates the \ac{GPU} market in desktop and \ac{HPC}/Cloud \cite{Forbes2021}. \ac{CUDA}s libraries make NVIDIAs \ac{GPU}'s power much more accessible to the scientific and engineering communities. CUBLAS, for example, is a basic linear algebra subroutine and CUSPARSE is a sparse matrix operation library. Both alleviate the burden of fine-tuning and scheduling these operations for \ac{GPU}. Libraries of several programming languages, such as scikit-learn for python, already use \ac{CUDA} in some of their functions and packages (scikit-CUDA). \ac{CUDA} also includes a Unified Shader pipeline that allows on-chip ALU with a program to perform general-purpose computation. \ac{CUDA} moved the limitation of using \ac{GPU}s for 3D graphics to general scientific computing such as fluid dynamics and financial modeling. 

Power system network nodes have very few interconnections, resulting in a sparse matrix system. That is a little problematic for \ac{GPU} processing as they are designed for dense systems associated with graphics processing. \ac{GPU}s use registries and cache memory, but those caches are not tailored to reduce latency and are not adjacent to cores. Moreover, \ac{GPU}'s different architecture may cause discrepancy and lower accuracy in results as floating points are often rounded in a different manner and precision than in \ac{CPU}s \cite{navarro2014}. Nevertheless, these challenges can be worked around with \ac{CUDA} and sparse techniques that reduce the number of ALUs required to achieve a massive speedup. Finally, GPUs can offer a huge advantage over CPUs in terms of energy efficiency and cost if their resources were used effectively and appropriately. 

\subsection{Other Hardware}

There are two more notable parallel devices to mention. One is the \ac{FPGA}. This chip consists of configurable logic blocks, allowing the complete user flexibility in programming the internal hardware and architecture of the chip itself. They are attractive as they are parallel, and their logic can be optimized for desired parallel processes. However, they consume a considerable amount of power compared to other devices, such as the Advanced RISC Machine. Those are processors that consume very little energy due to their reduced instruction set, making them suitable for portable devices and applications \cite{ARMLtd}.

\section{Aggregation and Paradigms}

In the late 70s, project ARPANET took place \cite{ARPANET} UNIX was developed \cite{UNIX}, and advancement in networking and communication hardware was achieved. The first commercial LAN clustering system/adaptor, ARCNET, was released in 1977 \cite{InternetTimeline_2016} and hardware abstraction sprung in the form of virtual memory, such as OpenVM, which was adopted by operating systems and supercomputers \cite{9388639}. Around that same time, the concept of computer clusters was forming. Many research facilities and customers of commercial supercomputers started developing their in-house clusters of more than one supercomputer, which leads us to \ac{HPC}.  

\begin{figure}[ht]
    \centering
    \includegraphics[width=0.8\textwidth]{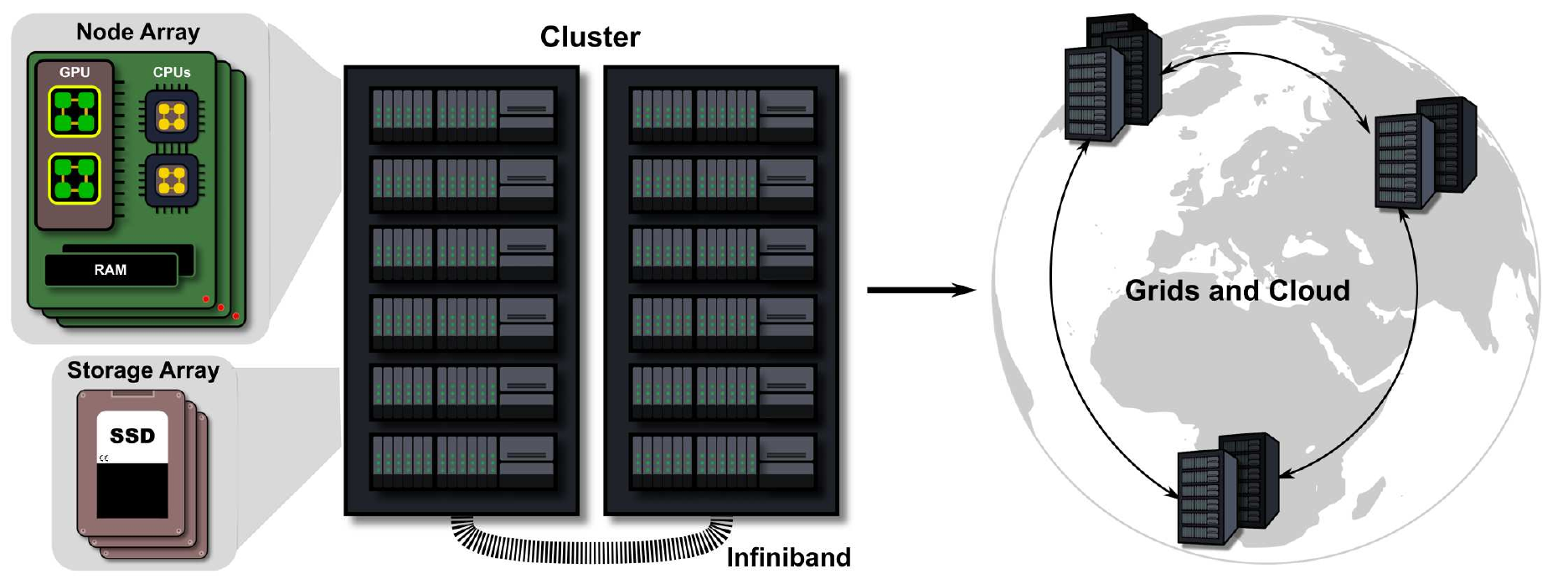}
    \caption{Powerful chips to Cloud computing}
   \label{fig4}
\end{figure}

HPC facilities are a network of high-performing aggregate computational hardware. They enable vast parallelism of computation, leading to higher computational capacity and speed, and their true potential lies in the distributed nature of hardware and memory. Fig. \ref{fig4} shows how aggregations starts with components on a compute node (\ac{CPU}, \ac{RAM} \& \ac{GPU}). Several nodes and big data storage are connected, forming a cluster through InfiniBand, and finally, clusters are connected to geographically separate clusters through the internet, forming a Grid and Cloud. The communication between processes through aggregate hardware is aided by high-level software such as \ac{MPI} available in various implementations and packages such as Mpi4py in python. Higher-level interfaces exist, such as Apache (Hadoop Airflow and Spark), Slurm, and mrjob, to aid in data management, job scheduling, and other routines. 

Specific clusters might be designed or equipped with components geared more towards specific computing needs or paradigms. \ac{HPC} usually includes tasks with rigid time constraints (minutes to Days, maybe weeks) that require a large amount of computation. The High Throughput Computing (HTC) paradigm, which has been introduced in 1996 \cite{AlanBeck1997}, involves tasks that require a large amount of computation, taking a very long time (months to years). An example would be chip simulations done by chip manufacturers or high-energy physics applications. Many Task Computing (MTC) paradigm involves computing various distinct \ac{HPC} tasks and revolves around the management of applications that require a high level of communication and data management and are not entirely parallel \cite{4777912}. The aggregation of heterogeneous clusters and supercomputers to form Grid or Cloud facilities provides users or participants the flexibility to perform tasks under any previously mentioned paradigms.

\subsection{Grid Computing}

The 90s was when the World Wide Web and the information age spurred, connecting computers worldwide. This set off the trend of wide-area distributed computing and ``Grid Computing". Ian Foster coined the term with Carl Kesselman and Steve Tuecke, who developed the Globus toolkit that provides grid computing solutions \cite{globus_2021}. In Grid Computing, massive heterogeneous aggregate computing resources are shared and utilized on demand by many organizations and coalitions such as NASA 3-EGEE, Open Science Grid, the European Grid Infrastructure I-Way, and EGI in Europe, Worldwide LHC Computing Grid, and WestGrid Canada. 

The computational resources of Grid computing facilities are distributed geographically and are not subject to a central controller or owner. The resources are managed and coordinated through \ac{VO}, which can be institutions or teams of individuals with shared competencies and interests. Namely, those are experts in ``Metacomputing," a field that emerged with grid computing. The scope, physical boundaries of control, and responsibility of \ac{VO}s are malleable and change over clusters.

Hardware virtualization and interactive applications are rare in Grids and usually discouraged by grid organizations for reliability/security reasons. Because grid computing is mainly adopted by research and scientific organizations, its infrastructure tends to be very secure and holds a private network of users sponsored by their organizations. The provenance of workflow and performance monitoring are also features of grids for scientific reproducibility. Computation services are usually executed on a batch job basis with queues and wait times, charging users or their groups by the \ac{CPU} hours or equivalent. Specific credit of \ac{CPU} hours is given for every user/group of users. The extent to which the credit is exhausted by a user/group plays a role in determining a requested job's priority and position in the job queue. Grids computing is suitable for sensitive closed-ended, and non-urgent applications as resources are not readily available on demand.

\subsection{Cloud Computing}

Cloud computing is essentially the commercialization and effective scaling of Grid Computing driven by demand, and it is all about the scalability of computational resources for the masses. The existence of grid computing, alongside the cheapening and development of high-speed networks, Solid State Drive storage,  high-speed internet, and multi-processors such as the IBM POWER4 and Intel core duo, paved the way for Cloud Computing in the mid-2000s. It mainly started with Amazon's demand for computational resources for its e-commerce activities, which precipitated amazon to start the first successful infrastructure as a service-providing platform with Elastic Compute Cloud \cite{EC2amazon} for other businesses that conduct similar activities. Later on, amazon also started hosting Platform-as-Service and Software-as-Service.

The distinct abstractions and concepts that define cloud computing are an implication of the business model for Grid Computing. Cloud computing is way more flexible and versatile than Grid in accommodating different customers and applications. Cloud computing relies heavily on virtualization (both Software and machine virtualization), allowing the sharing of resources and using resources for native interactive applications, hence easy commercialization. It makes Cloud inherently less secure, less efficient in performance than Grid, and a bit more challenging to manage, yet way more scalable, on-demand, and overall more resource-efficient. Users can customize or start pre-defined "Instances" of virtual hardware and are charged by the hour, for instance, running time, regardless of the actual usage of resources. Cloud achieves a delicate balance between efficiency and the cost of computation. 

Today, \ac{AWS}, Microsoft Azure, Oracle Cloud, Google Cloud, and many other cloud commercial services provide massive computational resources for online-based companies such as Netflix, LinkedIn, Facebook, application services such as Adobe, media companies such as BBC, ESPN, energy industry such as GE and Shell, tech Companies such as Apple and IBM, real estate and hospitality such as AirBnB, banks and financial services such as HSBC and Paypal, medical companies such as Jhonson and Jhonson and intelligence and defense such as NSA. The thirst for computational resources is ever-increasing in all sectors, and indeed, it is also growing in the electrical power sector. It only makes sense that the energy and electrical power industries are following the same path due to the widely massive adoption of \ac{HPC} and cloud services, especially in sensitive sectors such as financial and defense.

\subsubsection{Virtualization}

In 1984, the creators of OpenVM developed the VMScluster, which comprises several OpenVMS machines. The appearance of virtualization caused a considerable leap in massive parallel computing, especially after the software tool \ac{PVM} \cite{Geist1992} was created in 1989. It bridged a huge software gap that allowed heterogeneous hardware to be clustered together. Since then, tens and hundreds of virtualization platforms have been developed and are used today on the smallest devices with processing power \cite{UTM}. 
While virtualization is heavily adopted in Cloud, it is not strict to it. It can be implemented even on a single processor computer. Virtualization allows the sharing of resources in a pool. For example, one can emulate four fully functioning virtual laptops on the hardware of a single laptop. The only catch is that each virtual laptop cannot fully exhaust its ``promised" available resource simultaneously. 

Another example would be virtualizing a computer or machine (e.g., gaming console) on a completely different one (laptop). The underlying machine has the hardware capability of emulating all the processes of the virtualized one. Therefore, less hardware can be allocated or invested in Cloud computing for a more extensive user base. Often, the percentage of hardware being used is low compared to the requested hardware. Idle hardware is reallocated to other user processes that need it. The instances initiated by users float on the hardware like clouds shifting and moving, shrinking and expanding depending on the actual need of the process.

\subsubsection{Containers}

\begin{figure}[ht]
    \centering
    \includegraphics[width=0.4\textwidth]{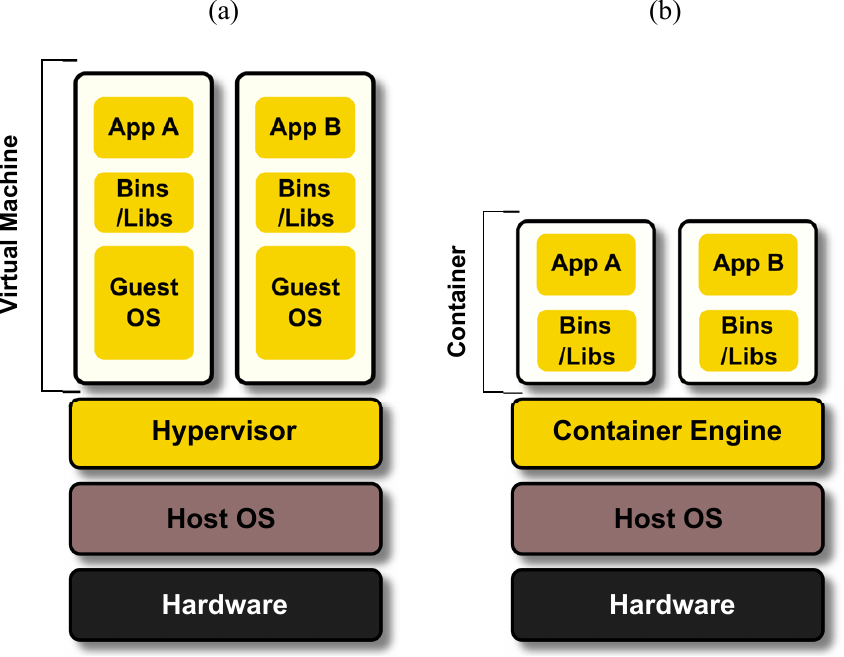}
    \caption{A comparison between Virtual Machine layers (a), Container Layers (b) }
   \label{fig5}
\end{figure}

While virtualization makes hardware processes portable, containers make software portable. Developing applications, software, or programs in containers allows them to be used on any \ac{OS} as long as it supports container engines. That means one can develop a Linux-based software  (E.g., that works on Ubuntu 20.04) in a container and run that same application on a machine with Windows \ac{OS} or iOS installed. Fig. \ref{fig5}  compares virtual machines with container layers. The software and libraries needed to support the application sit on the container engine installed on the Host computers \ac{OS}, which provides the needed compatibility by recognizing the application's native \ac{OS} in the container. In contrast, virtualization requires an additional layer of \ac{OS} to make to run the application.

Containers can also run on bare metal, which removes latency and reduces dependency and the complexity of application and software development. Containers are a big deal because they provide complete flexibility and interchangeability for service-based applications that utilize different high-performance computing facilities. An application can be developed once using containers and used by multiple clients on their on-premise cluster or a cloud service of their choice. Docker and Apptainer (formerly known as singularity) are commonly used containerization engines in Cloud and Grid receptively \cite{apptainer} \cite{Docker_2022}.

\subsubsection{Fog Computing}

Cloudlets, edge nodes, and edge computing are all related to an emerging IoT trend, Fog Computing. Fogs are computed nodes associated with a cloud that are geographically closer to the user end or control devices. Fogs mediate between extensive data or cloud computing centers and users. This topology aims to achieve data locality, giving several advantages such as low latency, higher efficiency, and decentralized computation. The fog computing paradigm has the disadvantages of low computational power, communication infrastructure requirement, and security risks.

\subsection{Volunteer Computing}

Volunteer computing is an interesting distributed computing model that originated in 1996 by Great Internet Mersenne Prime Search \cite{gimps} allowing individuals connected to the internet to donate their personal computer idle resources for a scientific research computing task. Volunteer computing remains active today with many users and various middleware and projects, both scientific and non-scientific, primarily based on BOINC \cite{boinc}, and in commercial services such as macOS Server Resources, \cite{appleserver}.

\subsection{Granularity}

Superficially,  the terms ``Fine-Grained" and ``Coarse-grained" algorithms describe how parts of a sequential code can be parallelized. However, the line which separates a fine-grained from a coarse one is blurry so is the convention of using the terms, making the associations below relevant. Fine-grained parallelism appears in algorithms that frequently repeat a simple homogeneous operation on a vast data set. They are often associated with embarrassingly parallel problems. The problems can be divided into many highly, if not wholly symmetrical simple tasks, providing high throughput. Fine-grained algorithms are also often associated with multi-threading and shared memory resources. 

Coarse-grained algorithms imply moderate or low task parallelism that sometimes involves heterogeneous operations. Today, coarse-grained algorithms are almost synonymous with Multi-Processing, where the algorithms use distributed memory resources to divide tasks onto different processors or logical \ac{CPU} cores. They can embed fine-grained algorithms, for example, at the solution task of an optimization problem, where the repetitive matrix and vector operations are performed. In decomposed problems, commercial solvers are often used to find the solution to each sub-problem. The solvers themselves may and often have embedded multi-threading parallelism. The steps of the algorithms used by the solver are also broken down and divided between threads to speed up the solution process. 

\subsection{Centralized vs Decentralized}

Centralized algorithms refer to problems with a single objective function, solved by a single processor, with data stored at a single location. When a centralized problem is decomposed into N subproblems, sent to N number of processors to be solved, and retrieved by the central controller to update variables, re-iterate, and verify convergence, the algorithm becomes a ``Distributed" algorithm. The terms distributed and decentralized are often used interchangeably and are often confused in the literature. There is an important distinction to make between them. A Decentralized Algorithm is where the decomposed subproblems do not have a central coordinator or a master problem. Instead, the processes responsible for the subproblems communicate with neighboring processes to reach a solution consensus. (several local subproblems with coupling variables where subproblems communicate without a central coordinator.) 

For small-scale problems, centralized algorithms are preferred, and distributed algorithms are not helpful mainly due to bottlenecks of communication overhead. In large-scale complex problems with many variables and constraints, distributed algorithms outperform centralized algorithms. The speedup keeps growing with the problem size if the problem has ``strong scalability." Distributed algorithms subproblems share many global variables. It means a higher communication frequency as all the variables need to be communicated back and forth to the central coordinator.
Moreover, In some real-life problems, central coordination of distributed computation might not be possible. For example, suppose a Multi-area \ac{ED} problem is decomposed to be solved by each power pool of independent system operators from different interconnected jurisdictions or countries separately. A central coordinator might not exist in that case, and it might be hard to assign such authority. 

\begin{figure}[ht]
\includegraphics[width=12cm]{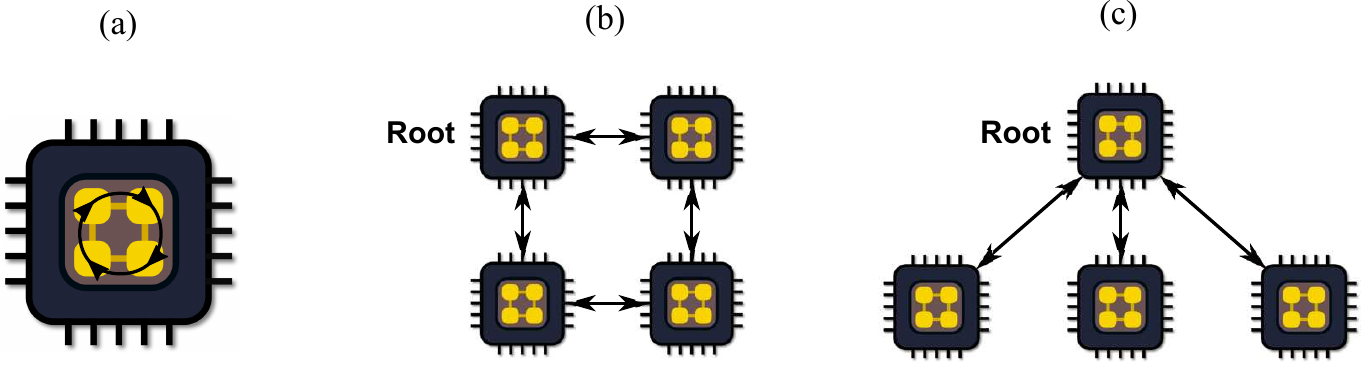}
\centering
\caption{Information exchange between processors in different parallel schemes, Centralized (a), Decentralized (b) Distributed (c)}
\label{fig6}
\end{figure}

Fully decentralized algorithms might solve this problem as their processes communicate laterally, and only neighboring processes have shared variables. They remove the need for a central coordinator, which helps achieve lower latency and fewer iterations since there are fewer global variables than distributed algorithms. Fig. \ref{fig6} illustrates the three schemes. 

The arrows point in the direction of communication. In a centralized case displaying a single general-purpose processor, parallelism could still exist in the form of multi-threading. However, all three paradigms can be executed on a single processor, limited by the number of possible threads.

\subsection{Synchronous vs Asynchronous}

Synchronous algorithms are ones where the parallelism steps are synchronized, and the overall sequential process does not move forward until all the parallel tasks are executed. Synchronous algorithms are more accurate and efficient if all the tasks have similar complexity and require similar execution times, which is often the case when the tasks and data are symmetrical. However, that is usually not the case in power system optimization studies. In decomposed power system problems, the subproblems of different regions or periods (depending on the problem and its decomposition) might be harder to solve than others. The computational power in certain regions might be more complex than in others. Suppose computation is geographically distributed, such as in fog computing. In that case, the difference in distance between regions creates different communication delays, which adds to the level of stochasticity of subproblem task time. In these cases, synchronous algorithms tend to be relatively inefficient. Even if almost all tasks are completed, a single delayed task can hold the process idle in the parallel step. 

Asynchronous algorithms specifically target this challenge as they allow the algorithm to proceed and idling workers take new tasks even if not all the adjacent processes are complete. Using asynchronous algorithms might be helpful, very efficient, and might outperform synchronous algorithms in many real-life cases with asymmetrical sub-problems. However, when the information processed in the parallel tasks is dependent on each other, such as in distributed and iterative algorithms, asynchronous algorithms compromise solution accuracy and cause an increase in the number of iterations. To achieve better results, often, the algorithm needs to ensure "Formation" is in place in the Asynchronous decomposition, meaning that while subproblems might have a deviation in the direction of convergence, they should keep a global tendency toward the solution, much like bird flocks heading in the same direction.

\subsection{Problem Splitting and Task Scheduling}

The above subsection speaks closely to and falls under the issue of problem splitting and parallel task scheduling. In Shared Memory parallelism or Multi-threading, synchronization is required to avoid ``Race Conditions" that cause numerical errors as multiple threads try to access the exact memory location or variable simultaneously. Hence synchronization does not necessarily imply that processes will execute every instruction simultaneously but rather in a coordinated manner. 

\begin{figure*}[ht]
\includegraphics[width=12cm]{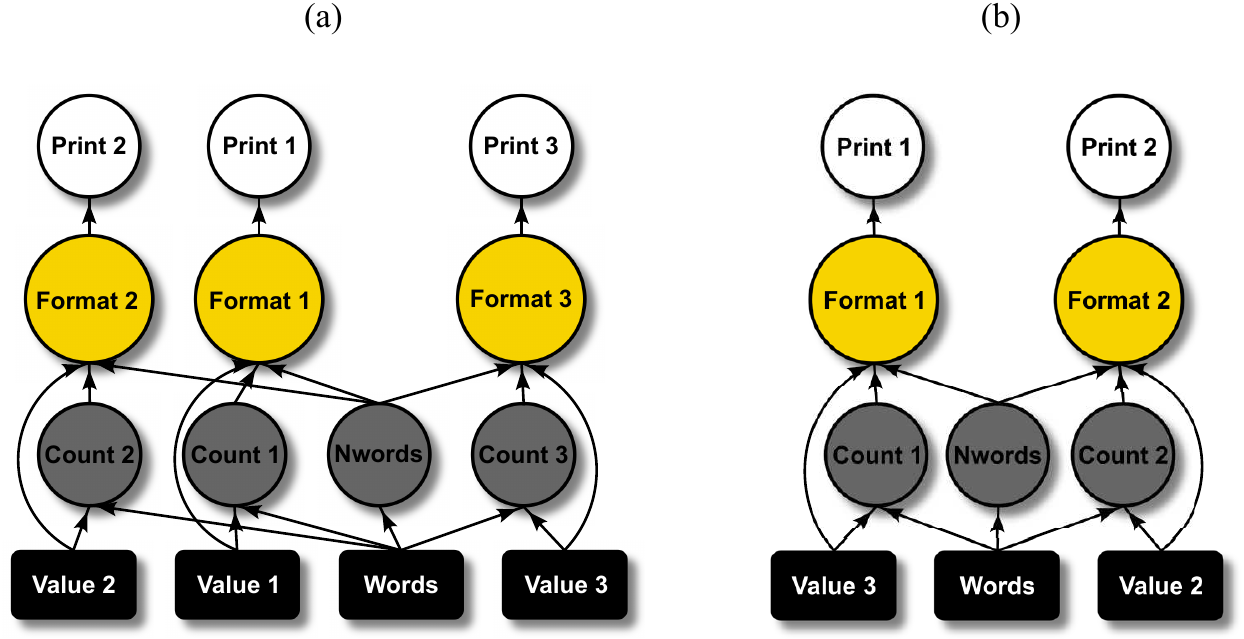}
\centering
\caption{Typical task graph generated and optimized by DASK \cite{DASKgraph}. The original schedule (a) the optimized schedule (b). }
\label{fig7}
\end{figure*}

Coordination mechanisms involve pipe-lining or task overlapping, which can increase efficiency and reduce the latency of parallel performance. For example, for synchronous algorithms with asymmetrical tasks, sub-tasks that take the longest time to process can be shared with the idle, less burdened workers if no dependencies or task interactions prevent such reallocation. Dependency analysis is occasionally carried out when splitting a problem or a task by structure or decoupling a system. For example, power system network models naturally produce block-diagonal-bordered sparse matrices, which can be factorized using LU factorization in order to parallelize their solution. In order to do that, however, dependencies must be found in an ordering step, where graph theory and Diakoptics are often used to highlight those dependencies and solve the independent blocks in parallel completely. In an elaborate parallel framework such as in multi-domain simulations or smart grid applications, task scheduling becomes its own complex optimization problem, often solved heuristically. However, there exist packages such as DASK \cite{DASKgraph} which can help in optimal task planning and parallel task scheduling as shown in Fig. \ref{fig7}.

DASK is a python library for parallel computing which allows easy task planning and parallel task scheduling optimization. The boxes in Fig. \ref{fig7} represent data, and circles represent operations. 
The management of tasks and process manipulation with schemes such as blocking and non-blocking communication that exits in \ac{MPI} and other parallel computing \ac{API}s allow enormous versatility in task scheduling and allocation. With a cleverly written sequential code, one can create an extremely efficient parallel code squeezing every bit of the available resources. 

\subsection{Parallel Performance Metrics}

The main driver behind using parallel computation is to significantly reduce solution time, sometimes to speed up an already simple problem, and other times to merely achieve tractability for a complex problem. In either case, both solution time and accuracy are sought as measures of the success of the parallel algorithm. 

According to the Amhals law of strong scaling, there is an upper limit to the speedup achieved for a fixed size problem. Dividing a specific fixed size problem into more subproblems does not result in a speedup linearly. However, if the parallel portion of the algorithm increases, then proportionally increasing the subproblems or the number of processors could continuously increase the speedup according to Gustafson's Law of strong scaling. The good news is that Gustafson's law applies to large decomposed power system problems. 

There are three types of metrics most frequently used in the literature. One is the speedup, which is the ratio of the serial algorithm latency to the parallel algorithm latency. The second is Efficiency, which is the speedup ratio to the number of processors used. The last one is the scalability, the ratio of the parallel algorithm latency on a single core to the parallel algorithm latency over N cores.  

This speedup metric is often used unfairly. For example, in decomposition algorithms, solving subproblems is parallelized, then the performance comparison is drawn against the same serial algorithm (i.e., the scalability equation used but referred to as speedup). The parallel algorithm must be compared to the best serial algorithm that can achieve the same task to have a truly fair comparison. This alludes to the fact that there are several approaches for the speedup metric, which should be discussed in parallel algorithm studies for higher transparency as done in \cite{Wu1995}. 

A Linear Speedup is considered optimal, while sublinear speedup is the norm because there is always a serial portion in a parallel code. However, there are cases where superlinear speedup is achieved in some studies. This is often due to a playfield change and deep modifications in algorithms' features, leading to unfair comparisons. For example, when the parallel algorithm's cache memory usage is optimized or distributed memory is used, allowing faster access to memory than serial processes. 

Energy efficiency in parallel coding is also an important factor, especially for users who use in-house \ac{HPC} facilities. Using multi-threading causes the energy consumption of multi-processors to increase drastically \cite{parallel02}. So in the case of resource abundance, it might make more sense to multi-process rather than multi-thread. 

Both sequential and parallel programs are vulnerable to major random errors caused by the Cosmic Ray Effect \cite{ADAMS198517} which has been known to cause terrestrial computational problems\cite{Flight72}. Soft errors might be of some concern regarding real-time power system applications. However, In parallel programming, especially in multi-processing, reordering floating-point computations is unavoidable; thus, a tiny deviation in accuracy from the sequential counterpart is expected and should be tolerated given that the speedup achieved justifies it.

When creating a parallel algorithm, emphasis on the quality of the serial portion of the algorithm must be taken, which is often not done. A parallel algorithm, after all, is contained and executed by a serial code. Thus, it is vital to have an excellent serial code focusing on data locality and a good parallel strategy.  

All the computing paradigms mentioned above are points to tweak and consider when creating and applying any parallel algorithm. Often compromises between speedup and accuracy are made; the final choice of parameters, in the end, depends on the need for the algorithm.

\section{Power Flow}

PF studies are central to all power system studies involving network constraints. The principal goal of \ac{PF} is to solve the network's bus voltages and angles to calculate the flows of the network. For some applications, \ac{PF} is solved using DC power flow equations, which are approximations based on realistic assumptions. Solving these equations is easy and relatively fast and results in an excellent approximation of the network \ac{PF} \cite{Conejo2019}. On the other hand, to obtain an accurate solution, non-linear full AC Power flow equations need to be solved, and these require numerical methods of approximation. The most popular ones in power system analysis are the \ac{NR} method and the \ac{IPM} \cite{FERC2013}. However, using these methods is computationally expensive and too slow for real-time applications, making them a target for parallel execution.  

\subsection{MIMD Based Studies}

Parallelism in \ac{PF} in some earlier studies was achieved by restructuring the hardware to suit the algorithm. In what might be the first parallel power flow implementation, Taoka H. et al. designed their own multi-processor system around the Gauss Iterative method in 1981, such that each i8085 processor in the system would solve the power flow for each bus \cite{Taoka1981}. Instead of modifying the algorithm, the hardware itself was tailored to it, achieving a linear speedup in comparison to the sequential implementation in \cite{Ward1956}. Similarly, a year after, S. Chan and J. Faizo \cite{Chan1982} reprogrammed the CDC supercomputer to solve the \ac{FDPF} algorithm in parallel. \ac{FPGA}s, were also used to parallelize LU decomposition for \ac{PF} \cite{Foertsch2005} \cite{Wang2007}. The hardware modification approach, while effective, is, for the most part, impractical, and it is common sense to modify algorithms to fit existing scalable hardware. 

The other way to parallelize \ac{PF} (or \ac{OPF}) is through network partitioning. While network partitioning usually occurs at the problem level in \ac{OPF}, in \ac{PF}, the partitioning often happens at the solution/matrix level. Such partitioning methods for \ac{PF} use sparsity techniques involving LU decomposition, forward/backward substitution and diakoptics that trace back to the late 60s predominantly by H. H. Happ \cite{Happ1966} \cite{Happ1967} for load flow on typical systems \cite{Carre1968} \cite{Andretich1968} and dense systems \cite{Wu1976}. Parallel implementation of \ac{PF} using this method started in the 80s on array processors such as the VAX11 \cite{Takatoo1985} and later in the 90s on the iPSC hypercube  \cite{Lau1991}. Techniques such as \ac{FDPF} were also parallelized on the iPSC using \ac{SOR} on Gauss-Sidel (GS) \cite{Huang1994}, and on vector computers such as the Cray X/MP using Newtons \ac{FDPF} \cite{Greenberg1990} \cite{Gomez1990}. \ac{PF} can also be treated as an unconstrained non-linear minimization problem, which is precisely what  E. Housos and O. Wing \cite{Housos1982} did in order to solve it using a parallelizable modified conjugate directions method. 

When general processors started dominating parallel computers, their architecture homogenized, and the enhancements achieved by parallel algorithms became comparable and easier to experiment with. It enabled a new target of optimizing the parallel techniques themselves. Chen and Chen used transputer-based clusters to test the best workload/node distribution on clusters, and \cite{Chen2000} and a novel BBDF approach for power flow analysis \cite{Chen2005}. The advent of \ac{MPI} allowed the exploration of scalability with the \ac{GMRES} in  \cite{Feng2002}, and the multi-port inversed matrix method \cite{Yalou2010} as opposed to the direct LU method. Beyond this point, parallel \ac{PF} shifted heavily towards using \ac{SIMD} hardware (\ac{GPU}s particularly) except few studies involving elaborate schemes. For example, Transmission/Distribution or smart grid \ac{PF} calculation \cite{Sun2015} or Optimal network partitioning for fast \ac{PF} calculation \cite{Su2018}. 

\subsection{SIMD Based Studies}
\subsubsection{Development} 

\ac{GPU} dominates recent parallel power system studies. The first power flow study implementation might have been by using a preconditioner Biconjugate Gradient Algorithm and sparsity techniques to implement the \ac{NR} method on a NVIDIA Tesla C870 \ac{GPU} \cite{Garcia2010}. Some elementary approaches parallelized the computation of connection matrices for networks where more than one generator could exist on a bus on a NIVIDIA \ac{GPU} \cite{Singh2010}. \ac{CPU}s were also used in \ac{SIMD}-based power flow studies since modern \ac{CPU}s exhibit multiple cores; hence multi-threading with OpenMP can be used to vectorize \ac{NR} with LU factorization \cite{Dag2011}. Some resorted to \ac{GPU}s to solve massive batches of \ac{PF} for \ac{PPF} or contingency analysis, thread per scenario, such as in \cite{Vilacha2011}. Others modified the power flow equations to improve the suitability and performance on \ac{GPU} \cite{Mei2012} \cite{XueLi2013}. 

While many papers limit their applications to NIVIDIA \ac{GPU}s by using \ac{CUDA}, OpenCL, a general parallel hardware \ac{API}, has also been used occasionally \cite{Ablakovic2012}. Some experimented with and compared the performance of different \ac{CUDA} routines on different NIVIDIA \ac{GPU} models \cite{Blaskiewicz2015}. Similar experimentation on routines was conducted to solve \ac{ACOPF} using  \ac{FDPF} \cite{Huang2017}. In  \cite{Guo2012}, \ac{NR}, Gauss Sidel, and Decoupled \ac{PF} were tested and compared against each other on \ac{GPU}. Improvement on Newtons Method and parallelizing different steps of it were performed previously \cite{Wang2017}. Asynchronous \ac{PF} algorithms were applied on \ac{GPU}, which sounds difficult as the efficiency of \ac{GPU}s depends on synchronicity and hegemony.\cite{Marin2017}. Even with the existence of \ac{CUDA}, many still venture into creating their routines with OpenCL \cite{Gnanvignesh2019} \cite{Tang2019} or direct C coding \cite{ARAUJO2019} of \ac{GPU} hardware to fit their needs for \ac{PF} . Very recently a few authors made thorough overviews for parallel power flow algorithms on \ac{GPU} covering general trends \cite{DongHee2020} \cite{Daher2022} and specifically AC power flow \ac{GPU} algorithms \cite{Abhyankar2021}. In the State of the Art subsection, the most impactful work is covered. 

\subsubsection{State of the Art}

A lot of the recent work in this area focuses on pre-conditioning and fixing ill-conditioning issues in iterative algorithms for solving the created \ac{SLS}. An ill-conditioned problem exhibits a huge change in output with respect to a minimal change in input, making the solution to the problem hard to find iteratively. Most sequential algorithms are LU-based direct solvers as they do not suffer from ill-conditioning. However, Iterative solvers such as the Conjugate Gradient method, which have been around since the 90s \cite{Dag1999}, are re-gaining traction for their scalability and parallel computing advancement. 

The \ac{DCPF} problem was solved using Chebyshev pre-conditioner and conjugate gradient iterative method in a \ac{GPU} (448 cores Tesla M2070 ) implementation in  \cite{Li2014} \cite{Li2015}. The vector processes involved are easily parallelizable in the most efficient way with \ac{CUDA} libraries such as  CUBLAS and CUSPARSE, which are Basic Linear Algebra Subroutine and  Sparse Matrix Operation Libraries. In this work, comparisons of sparse matrix storage formats were used, such as the Compressed Sparse Row (CSR) and Block Compressed Sparse Row (BSR). On the largest system size, a speedup of 46x for the pre-conditioning step and 10.79x for the conjugate gradient step was achieved compared to a sequential Matlab implementation (8-core \ac{CPU}). 

Later, the same author went on to Parallelize the \ac{FDPF} using the same hardware and pre-conditioning steps  \cite{Li2017a}. Two real systems were used, the Polish system, which had groups of locally connected systems, and the Pan-European system, which consisted of several large coupled systems. This topology difference results in a difference in the sparsity patterns of the \ac{SLS} matrix, which brings a unique perspective. Their proposed \ac{GPU}-based \ac{FDPF} was implemented with Matlab on top of MatPower v4.1. In their algorithm, the \ac{CPU} regularly corresponds with the \ac{GPU}, sending information back and forth over one iteration. Their tests showed that the \ac{FDPF} performed better on the Pan-European system because the system's connections were more ordered than the Polish system. \ac{CPU}-\ac{GPU} communication frequently occurred in their algorithm steps, most likely bottlenecking the speedup of their algorithm (less than 3x achieved compared to \ac{CPU} only). 

Instead of adding pre-conditioning steps, M. Wang et al. \cite{Wang2018}  focus on improving the continuous Newtons method such that a stable solution is found even for an ill-conditioned power flow problem. For example, if any load or generator power exceeds 3.2pu in the IEEE-118 test case, the \ac{NR} method fails to converge; even if the value is realized in any iteration, their algorithm will still converge to the solution with their method. This was achieved using different order numerical integration methods. The \ac{CPU} loads data into \ac{GPU} and extracts the results upon convergence only, making the algorithm very efficient. The approach substantially improved over the previous work by removing the pre-conditioning step and reducing \ac{CPU}-\ac{GPU} communication (speedup of 11x compared to \ac{CPU}-only implementation). 
 
Sometimes, dividing the bulk of computational load between the \ac{CPU} and \ac{GPU}  (a hybrid approach) can be more effective depending on the distribution of processes. In one hybrid \ac{CPU}-\ac{GPU} approach, a heavy emphasis on the sparsity analysis of  \ac{PF} generated matrices were made in  \cite{Su2020}. When using a sparse technique, the matrices operated on are reduced to ignore the 0 terms. For example, the matrix is turned into a vector of indices referring to the non-zero values to confine operations to these values. Seven parallelization schemes were compared, varying the techniques used (Dense vs. Sparse treatment), the majoring type (row vs. column), and the threading strategy. Row/column-major signifies whether the matrix same row/column data are stored consecutively. The thread invocation strategies varied in splitting or combining the calculation of P and Q of the mismatch vectors. In this work, two sparsity techniques were experimented with, showing a reduction of operations down to 0.1\% of the original number and two or even three orders of magnitude performance enhancement for power mismatch vector operations. In 100 trials, their best scheme converged within six iterations on a 4-core host and a GeForce GTX 950M \ac{GPU}, with a small deviation in solution time between trials. \ac{CPU}-\ac{GPU} communication took about 7.79\%-10.6\% of the time, a fairly low frequency. However, the proposed approach did not consistently outperform a \ac{CPU}-based solution with all of these reductions. The authors suggested that this was due to using a higher grade \ac{CPU} hardware than the \ac{GPU}. 

Zhou et al. might have conducted the most extensive research in \ac{GPU}-accelerated batch solvers in a series between 2014-2020. They fine-tune the process of solving \ac{PPF} for \ac{GPU} architecture in \cite{ZHOU2019} \cite{Zhou2018}. The strategies used include Jacobian matrix packaging, contiguous memory addresses, and thread block assignment to avoid divergence of the solution. Subsequently, they use the LU-factorization solver from previous work to finally create a batch-DPF algorithm \cite{Zhou2017}. They test their batch-DPF algorithm on 3 cases: 1354-bus, 3375-bus, and 9241-bus systems. For 10,000 scenarios, they solved the largest case within less than 13 seconds, showing the potential for online application. 

Most of the previous studies solve the \ac{PF} problem in a bare and limited setup when compared to the work by J. Kardos et al. \cite{Kardos2020} that involves similar techniques in a massive \ac{HPC} framework. Namely, preventative \ac{SCOPF} is solved by building on an already existing suite called BELTISTOS \cite{beltistos}. BELTISTOS specifically includes \ac{SCOPF} solvers and has an established Schurs Complement Algorithm that factorizes the KKT conditions, allowing for a great degree of parallelism in using  \ac{IPM} to solve general-purpose \ac{NLP} problems. Thus, the main contribution of this work is in removing some bottlenecks and ill-conditioning that exist in Schur's Complement steps introducing a modified framework  (BELTISTOS-SC). The parallel Schur algorithm is bottlenecked by a dense matrix associated with the solution's global part. This matrix is solved in a single process. Since \ac{GPU}s are meant to be used for dense systems, they factorize the system and apply forward-backward substitution, solving it using cuSolve, a \ac{GPU} accelerated library to solve dense linear algebraic systems. 

They performed their experiments using a multicore Cray XC40 computer at the Swiss National Supercomputing Centre. They used 18 2.1Ghz cores, and NVIDIA Tesla P100 with 16GB memory, and many other BELTISTOS and hardware-associated libraries. They tested their modification on several system sizes from PEGASE1354 to PEGASE13659. Their approach sped up the solution of the Dense Schur Complement System by 30x for the largest system over \ac{CPU} solution of that step, achieving notable speed up in all systems sizes tested. They later performed a large-scale performance study, where they increased the number of computing cores used from 16 to 1024 on the cluster. The BELTISTOS-SC augmented approach achieved up to 500x speedup for the PEGASE1354 system and 4200x for the PEGASE9241 when 1024 cores are used, demonstrating strong scalability up to 512 cores.

\section{Optimal Power Flow}

Like \ac{PF}, \ac{OPF} studies are the basis of many operational assessments such as \ac{SSA}, \ac{UC}, \ac{ED}, and other market decisions \cite{Conejo2019}. Variations of these assessments include Security Constrained Economic Dispatch (SCED) and \ac{SCOPF}, both involving contingencies. \ac{OPF} ensures the satisfaction of network constraints over cost or power loss minimization objectives. The full \ac{ACOPF} version has non-linear, non-convex constraints, making it computationally complex and hard to reach a global optimum. \ac{DCOPF} and other methods such as decoupled \ac{OPF} linearize and simplify the problem, and when solved, they produce a fast but sub-optimal solution. Because \ac{DCOPF} makes voltage and reactive power assumptions, it becomes less reliable with increased \ac{RE} penetration. \ac{RE} deviates voltages and reactive powers of the network significantly. This is one of the main drivers behind speeding up \ac{ACOPF} in real-time applications for all algorithms involving it. The first formulation of \ac{OPF} was in 1962 by J. Carpenter \cite{CARPENTIER19793}, which was followed by an enormous volume of \ac{OPF} formulations and studies as surveyed in \cite{Huneault1991}.

\subsection{MIMD Based Studies}

\subsubsection{Development} 

\ac{OPF} and \ac{SCOPF} decomposition approaches started appearing in the early 80s using P-Q decomposition \cite{Shoults1982} \cite{Talukdar1983} and including corrective rescheduling \cite{Pereira1987}. The first introduction to parallel \ac{OPF} algorithms might have been by Garng M Huang and Shih-Chieh Hsieh in 1992 \cite{Huang1992}, who proposed a ``textured" \ac{OPF} algorithm that involved network partitioning. In a different work, they proved that their algorithm would converge to a stationary point and that with certain conditions, optimality is guaranteed. Later they implemented the algorithm on the nCUBE2 machine \cite{Huang1995} showing that both their sequential and parallel textured algorithm is superior to non-textured algorithms. It was atypical for studies at the time to highlight portability, which makes Huangs work in \cite{Huang1994} special. It contributed another \ac{OPF} algorithm using Successive Overrelaxation by making it ``Adaptive," reducing the number of iterations. The code was applied on the nCUBE2 and ported to Intel iPSC/860 hypercube demonstrating its portability. 

In 1990 M.Teixeria et al. \cite{Teixeira1990} demonstrated what might be the first parallel \ac{SCOPF} on a 16-\ac{CPU} system developed by the Brazilian Telecom R\&D center. The implementation was somewhat ``makeshift" and coarse to the level where each \ac{CPU} was installed with a whole MS/DOS \ac{OS} for the multi-area reliability simulation. Nevertheless, it outperformed a VAX 11/780 implementation and scaled perfectly, was still 2.5 times faster than running on, and exhibited strong scalability. 

Distributed \ac{OPF} algorithms started appearing in the late 90s with a coarse-grained multi-region coordination algorithm using the \ac{APP} \cite{Kim1997}\cite{Kim1999}\cite{Baldick1999}. This approach was broadened much later by \cite{Anese2013} using Semi-Definite Programming and \ac{ADMM}. Prior to that, \ac{ADMM} was also compared against the method of partial duality in \cite{Liu2011}. The convergence properties of the previously mentioned techniques and more were compared comprehensively in \cite{Kim2000}. 

\begin{figure*}[ht]
\includegraphics[width=14cm]{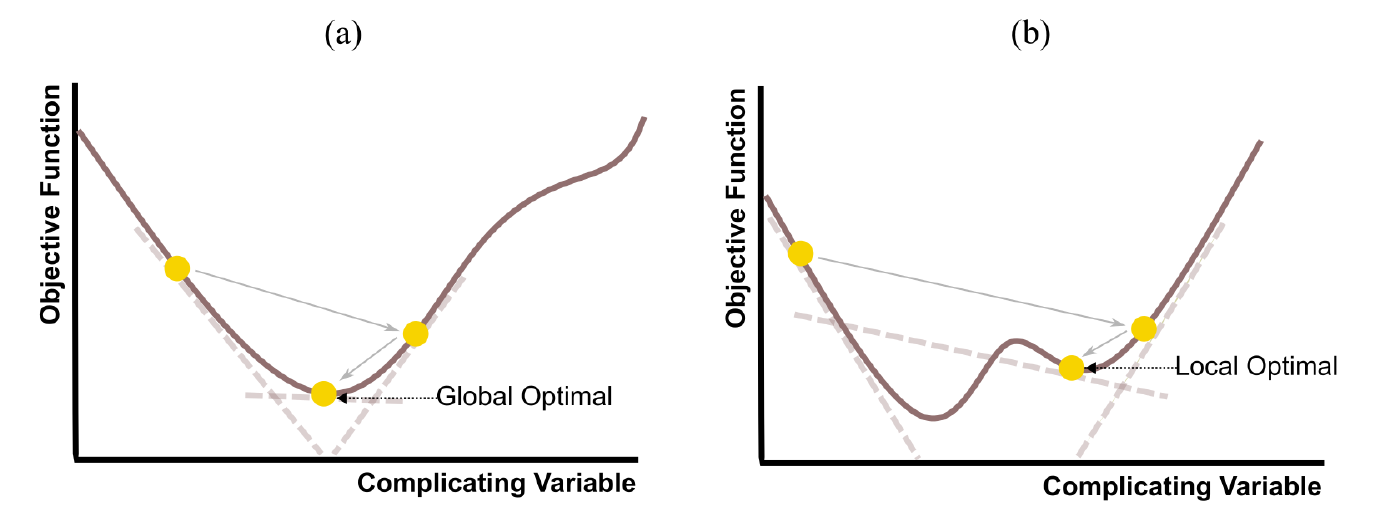}
\centering
\caption{The conversion steps of the complicated variable when its Convex (a) and Non-Convex (b) with respect to the objective function}
\label{fig8}
\end{figure*}

The asynchronous parallelization of \ac{OPF} first appeared on preventative  \cite{Talukdar1994} and corrective \ac{SCOPF} \cite{Rodrigues1994} targeting online applications \cite{Ramesh1996} motivated by the heterogeneity of solution time of different scenarios. Both \ac{SIMD} and \ac{MIMD} machines were used, emphasizing portability as ``Getsub and Fifo" routines were carried out. On the same token, \ac{MPI} protocols were used to distribute and solve \ac{SCOPF} decomposing the problem with GRMES and solving it with the non-linear  \ac{IPM} method varying the number of processors \cite{Wei2005}. Real-time application potential was later demonstrated by using Benders Decomposition instead for distributed \ac{SCOPF} \cite{Borges2007}. Benders decomposition is one of the most commonly used techniques to create parallel structures in power system optimization problems and shows up in different variations in the present literature, as illustrated in Fig. \ref{fig8} shows. Benders Decomposition is applied by fixing the complicating variables of the objective function to a different value in every iteration and constructing a profile with benders cuts to find the minimum objective function value with respect to the complicating variables. If the profile is non-convex, then optimality is not guaranteed since the value is changed in steps descending the slope of the cuts.

\subsubsection{State of the Art} 

As mentioned previously, involving AC equations in large-scale power system studies is crucial and might soon become the standard. The difficulty of achieving this feat varies depending on the application. For example, precise nodal price estimation is attained by solving \ac{ACOPF} multiple times under different possible system states (Probabilistic \ac{ACOPF}). This can be effortlessly scaled as each proposed scenario can be solved independently, but a large number of scenarios can exhaust available resources. This is when researchers resort to scenario reduction techniques such as Two-Point Estimation  \cite{Yuan2016}. In this study, it was applied on 10k \ac{MC} scenarios, and the reduced set was used to solve a conically relaxed \ac{ACOPF} model following the approach in \cite{Farivar2013}. Using 40 cores from the \ac{HPC} cluster of KTH Royal Inst of Tech, the approach resulted in an almost linear speedup with high accuracy on all test cases.

In contrast, parallelizing a single \ac{ACOPF} problem itself is much more complicated. However, the same authors did it readily since their model was already decomposable due to the conic relaxation \cite{Yuan2019}. Here, the choice of network partitions is treated as an optimization problem to realize the least number of lines between sub-networks. A graph partitioning algorithm and a modified benders decomposition approach were used, providing analytical and numerical proof that they converge to the same value as the original benders. This approach achieved a lower-upper bound gap of around 0-2\%, demonstrating scalability. A maximum number of 8 partitions (8 subproblems) were divided on a 4-core 2.4GHz workstation. Beyond four partitions, hyperthreading or sequential execution must have occurred. This is a shortcoming as only four threads can genuinely run in parallel at each time. Hyper-threading only allows a core to alternate between two tasks. Their algorithm might have even more potential if distributed on an \ac{HPC} platform. 

The \ac{ACOPF} formulation is further coupled and complicated when considering Optimal Transmission Switching (OTS). The addition of binary variables ensures the non-convexity of the problem turning it from a \ac{NLP} to a \ac{MINLP}. In Lan et al. \cite{Lan2017} this formulation is parallelized for battery storage embedded systems, where temporal decomposition was performed recording the State of Charge (SoC) at the end of each 6 hours (4 subproblems). They employed a two-stage scheme with an \ac{NLP} first stage to find the \ac{ACOPF} of a 24-hour time horizon and a transmission switching second stage. The recorded SOCs of the first stage are added as constraints to the corresponding subproblems, which are entirely separable. They tested the algorithm IEEE-188 test case and solved it with Bonmin with GAMS on a 4-core workstation. While the coupled  \ac{ACOPF}-OTS formulation achieved a 4.6\% Optimality gap at a 16h41m time limit, their scheme converged to a similar gap within 24m. The result is impressive considering the granularity of the decomposition. This is yet another example where a better test platform could have shown more exciting results as the authors were limited to parallelizing four subproblems. Algorithm-wise, an asynchronous approach or better partition strategy is needed as one of the subproblems took double the time of all the others to solve. 

The inclusion of voltage and reactive power predicate the benefits gained by \ac{ACOPF}. However, it is their effect on the optimal solution that matters, and there are ways to preserve that while linearizing the \ac{ACOPF}. The \ac{DCOPF} model is turned into a \ac{MILP} in  \cite{Zhang2018} by adding on/off series reactance controllers (RCs) to the model. The effect of the reactance is implied by approximating its value and adding it to the DC power flow term as a constant without actually modeling reactive power. The binary variables are relaxed using the Big M approach to linearize the problem, derive the first-order KKT conditions, and solve it using the decentralized iterative approach. Each node solved its subproblem, making this a fine-grained algorithm, and each subproblem had coupling variables with adjacent buses only. The approach promised scalability, and its convergence was proven in \cite{Mohammadi2015}. However, it was not implemented in parallel, and the simulation-based assumptions are debatable. 

Decentralization in that manner reduces the number of coupling variables and communication overhead. However, this also depends on the topology of the network, as shown in \cite{Wang2017}. In this work, a stochastic \ac{DCOPF} formulation incorporating demand response was introduced. The model network was decomposed using \ac{ADMM} and different partitioning strategies where limited information exchange occurs between adjacent subsystems. The strategies were implemented using MATLAB and CPLEX on a 6-bus to verify solution accuracy and later on larger systems. The \ac{ADMM}Based Fully Decentralized \ac{DCOPF} and Accelerated \ac{ADMM} for Distributed \ac{DCOPF} were compared. Recent surveys on Distributed \ac{OPF} algorithms showed that in \ac{OPF} decomposition and parallelization, \ac{ADMM} and \ac{APP} are preferred in most of the studies as a decomposition technique \cite{Sadnan2020} \cite{Yuan2019}. The distributed version converged faster, while the decentralized version exhibited better communication efficiency. More importantly, a separate test showed that decentralized algorithms work better on subsystems that exhibit less coupling (are less interconnected) and vice versa. Breeding the idea that decentralized algorithms are better suited for ring or radial network topologies while distributed algorithms are better for meshed networks \cite{Molzahn2017} \cite{Sadnan2020}.

Distribution networks tend to be radial. A ring topology is rare except in microgrids. Aside from their topology, they have many differences compared to transmission networks causing the division of their studies and OPF formulations. OPF for Transmission-Distribution co-optimization makes a great case for \ac{HPC} use in power system studies as co-optimizing the two together is considered peak complexity. S. Tu et al.\cite{Tu2020} decomposed a very large-scale \ac{ACOPF} problem in the Transmission-Distribution network co-optimization attempt. They devised a previously used approach where the whole network was divided by its feeders, where each distribution network had a subproblem \cite{Sun2015}\cite{DeMiguel2006}. The novelty in their approach lies in a smoothing technique that allows gradient-based non-linear solvers to be used, particularly the Primal-Dual-Interior Point Method, which is the most commonly used method in solving \ac{ACOPF}. Also, a two-stage stochastic algorithm was implemented to account for the uncertainties in distributed energy generation. 

S. Tu et al. used an augmented IEEE-118 network, adding distribution systems to all busses, resulting in 9206 buses. Their most extreme test produced 11,632,758 bus solutions (1280 scenarios). Compared to a generic sequential PDIPM, the speedup of their parallelized approach increased linearly with the number of scenarios and scaled strongly by increasing the number of cores used in their cluster. In contrast, the serial solution time increased superlinearly and failed to converge within a reasonable time in a relatively trivial number of scenarios. While their approach proved to solve large-scale \ac{ACOPF} much faster than a serial approach, it falls short in addressing Transmission-Distribution co-optimization because it merely considered the distribution network as sub-networks with the same objective as the transmission, which is unrealistic.

\section{Unit Commitment}

The \ac{UC} problem goes back to the 60s \cite{Kerr1966}. In restructured electricity markets, Security Constrained Unit Commitment ( \ac{SCUC}) is used to determine the generation schedule at each time point at the lowest cost possible while maintaining system security. A typical formulation used in today's industry can be found in \cite{Midwest2009}. Often, implementations use immensely detailed stochastic models involving N-1 or N-2 contingencies \cite{Shiina2003}, \ac{ACOPF} constraints, and incorporating \ac{RE} resources \ac{DER}  \cite{Papavasiliou2013} and distribution networks \cite{Ji2021}. This leads to a large number of scenarios and a tremendously complex problem. Decomposition of the problem using Lagrangian Relaxation (LR) methods is very common \cite{E.M.Constantinescu2011}\cite{L.Wu2007} and many formulations are ready to segue for \ac{HPC} parallel implementation. This includes global optimal solution methods for AC- \ac{SCUC} as in \cite{Liu2019AC}. Recent literature on parallel \ac{UC} is abundant, making this section the largest in this review. 

\subsection{MIMD Based Studies}

\subsubsection{Development} 

Simulations of Parallel environments to implement parallel \ac{UC} algorithms started appearing around 1994 modeling hydrothermal systems \cite{Wong1993} and stochasticity \cite{Numnoda1996} on supercomputers \cite{Misra1994} workstation networks \cite{Lau1997}. The earliest \ac{UC} implementations on parallel hardware used embarrassingly parallel metaheuristics, such as simulated annealing \cite{Numnonda1996} and other genetic algorithms \cite{Yang1995} \cite{Yang1997}. However, the first mathematical programming approach might have been by Misra in 1994 \cite{Misra1994} using dynamic programming and vector processors. Three years later, K.K. Lau and M.J. Kumar also used dynamic programming to create a decomposable master-slave structure of the problem. It was then distributed over a network of workstations using  \ac{PVM} libraries, with each subproblem solved asynchronously \cite{Lau1997}. These, however, were not network-constrained problems.  

Only in 2000, Murillo-Sanchez and Robert J. Thomas \cite{Murillo-s2000} attempt full non-linear AC-\ac{SCUC} in parallel by decomposing the problem using APP, but failed to produce any results, upholding the problems complexity. In quite an interesting case, volunteer computing with the BOINC system was used to parallelize the \ac{MC} simulations of stochastic load demand in \ac{UC} problem with hydro-thermal operation \cite{Baslis2009}. However, that did not include network constraints either. There has been some work in decomposition algorithms for network constrained \ac{UC} but rarely applied on a practical parallel setup. Most parallel implementations happened in the last decade.

\subsubsection{State of the Art}

Papavasillio et al. \cite{PapavasiliouA;OrenS.S;ONeill2011}, set up the framework of a scenario-based two-stage stochastic framework of \ac{UC} with reserve requirement for wind power integration emphasizing wind forecasting and scenario selection methodology. In later work, Papavasiliou compared a benders decomposition approach that removes the second stage bottlenecking and a Lagrangian Relaxation (LR) algorithm based on \cite{L.Wu2007} where the impact of contingencies on decisions was implied in the constraints \cite{Papavasiliou2013}. The LR approach proved to be more scalable for that formulation. As a result, Papavasillo chose LR approach to solving the same formulation by adding DC network constraints \cite{Papavasiliou2015}. Even though the wind scenarios were carefully selected, there existed instances where specific subproblems took about double the time of the following most complex subproblem. In a follow-up work, Aravena and Papavasillio resorted to an asynchronous approach in \cite{aravena2015} that allows time-sharing. This work showed that the synchronous approach to solving subproblems could be highly inefficient as the idle time of computational resources can reach up to 84\% compared to the asynchronous algorithm. 

While many  \ac{SCUC} parallel formulations use DC networks, the real challenge is using \ac{ACOPF}, as exhibited in earlier failed attempts \cite{Murillo-s2000}. In \cite{Bai2015} the conic relaxation approach mentioned earlier \cite{Farivar2013}  was used to turn the AC- \ac{SCUC} into a Mixed Integer Second Order Conic (MISOC) program. It allowed the decomposition of the problem to a master problem, where \ac{UC} is determined, and a subproblem where the \ac{ACOPF} is solved iteratively. In their approach, the active power is variable in the master problem is fixed, and only the reactive power is solved to check if the commitment is feasible. Fixing the active power allows for time decomposition since ramping constraints no longer apply in the subproblems. They compared the computational efficiency of their approach against a coupled DC- \ac{SCUC} and AC- \ac{SCUC}, solved using commercial solvers such as GAMS, DICOPT, and SBB. Their approach took only 3.3\% of the time taken by previous similar work \cite{BAI2008} to find a solution at a 0.56\% gap. However, their approach faced accuracy and feasibility issues, and the parallelism strategy was unclear since they created eight subproblems while using three threads.  

Temporal decomposition was also used on a unique formulation, Voltage Stability Constrained  \ac{SCUC} (VSC- \ac{SCUC}) \cite{Khanabadi2016}. The problem is an \ac{MINLP} with AC power flow constraints and added voltage stability constraints that use an index borrowed from \cite{KARGARIAN2011}. \ac{APP} decomposition was used to decompose the model into 24 subproblems, and it was compared against conventional AC- \ac{SCUC} on several test cases. It converged after 55 iterations compared to 44 by the AC- \ac{SCUC} solution on the IEEE-118 case. The structure and goals achieved by VSC- \ac{SCUC} make tractability challenging, deeming the approach itself promising. However, the study fell short of mentioning any details about the claimed parallel routine used.  

Nevertheless,  \ac{SCUC} decentralization is valued for more than performance enhancement. It can help achieve privacy and security, and it fits the general future of the smart grid, IoT, and market decentralization. In a market where the ISO sets prices of energy and generators are merely price takers, a decentralization framework called ``Self-Commitment" can be created from the \ac{UC} formulation \cite{SIOSHANSI2008245}. Inspired by self-commitment, Feizollahi et al. decentralize the  \ac{SCUC} problem relevant to bidding markets, including temporal constraints  \cite{Feizollahi2015}. They implement a ``Release-and-Fix" process which consists of three \ac{ADMM} stages of decomposing the network. The first stage finds a good starting point by solving a relaxed model. The second and third stages are iterative, where a feasible binary solution is found, followed by a refinement of the continuous variables. They used two test cases (3012 and 3375-bus) and applied different partitions from 20 to 200 sub-networks with different root node (co-ordinator node) combinations. They also varied the level of constraints in 3 cases, from no network up to AC network and temporal constraints. A sub 1\% gap was achieved in all cases, outperforming the centralized solution in the most complicated cases and showing scalability where the centralized solution was intractable. The scalability saturated, however, at 100 partitions, and one of the key conclusions was that the choice of the root node and partition topologies are crucial to achieving gains. 

Multi-Area formulations often involve \ac{ED}, but rarely \ac{UC}, as done in \cite{Doostizadeh2016}. The \ac{UC} formulation in this study includes wind generation. The wind uncertainty is incorporated using  Adjustable Interval Robust Scheduling of wind power, a novel extension of Interval Optimization Algorithms, a chance-constrained algorithm similar to Robust Optimization. The resulting Mixed-Integer Quadratic Programming (MIQP) model is decentralized using an asynchronous consensus \ac{ADMM} approach. They verified the solution quality on a 3-area 6-bus system (achieving a 0.06\% gap) and then compared their model against Robust Optimization and Stochastic Optimization models on a 3-area system composed of IEEE-118 bus system each. For a lower \ac{CPU} time, their model achieved a much higher level of security than the other mentioned models. The study mentions that the parallel procedure took half the time the sequential implementation did, promising scalability. However, no details were given regarding the parallel scheme used and implementation. 

Similarly, Ramanan et al. employed an asynchronous consensus \ac{ADMM} approach to achieve multi-area decentralization slightly differently as their formulation is not a consensus problem \cite{Ramanan2017}. Here, the algorithm is truly decentralized as the balance in coupling variables needs only to be achieved between a region and its neighboring one. The solution approach is similar to that of the one from \cite{PapavasiliouA;OrenS.S;ONeill2011} where \ac{UC} and \ac{ED} are solved iteratively. They divided the IEEE 118 bus into ten regions; each region (subproblem) was solved by one intel Xeon 2.8Ghz processor. They ensure asynchronous operation by adding 0.2 seconds of delay for some subproblems. The mean results of 50 runs demonstrated the time-saving and scaling potential of the asynchronous approach that was not evident in other similar studies \cite{Shi2014}. However, the solution quality significantly varied and deteriorated, with an optimality gap reaching 10\% for some runs, and no comparison was drawn against a centralized algorithm. 
 
In later work, the authors improved the asynchronous approach by adding some mechanisms, such as solving local convex relaxations of subproblems while consensus is being established. This allowed the subproblem to move to the next solution phase if the binary solution is found to be consistent over several iterations \cite{Ramanan2019}. In addition, they introduced a globally redundant constraint based on production and demand to improve privacy further. Moreover, they used point-to-point communication without compromising the decentralized structure. They implemented their improved approach on an IEEE-3012 bus divided into 75, 100, and 120 regions. A 2.8Ghz core was assigned to solve each region and controller subproblem. They compared their approach this time against Feizollahis implementation from 2015 \cite{Feizollahi2015} and a centralized approach. The idle time of the synchronous approach was higher than the computation time of the Asynchronous approach,  doubling the scalability for higher region subdivisions. The gap achieved in all cases was larger than that of the Centralized solution by around 1.5\%, which is a huge improvement considering the previous work and the 18x speedup achieved. 

Consensus \ac{ADMM} methods typically do not converge for \ac{MILP} problems like \ac{UC} without a step size diminishing property \cite{Boyd2011}. Lagrangian methods, in general, are known to suffer from a zigzagging problem. To overcome that issue, the \ac{SLR} algorithm was used in \cite{Bragin2017}  to create a distributed asynchronous \ac{UC}. In later work, their approach was compared against a Distributed Synchronous \ac{SLR}, and a sequential \ac{SLR} \cite{Bragin2021} using four threads to parallelize the subproblems. With that, better scalability against the synchronous approach was demonstrated, and a significant speedup was achieved (12x speed up to achieve a 0.03\% duality gap in one instance). 

To avoid the same zigzagging issue, but for Multi-Area  \ac{SCUC}, Kargarian et al. opted for \ac{ATC} since multiple options exist for choosing the penalty function in \ac{ATC} \cite{Kargarian2018}. They take the model from \cite{Fu2013} and apply \ac{ATC} from \cite{Kargarian2015} to decompose the problem into a distributed bi-level formulation with a central co-ordinator being the leader and subproblems followers. In this work, they switched the hierarchy by putting the co-ordinator, making it the follower instead of the leader. This convexified the followers' problem, allowing the use of  KKT conditions, turning it into a Mixed Complementarity Problem (MCP). Those steps turned the formulation into a decentralized one as only neighboring subproblems became coupled. They numerically demonstrated that with their reformulation, the convergence properties of \ac{ATC} still upheld virtually and that the convex quadratic penalty functions act as local convexifiers of the subproblems. Moreover, they demonstrated numerically how the decentralized algorithm is less vulnerable to cyber attacks. Unfortunately, the approach was not implemented practically in parallel; rather, the parallel solution time was estimated based on the sequential execution of the longest subproblem. 

In a similar work tackling Multi-Area  \ac{SCUC}, a variation of ACT is used \cite{MingZhou2018} where the master problem determines the daily transmission plan, and each area becomes an isolated  \ac{SCUC} subproblem. This problem is much more complicated as it involves AC power flow equations, HVDC tie-lines, and wind generation. The power injection of the tie-lines is treated as a pseudo generator with generation constraints that encapsulate the line flow constraints. This approach removes the need for consistency constraints used in traditional \ac{ATC}-based distributed  \ac{SCUC} like the ones used in \cite{Kargarian2015}. In the case study, they subdivide several systems into 3-Area networks and split their work into three threads. Comparisons were drawn against a centralized implementation, the traditional distributed from, and four different tie-line modes of operation varying load and wind generation. Their approach consistently converged at lower times than the traditional \ac{ATC} algorithm. It slightly surpassed the centralized formulation on the most extensive network of 354-bus, which means the approach has the potential for scalability. 

Finding the solution of a \ac{UC} formulation that involves the transmission network, active distribution network, microgrids, and \ac{DER} is quite a leap in total network coordination. This challenge was assumed by  \cite{Ji2021} in a multi-level interactive \ac{UC} (MLI-UC). The objective function of this problem contained three parts: the cost of \ac{UC} at the transmission level, the cost of dispatch at the distribution level, and the microgrid level. The three levels' network constraints were decomposed using the \ac{ATC} algorithm, turning it into a multi-level problem. A few reasonable assumptions were made to aid in the tractability of the problem. The scheme creates a fine-grained structure at the microgrid level and a coarser structure at the distribution level, both of which were parallelized. The distribution of calculation and information exchange between the three levels provides more information regarding costs at each level and the Distribution of Locational Marginal Pricing (DLMP). 

Most of the previous work in power system problems - apart from \ac{UC} - focuses on the solution process rather than the database operations involved. In \cite{Wei2020}, a parallel  \ac{SCUC} formulation for hydro-thermal power systems is proposed, incorporating pumped hydro. This paper uses graph computing technologies and graph databases (noSQL) rather than relational databases to parallelize the formulation of their MIP framework. Their framework involves Convex Hull Reformulation and Special Ordered Set method to reduce the number of variables of the model, constrained relaxation techniques \cite{YongFu2007}, and LU decomposition. The graph-based approach showed significant enhancement over speedup over a conventional MIP solution method on a Tigergraph v2.51 database. Similar applications of noSQL were explored in other power system studies \cite{FengWei2018,Yuan2020,ShiQingxin2019}.

\begin{figure}[ht]
\includegraphics[width=12cm]{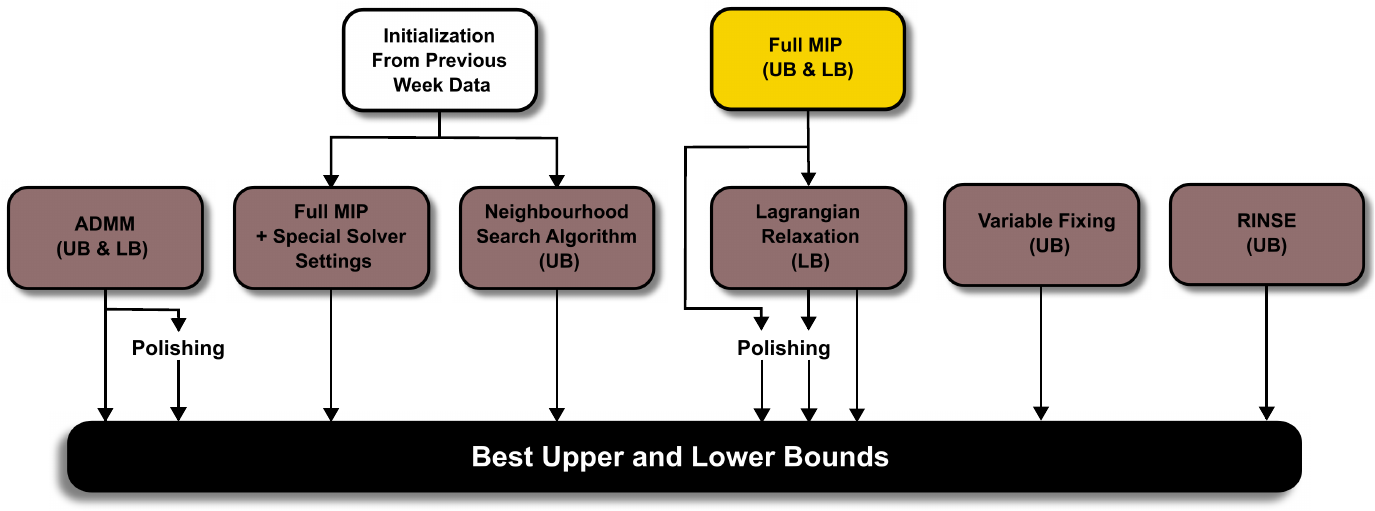}
\centering
\caption{HIPPO Concurrent Optimizer (vertically aligned processes run in parallel)}
\label{fig9}
\end{figure}

In real industrial applications,  there is a lower emphasis on the accuracy of the solution, and a high-speed ``good enough" policy is adopted, often using heuristics extensively. Midwest ISO (MISO) published a few papers showing the development of their Day-Ahead DC \ac{UC} decision making  in 2016 \cite{Chen2016}, 2020 \cite{Chen2020} and 2021\cite{Chen2021}. Their \ac{HPC} parallel approaches introduce novel strategies as part of the HIPPO project \cite{HIPPO} focusing on finding smart heuristics to speed up  \ac{SCUC} decomposition and distributed methods. MISO has been using CPLEX to solve day Ahead  \ac{SCUC} and SCED for 50,000 binary variables and 15,000 transmission constraints over 36 hourly intervals, and they limit the day ahead of  \ac{SCUC} to 1200s. Fig. \ref{fig9} illustrates the processes run by the HIPPO system in parallel. They run several algorithms in parallel to ensure continuity. If the most accurate fails to converge within their time limit, the solution of the next most accurate algorithm is taken instead, such as SCED. The convergence criterion of optimality gap is 0.1\%  which amounts to about \$24,000. They use surrogate ADMM,  Lazy transmission constraints from experience, and a Polishing-E method, which reduces the set of possible generators and an Uplift function to choose a good set of generators.

In their latest work \cite{Chen2021} they introduce a neighboring search algorithm that improves their E-polishing algorithm and the selection of a set of lazy constraints. This HIPPO system uses 18 nodes 24 2.3GHz  64GB \ac{RAM} on the \ac{PNNL} \ac{HPC} cluster and uses Gurboi to solve their problems. They compared in their study the time it takes to sequentially solve 110 different  \ac{SCUC} problems of different complexities against using HIPPO. They showed that problems that take a long time sequentially experience a more drastic speed up with HIPPO, meaning their system scales efficiently. One problem that took 2000 seconds in full sequential MIP was solved in 200s with HIPPO. For all the tested cases, the highest solution time on the MISO network using HIPPO was 633s, under the required standard time limit for finding solutions. In addition, they explore the possibility of solving 15-minute rather than an hourly interval as it is more appropriate for \ac{RE} generation. They solve the MIP for one hour and then feed that solution to the 15-Minute interval problem as an initial point. Compared to using root relaxation for solving 15-minute intervals, a huge jump in speed up is achieved once again for the more complex problems.

\section{Power System Stability}

Power system stability studies in this section include Static, Transient, and Dynamic Stability. A power system is considered stable if it can regain operating equilibrium with the entire system intact after being subjected to a disturbance. Depending on the type of study, the stability metric could be the line flows, the generator rotor angle, or bus voltage and frequency \cite{stott1987} \cite{Balu1992}. Parallelism of those types of power system studies has been one of the most abundant and earliest to investigate, especially in transient stability. Most of these studies are performed offline, and the goal is to speed them up to solve them in real-time.

\subsection{MIMD Based Studies}

\subsubsection{Development}

Power system operators need to detect system states or schedules that carry the risk of steady-state emergency if single or multiple equipment failures occur. This is to alleviate that risk either by changing the system state to avoid such an emergency (preventative measure) or by employing a form of a control strategy that would mitigate that emergency if it occurs (corrective measure) \cite{stott1987}. Contingency screening, in particular, is an embarrassingly parallel task and one of the easiest to parallelize, as it involves splitting several conventional power flow problems equivalent to the number of possible contingencies over multiple job arrays. 

In earlier studies, the parallel steady-state analysis involved parallelizing the different contingency cases and not the power flow algorithm. The first successful attempt at distributed contingency screening might have been in \cite{Hao1995} where the process was distributed over four DN425 processors. By adding pre-filtering schemes and strategies to reduce the computational burden, real-time static security assessment was already achieved in the early 2000s using multi-processing  \cite{Mendes2000}  \cite{Balduino2004}. Further studies in enhancing the allocation of resources and dynamic load balancing in order to reduce the idling time of processors were done by \ac{PNNL} on their local cluster with the aid of \ac{MPI} \cite{Huang2009} \cite{Huang2010}. The same research team followed up the work by applying parallel betweenness centrality to identify the high-impact transmission lines in the screening process. 

Some work in the area emphasized processor load balancing to task scheduling for contingency screening, from master-slave scheduling to proactive task scheduling. Incorporating both multi-processing and multi-threading on various systems and using various concurrent programming languages such as D \cite{Khaitan2013} and X10  \cite{Khaitan2014}. Most of the work today for static security involves improvement in whole EMS systems and software, and most of the modern work involves the efficient allocation of cloud-based resources. However, more elaborate schemes are appearing that involve parallelizing the OPF within the contingency analysis, which opens room for improvement, especially when considering more complicated OPF and post contingency network models. 

Dynamic stability is another form of steady-state analysis that evaluates the system condition and oscillatory behavior after very small signals and disturbances that last for up to 30 seconds due to fluctuations in generation and load levels or controllers. It is an obvious parallelism candidate since it is the most intensive computational task in power system studies, as it models the electro-mechanical interaction between system components and their controllers. It is also one of the most important and directly related to secure operation.

\begin{figure}[ht]
\includegraphics[width=6cm]{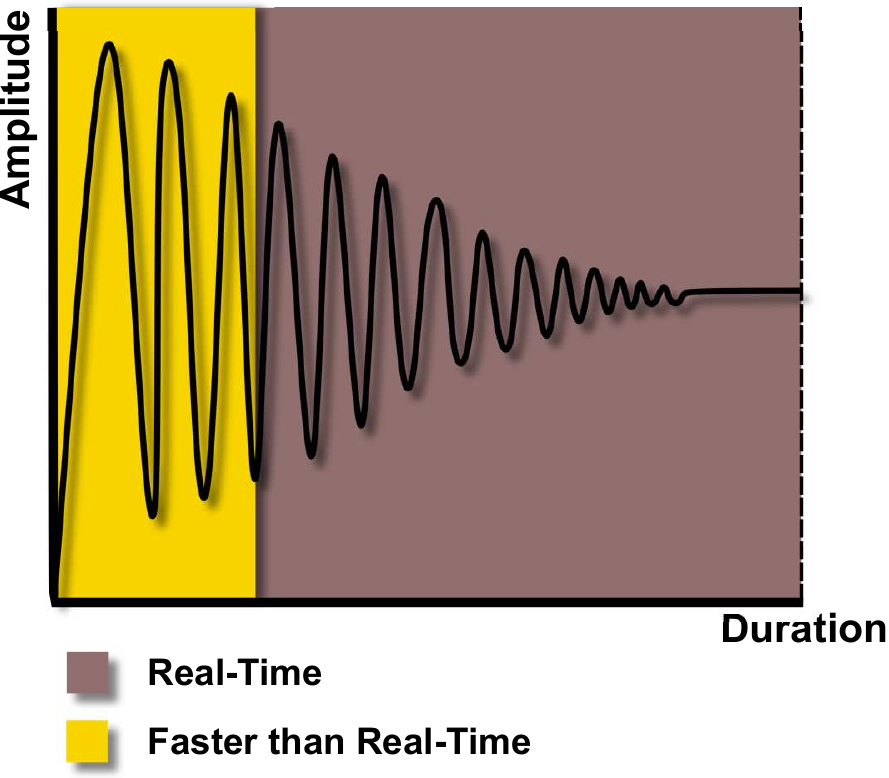}
\centering
\caption{Real-time simulation vs Faster than Real-Time simulation}
\label{fig10}
\end{figure}

Fig. \ref{fig10} illustrates the two goal-posts of dynamic simulation during its development. Real-time simulation means the computational time matches the duration of the simulated interaction. Faster than real-time simulation means the computational time is lower than the duration of the simulated interaction. The goal for achieving practical real-time parallel dynamic simulations was already being set in the 90s \cite{SEKINE1994}. Pioneering applications used the Conjugate Gradient method on the Cray Y-MP, iPSC/860, and IBM 3090 mainframe \cite{Pai1992} \cite{Decker1996}. Parallel dynamic simulation studies quickly moved to large system implementations emphasizing balanced network partitioning and computational load and creating parallel software tools in the 2000s\cite{Xue2005}. Faster than real-time simulations became the default, and the first faster than real-time parallel dynamic simulation was achieved on the WECC system for the first time in 2013 \cite{Jin2013}.  

Transient stability studies are made to ensure that after large disturbances such as circuit breaker trips or load loss, the system remains synchronized and can return to normal conditions. Parallel transient stability was explored in abundance as they are critical, and the trapezoidal integration method makes them disposed to parallelism. Fernando L. Alvarado, in an impactful paper \cite{Alvarado1979}, demonstrated analytically that for time T and with some $\frac{T}{2}$ processors, transient stability differential equations could be solved in time order of  $log_2 T$ using the trapezoidal method with potentially better convergence properties than serial implementation. The Electric Power Research Institute made a report in 1977 exhibiting various works exploring potential parallel applications for power system analysis \cite{P.M.Anderson}. 

Most of the work until that point discussed potential parallel methods for to apply on parallel machines such as Cray-1, CDC STAR-100, IBM 370-195, and ILLIAC. Moreover, much like power flow development, a lot proposed parallel architectures that would exploit the parallelism in using methods such as Chaotic Relaxation, BBDF Gauss-Seidel, and Newtons method  \cite{Hatcher1977} \cite{Hatcher1976}. Some simulated the performance of new microcomputer and processor architectures using networks of computers or existing supercomputers, such as the  CDC 6600 \cite{Brasch1979} \cite{Brasch1982}. 

The first parallel implementation of transient stability might have been the first parallel power system study implementation by F. Orem and W. It was solved using a CDC 6500 equipped with an AP-120B array processor hosted on a computer. Comparisons on different hardware such as vector processors vs. array processors or Cray-1 vs. IBM-3081D were made using the trapezoidal integration method with linear and non-linear loads   \cite{Taoka1983} \cite{Taoka1984}. Different decomposition methods started being introduced in the literature, including parallel-in-space and parallel-in-time approaches. The combination of the two was achieved while using ``Nested Iteration" and time-windowing to enhance convergence in \cite{LaScala1990},\cite{Scala1990gaussi} and \cite{Scala1991}. 

The variety of algorithms used and the discovery of different parallelism paradigms in transient stability simulation were promising for parallelism in power system studies in general. \ac{SOR} and The Maclaurin-Newton Method (MNM) were used on the Alient, and IPSC machines in \cite{Zhu1992} \cite{Chat1993}. The waveform relaxation method (WRM) was proposed by \cite{Spong1987} to decompose the non-linear system into several dynamical subsystems to be solved in parallel, followed by \cite{Scala1990} and \cite{Crow1990}. A Parallel-in-Frequency paradigm was introduced with a demonstration of the possibility of vector processing coarse-grained algorithm \cite{Tylavsky1993}. More complicated fine/coarse-grained frameworks combining general processing (CRAY Y-MP9/464) and vector processing \cite{Granelli1994}. 

The possibility for real-time simulation was first demonstrated on the NCube2 HypeCube \cite{Taoka1992}. HyperCube machines were popular in the early 90s for transient studies exhibiting various techniques such as \ac{SOR} \cite{Chai1991} and LU factorization path trees with different communication scheduling techniques \cite{Lee1991} \cite{Lau1991}. 

Message passing in heterogeneous clusters started appearing in the late 90s. Real-time Contingency and Transient stability with PMV  \cite{Aloisio1997} and other formulations with \ac{MPI} \cite{Hong1997} \cite{Hong2000}. Faster than real-time transient stability simulation was achieved combining \ac{MPI} and multi-threading techniques with network and time-domain decomposition in  \cite{Xue2004},\cite{Xue2005} and \cite{Jikeng2008}. Later algorithms were used \cite{Jikeng2009}. A full rundown on most of the techniques used until that point for large-scale transient stability studies can be found in \cite{Jalili2010}. 

Electromagnetic Transient studies (EMT) are transient studies that assess systems overcurrents and overvoltages due to fault conditions or large disturbances. \ac{EMT} simulations are the most accurate tools to describe fast dynamics of power systems, hence most computationally intensive. Software such as PSCAD uses an EMTP-type program or EMtp, which is the most wildly used algorithm for this type of study. Parallel \ac{EMT} studies were first proposed in the early 90s on \ac{MIMD} hardware \cite{Werlen1993} and implemented using network partitioning techniques on a hypercube machine with care to load balancing \cite{Falcao1993}. Since then, various parallel approaches have been used, parallelizing the problem in space and time. The Very Dishonest Newton method as implemented on multi-computer setup \cite{Morales2001}, on \ac{FPGA}s real-time simulation of  \ac{EMT} was aimed for \cite{Dufour2012}. Techniques such as system partitioning and solving short time-steps for more dynamic parts and long time-steps for others in combination with multi-processing have been proposed in\cite{Peng2017} but with no actual implementation. 

Most of the recent work today in power system stability studies, particularly transient stability, takes place on \ac{SIMD} architecture (\ac{GPU}s specifically). The traditional way to perform the study is similar to that of regular transient studies. Thus, in-space-in-time decomposition can be achieved with similar techniques such as  LU factorization \cite{Hong2000}, forward and backward substitutions \cite{Morales2001}, and graph theory \cite{Beaudin2003} to be solved in parallel.  

\subsubsection{State of the Art}

When it comes to probabilistic studies performed in contingency analysis,  not many studies apply the N-2 criterion. Duan et al. performed N-2 contingency analysis generating with Probabilistic Power Flow (PPF) for each contingency case \cite{Duan2018}. With an IEEE-300 test case, using AC power flow equations and the \ac{NR} method, and 1000 \ac{MC} system state scenarios, the solution was distributed on the Danzek \ac{HPC} cluster at Manchester. The solution time was 168 hours. The only thing that this work offers is a testimony to the complexity of full AC Power flow equations. N-2 criterion merely increases the number of embarrassingly parallel tasks. Any real improvement needs to be made either with a finer decomposition of the embedded power flow problems or by incorporating complexity formulations used. 

In another N-2 reliability contingency analysis, transmission switching was incorporated for corrective action \cite{Li2017}. This work performed a dynamic stability analysis of the corrective transmission switching action to ensure its viability. AC power flow equations were used, and the analysis was conducted on the PJM interconnections (15.5k bus system). Both transmission and generation contingencies were considered creating around 1.4m contingencies. Heuristics were used to reduce the additional computation. The approach was very effective as the solution time was reduced to 96s using 128 threads ( compared to 999s with eight threads). This contrast between this result and the previous study demonstrates the importance of having a good parallel scheme. 

One of the major challenges that face current implementations is the lack of standardization, as reflected in studies such as S. Jin et al. \cite{Jin2017} which compared four different parallel implementations of the dynamic model. Some implementations were run on the IEEE-288 bus system and others on the WECC system, using different supercomputers with different hardware allocations for each implementation. This variety resulted in valuable but difficult to compare observations due to the unequal testing grounds. The work concludes that the direct integration method with a fast direct LU solver is the recommended approach for \ac{HPC} dynamic simulation as it enables faster than a real-time solver. Nevertheless, the recommendation is based on trials that are hard to compare fairly.

In an impactful series of work, P. Aristidou et al. implemented a parallel dynamic simulation on a transmission-distribution network using the Very Dishonest Newtons Method (VDHN) and Schur-Complement based decomposition in 2014 \cite{Aristidou2014} 2015 \cite{ARISTIDOU2015} and 2016 \cite{Aristidou2016}. Their most significant test case included 15226 buses, 21765 branches, and detailed 3483 generator models, including their excitation systems and governors, voltage regulators, and power system stabilizers. The software was written using standard Fortran language with OpenMP directives. The authors were very thorough in analyzing the performance, as their approach was compared against several, including fast and widely implemented sequential algorithms in terms of speedup and scalability. They tested their algorithm on various platforms, with the highest speedup achieved being x4.7 on 24-core AMD processor-based systems. Faster than real-time dynamic simulation of the entire WECC system, interconnection (17000 buses) was achieved in \cite{Huang2017}. The simulation included machine models, exciter, governor, relay, and network models. With 16 core cores, they simulated a 0.05s 3-phase fault lasting for the 20s, with 0.005s time steps, within 15.47s. They achieved this feat using an open-source \ac{HPC} framework called GridPack, which they developed. GridPack is further explored in the discussion section of this review.

\subsection{SIMD Based Studies}

\subsubsection{Development}

\ac{SIMD} architecture has long been used for transient and dynamic simulations to solve the non-linear differential-algebraic equations (DAE) and parallelize the trapezoidal rule, and \ac{GPU}s have been used in power system studies as early as 2007 \cite{Gopal2007}. 

The first paper using \ac{GPU} for static stability was published the same year \ac{CUDA} was released, where DC Power flow contingency analysis was performed to solve the Gauss-Jacobi algorithm \cite{Gopal2007}. The paper used a NIVIDIA 7800 GTX card and direct instructions as opposed to a \ac{CUDA} interface. \ac{GPU}-based contingency analysis studies have recently been enhanced with pre-conditioning methods such as the Krylov theory\cite{Tang2018}. Pre-conditioned Conjugate Gradient method \cite{Fu2020} and compensation method and \ac{FDPF} to parallelize \ac{ACPF} within contingency analysis\cite{Huang2018a}. 

For transient stability, A parallel program was developed and used by engineers in Hydro-Quebec in 1995, which simulates transient stability by parallelizing the Very Dishonest Newtons Method (VDHN) with LU Decomposition on the shared memory machine Sequent Symmetry S81. \cite{Wu1995}. Waveform-Relaxation method was used later on the same machine \cite{Hou1997}. The first re-purposing of non-\ac{GPU} image processing hardware for power system studies might have been in 2003 in \cite{Beaudin2003}, where PULSE IV image processor was used to achieve real-time transient stability simulation on WSCC 9 bus test case. 

The first huge leap in performance was by Jalili-Marandi et al., in a hybrid approach achieving a speedup of up to x344 for a  1248-bus system compared to \ac{CPU} only approach \cite{Jalili-marandi2009}. Later Jalili-Marandi et al. refined the algorithm to perform all the calculations on \ac{GPU} on an almost purely \ac{SIMD}-based approach. Which proved to be more effective beyond 500-bus size systems \cite{Jalili-marandi2010}. In their last work, they showed the potential of \ac{GPU} augmentation to enhance the inner solution performance. An instantaneous relaxation technique involving full newton iteration, implicit integration, and a sparse LU linear solver was tailored to run simultaneously on four T10 \ac{GPU}s. A 9985 bus system generates a 22272x22272 matrix and 99.69\% sparsity, which was solved within 5 minutes \cite{Jalili-marandi2012}.

Yu et al. \cite{Yu2014} performed another hybrid-based transient stability study by constructing a Jacobian Free Newton Generalized Minimal Residual method, which approximates the Jacobian vector products using the finite difference technique. While it eliminates a jacobian matrix step, it still requires heavy matrix-vector multiplication for a pre-conditioning step, making it suitable for \ac{GPU}. This approach proved scalability starting from a 216-bus sized system, and it outperformed the newton based transient simulation solver PSAT. This approach showed better consistency performance enhancements for various sized systems compared to a similar work, where a pre-conditioning step was parallelized prior to the \ac{GMRES} method \cite{Baijian2012}. Yet, using pre-conditioned GRMES in a combination of in-time coarse-grained schemes and in-space fine-grained schemes with \ac{GPU} acceleration as in \cite{Liao2016} showed universal scalability whether the number of \ac{GPU}s used or the problem size increased.

There have been a few \ac{GPU}-based parallel EMTP-type simulators for \ac{EMT} studies. Some integrating wind farms to the models \cite{Gao2014}. Others added complicated controls to large-scale systems with PV, transformers, and reactive components. \cite{Song2017}. \ac{GPU}s were used primarily to solve the linear algebraic equations associated with the algorithm, while the \ac{CPU} performed most of the other parts. However, ``Full" \ac{GPU}-based parallel solvers that parallelize numerous other steps of the algorithm were developed recently \cite{Song2018}.

\subsubsection{State of the Art} 

\begin{figure}[ht]
\includegraphics[width=14cm]{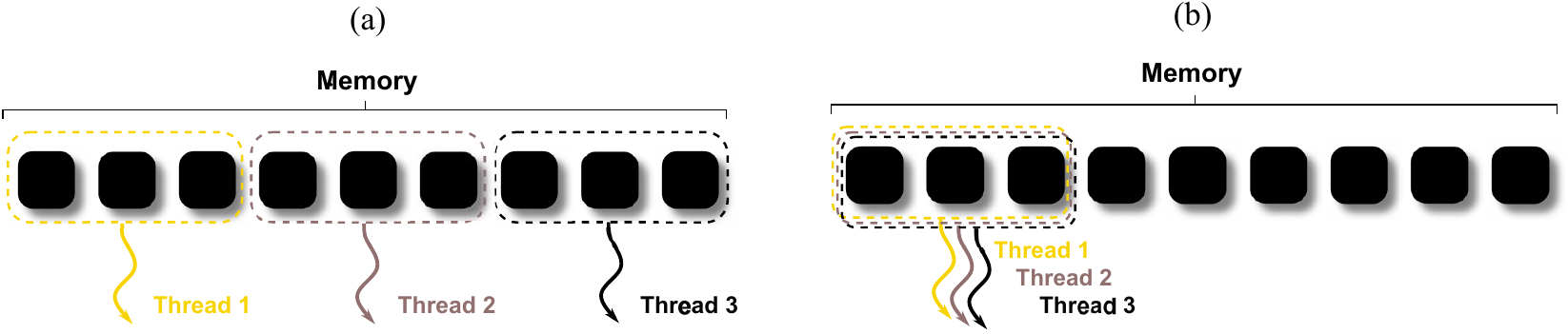}
\centering
\caption{Illustration of strided memory access (a) and Coalesced memory access (b)}
\label{fig11}
\end{figure}

The most relevant recent work in power system stability mainly involved \ac{SSA} and \ac{EMT}. In \cite{Zhou2016}, Zhou et al. continue their work in N-1 \ac{SSA}, this time treating the DC power flow contingency screening part, where only branch thermal violation is accounted for. This study tries to apply the Critical Contingency list contingency screening on \ac{GPU}. \ac{DCPF}-based Contingency screening involves dense vector operations. This paper claims to be the first of its kind to present a novel \ac{GPU}-accelerated algorithm for DC contingency screening, where they optimize the data transmission, parallelization, memory access, and \ac{CUDA} streams in their algorithm. The presented algorithm was tested on 300-bus, 3012-bus, and 8503- bus systems. The hardware used was A Tesla K20C GPU, and the host was Intel Xeon E5-2620 2 GHz \ac{CPU}. They compared the performance against a single-threaded \ac{CPU}-based algorithm implemented on a higher-end Intel Core i7-3520M 2.90 GHz \ac{CPU} and 4 GB memory notebook. Their largest test case exhibited a speedup of 47x over the sequential case, demonstrating their approach's scalability. They achieved it by reducing the data transmission by x50, further optimizing task allocation and thread/block redistribution, and using memory coalescing to enhance memory access. The last improvement is particularly important as memory handling is often ignored in the field, yet it is very crucial. Fig. \ref{fig11} demonstrates the impact category of the coalescing strategy used. Single threads often access different chunks of memory locations at the same time; when GPUs are most efficient when multiple threads access contiguous memory locations at the same time, this is called coalesced memory access.  

To tackle the same problem by Zhou et al., Chen et al. \cite{Chen2017a} uses a slightly different \ac{GPU} implementation, which exhibits a pipelined fashion. They employ a two-layered parallel method. In the first layer, they apply a hierarchial path tree parallel LU decomposition for all contingency cases. In the second layer, they solve the decomposed problems for every contingency in parallel. I.e., the first layer process is repeated for every contingency case in parallel, sending groups of threads for each contingency. This means that the same process will run for several contingencies simultaneously, subject to the \ac{GPU}'s number of thread blocks allowed to run simultaneously. They employed an asynchronous scheme, where the \ac{CPU} performs convergence checks. It receives the output convergence of three cases simultaneously; if one contingency calculation converges, the next one in line is sent to that available thread block. In their work, they pay attention to data management and utilize the cache architecture of \ac{GPU} to improve their process. They compared their \ac{GPU} approach against a KLU-based commercial suit that solves the problem on \ac{CPU}. On the largest case and using 32 \ac{GPU} thread groups, their algorithm showed a 9.22x speedup over a single thread \ac{CPU} solution and a 3x speedup on a 4-threaded \ac{CPU} solution. In this study, the importance of matching thread number to warp number is demonstrated because using 32 thread groups did not make a massive difference to the speedup against 16 thread groups. 

In \cite{Zhou2017a}, Zhou et al. expand on their previous work and attempt to skip the screening step, directly solving \ac{ACPF} for all cases in a batch-ACPF solver. Their framework packages the batch-ACPF into a new problem through which a high degree of parallelism can be achieved. They created dependency graphs and used QR left-looking algorithm for its numerical stability compared to LU decomposition. They compared their solver against commercial multi-threaded \ac{CPU}-based \ac{SLS} solvers KLU and PARADISO in several test cases. On an 8503-bus system, their solver took 2.5 seconds vs. 9.9s (4x speedup) on a 12-threaded KLU solution and  144.8 seconds (57.6x speedup) on a sequential single-threaded solution. At face value, a \ac{GPU} approach is superior. However, due to memory bandwidth limitations, adding more than ten \ac{CPU} solvers would not have enhanced the commercial solver performance. Thus performance-wise (core to core), this conclusion cannot be drawn. Economically speaking, the \ac{GPU} is far more superior, as adding a single \ac{GPU} is much cheaper than several equivalents compute nodes. 

Following the trend of resurrecting iterative solvers for \ac{PF} applications mentioned earlier, in 2020, Zhou et al. designed a \ac{GPU}-accelerated Preconditioned Conjugate Gradient solver to solve the same \ac{ACPF} for N-1 \ac{SSA}. They test their \ac{GPU}-designed algorithm on 118, 1354, 2869, and 10828 bus systems, the last system being the east china power grid. The number of contingencies was respectively 177, 1728, 4204, and 11062. They compared their approach to 2 solution algorithms: 1- Complete LU \ac{SSA} solution, which was implemented on a single core. 2-  Rank-one update-based solution implemented on a single, 4, and 8 cores supported by a multi-threaded solver. The hardware specs were similar to previous studies. With their \ac{GPU} implementation of this algorithm on the east china power grid, the solution speedup was 4.9x compared to the eight-core multi-threaded \ac{CPU} implementation. Zhou's works show that solvers for power system-related problems tailored for \ac{GPU}s have considerable potential. 

When it comes to \ac{EMT}, the first record of \ac{GPU} use is in \cite{Debnath2011}, where co-processing of vector operations for 117 bus networks on \ac{GPU} had double the speed of PSCAD, a \ac{CPU} based software. In later work, they reduced the communication between the \ac{CPU} and \ac{GPU} and derived a \ac{GPU}-specific algorithm to achieve close to x40 speedup on a 3861 bus network \cite{Debnath2016}. A similar implementation that aimed to reduce communication time, fully and efficiently exploit the \ac{SIMD} architecture in \ac{EMT} was also conducted in \cite{Song2014}. 

Earlier work by Zhou et al. was a \ac{GPU} implementation of Electromagnetic Transient Simulation \cite{Zhou2014}. They utilize both \ac{SIMD} and scalar operations and emphasize the importance of avoiding simultaneous memory access when parallel processing. The homogeneity of tasks was ensured where elements that are modeled similarly were grouped; for example, all RLC elements were processed in the same kernel with a unified lumped model. Separate Kernels were made for the Universal Line Model (ULM) in four stages, where every stage kernels were executed simultaneously. The Unified Machine Model (UMM) simulation, the third ubiquitous task, was also divided and managed in detail. The level of parallelism in this work is massive and attempts to squeeze every parallel structure in the problem and ounce of performance from the \ac{GPU}. The algorithm was tested on 8 test cases from 40 bus up to 2458 bus systems. They used a NIVIDIA Tesla C2050 \ac{GPU} with 448 cores and 3GB memory and an AMD Phenom 4 core 3.2GHz \ac{CPU}. The total simulation time was 100 milliseconds at 20-microsecond. On that setup, the speedup of 5.63x was achieved compared to an optimized commercial software EMTP-RV. 

Finally, in one of the most impactful papers by the same authors, a huge test case containing 240,000 buses was decomposed with propagation delay  \cite{Zhou2017c}. The system was divided into linear, non-linear, and control subsystems. Also, the jacobian domain and the voltage calculations were parallelized, creating another decomposition layer, resulting in a highly fine-grained problem. Two GP104 \ac{GPU}s were used where all the iterations were processed, and convergence was checked. Using a single \ac{GPU} against the EMTP-RV resulted in a 15x speedup. Moreover, \ac{GPU} linear scaling was demonstrated by achieving double the speedup (30x) by adding the second \ac{GPU}. However, it is important to note that the 240,000 bus system was an augmentation of the IEEE-39 bus system.

\section{ System State Estimation} 

Power \ac{SSE} is a centerpiece of control centers. Thousands of voltage measurements are collected from  SCADA systems and \ac{PMU} and processed to understand the conditions of the grid better. \ac{SSE} studies provide accurate and reliable estimates of the phase angle and bus voltages from incomplete system measurements. 

\subsection{MIMD Based Studies}

\subsubsection{Development}

Parallelizing \ac{SSE}, much like other studies, started with simulations. Y Wallach and E. Handschin proposed a distributed master-slave topology \cite{Wallach1981}, showing that merely partitioning the network would achieve speed gains. Later in 1982, C. W. Brice and R. K. Cavin Simulated the potential performance of distributed and decentralized algorithms on parallel hardware for state estimation, where one is communication-intensive, the other is computationally intensive  \cite{Cavin1982}. Different \ac{SSE} decomposition techniques and parallel simulations were carried out over the 80s and 90s. Those include the block partitioning to decompose the state estimation problem by the network simulating the performance on a \ac{MIMD} machine \cite{Aoki1987},  parallel forward-backward substitution \cite{ABUR1990}, recursive quadratic programming \cite{Lin1992}, Dantzig-Wolfe Decomposition Algorithm \cite{El-keib1992}, and other simulations  \cite{Falcao1995}. The first practical implementation was in the year 2000 using \ac{APP}\cite{Ebrahimian2000} \cite{Carvalho2000}. 

The \ac{WLS} algorithm is the most commonly used in \ac{SSE}. \ac{WLS} contains a matrix inversion step, which can be solved using LU decomposition. The first parallel \ac{WLS} solver implementation might have been in \cite{Niepolcha2006} where shared memory vs. \ac{MPI} schemes solving the linear system of equations were compared. The exploitation of parallelism in the Khan Filtering Method only showed up in 2009  \cite{Schneider2009}, when its been in use since the 70s to improve the prediction aspect of \ac{SSE}. 

\subsubsection{State of the Art}

The most relevant work made in this area on \ac{MIMD} is by Korres et al.. They used an efficient distributed \ac{WLS} algorithm to perform multi-area state estimation in parallel using \ac{MPI} \cite{KORRES2013}. They tested the algorithm with several processors from 1 up to 60, solving problems using the scientific toolkit PETSc which contains parallel optimization linear and non-linear solvers. They employed different communication/coordination strategies and control area partition numbers and sizes to estimate the state of an 1180 bus system (10x IEEE-118 system) in two cases, where case 2 exhibited more interconnections between ``slave" areas. The algorithm was implemented on the National Technical University of Athens cluster consisting of 11 Intel Core 2 Duo E8200 PC nodes. However, in this work, scalability was demonstrated, but it lacked a speedup comparison against the fastest sequential algorithm.

\subsection{SIMD Based Studies}

\subsubsection{Development}

Most of the development of state estimation occurred over \ac{MIMD} studies. The earliest significant \ac{SIMD} application occurred in\cite{Beaudin2003} where the PULSE IV, a scalable \ac{SIMD} Chip, was designed to help achieve faster than real-time application in 2003. 

\subsubsection{State of the Art}

Several uncommon \ac{SSE} techniques were customed to \ac{GPU}, such as the Fast Decoupled State Estimation \cite{Xia2017} and selective column identification in the numerical differentiation  \cite{Magana-Lemus2013} \cite{Magana-Lemus2015}. The \ac{WLS}) algorithm is the most commonly used for electrical \ac{SSE}. \ac{WLS} contains a matrix inversion step, which can be solved using LU decomposition. That is what Karimipour et al. did in \cite{Karimipour2013} where they implemented all the steps of the solution algorithm from the Admittance matrix formation to the convergence check on \ac{GPU}. The \ac{GPU} used was a 512 core NIVIDIA with double-precision peak performance and the Intel Xeon E5-2610 2GHz \ac{CPU} as a host. The test system used was an IEEE-39-bus system, duplicated and interconnected to create larger systems of up to 4992 buses. They achieved a speedup of up to 38 for the largest system compared to a sequential \ac{CPU} implementation. The algorithm exhibited strong scalability, and they estimated that the maximum theoretical speedup achievable by that \ac{GPU} is 312x for this algorithm. Notably, they also address the issue of solution discrepancy due to the hardware architectural difference. They did it by considering both Correlated and Uncorrelated Gaussian Noise in the measurement samples (to consider bias) in small test cases, which is supposed to lead to larger errors in the final result. The Errors that occurred on the \ac{GPU} solution matched those in the \ac{CPU} solution, which confirmed the robustness of their algorithm. 

Later Karimipour et al. produced a \ac{GPU} parallelized Dynamic State Estimation based on Kalman Filters \cite{Karimipour2014} on a Tesla S2050 \ac{GPU}. Compared to a quad-core \ac{CPU}, the approach achieved a speedup of x10 for a 19968 bus system with 5120 generators exhibiting close to 0 error of estimation. They extended and refined the approach by increasing the \ac{GPU} portion of work and utilizing PMUs, and SCADA measurements \cite{Karimipour2015}. For a smaller system (4992-bus), they achieved a higher speedup (15x) with high voltage precision (0.002 p.u. and 0.05 rad error). Finally, they made the algorithm robust against coordinated false data injection enabling its detection through the parallel algorithm and optimized, secure PMU installation \cite{Karimipour2017}. 

In \cite{Rahman2016} the Dishonest Gauss Method in the \ac{WLS} algorithm is used, where the Jacobin update is not executed at every iteration. The algorithm was implemented on a Tesla K20c \ac{GPU}, fragmenting the original by vectorizing multiplication and multi-threading addition processes. They investigate the method's accuracy and show that to get an accuracy of 100\%, the Jacobian needs to be updated every seven iterations. They also demonstrate the algorithm's robustness by applying different noise levels to see its effect on accuracy, showing the method ranges between 98\% - 100\% at different levels of noise. A complete mathematical analysis of the convergence of their method was conducted in \cite{Rahman2017}. Real-Time Digital Simulators measurements were used while adding errors that vary from 1-15\%, and 30 samples per second were taken, a typical \ac{PMU} sampling Rate. On an IEEE 118 system and a 3-second duration, their algorithm achieved a 15x speedup over what is claimed to be the best existing \ac{CPU} implementation. 

\section{Unique Formulations \& Other Studies}

\subsection{SIMD Based Studies}

Almost all of the previous studies involve optimization problems, and those can be augmented and combined in various ways to achieve various objectives. Thus, a few unique and computationally expensive formulations in the literature sprouted with parallel implementation. A common one is an OPF problem that ensures the voltage stability in the solution, the \ac{TSCOPF}. Adding such constraints complicates the problem, but it was shown that by using techniques such as Benders decomposition \cite{Kim2011} and reduced space \ac{IPM} \cite{Geng2012}  a remarkably faster solution can be achieved prior to parallelizing the subproblems. 

In one formulation by \cite{Jiang2013}, the \ac{TSCOPF} is combined with \ac{UC} to create a \ac{TSCUC}. This formulation was decomposed temporally using \ac{APP}, and 24 cores were used to solve it on the IEEE-300 system. The model showed notable scalability; the solution time was reduced from 16h to 1h. Furthermore, using the same pre-defined contingency, transient stability of the first-hour interval of the solution was examined and compared to a standard  \ac{SCUC} solution. The \ac{TSCUC} solution maintained whole system stability without any alteration, whereas the  \ac{SCUC} solution failed to regain stability even by modifying the power output post solution. Meaning the proposed model guarantees stability as opposed to the conventional one. 

Deviating slightly to the security side, interesting work by \cite{Gong2020} investigates mitigating the disturbance effect of geomagnetic storms on power systems. Those manifest as low-frequency, quasi-DC, high-impact extreme events. Tools for large and complex power systems to mitigate this problem do not exist; thus, this paper proposes a parallel solution approach to conduct optimal secure operation planning considering geomagnetic storm disturbances (GMD). They include Geomagnetic Induced Current (GIC) into an \ac{ACOPF} model and turn it into an \ac{MINLP} by modeling GIC blocking components holding three states, bypass, resistor, and capacitor. \ac{APP} is used to decompose the problem into two problems, an \ac{NLP}, solving for \ac{ACOPF}, and a \ac{MILP} which determines the state of switchable network components and calculates the GIC flow. The two subproblems can be solved independently and in parallel-coupled by a third model. Tests were performed on a multi-core workstation showing the utility of this approach in mitigating GIC for a 150-bus network. However, the authors did address that their extensive experience with the method guarantees that their \ac{APP} approach results in an effective solution. 

\subsection{MIMD Based Studies}

There have been a few venturesome studies, such as in \cite{D.Vasquez2019} which used \ac{GPU}s to accelerate a stability analysis involving both transmission and distribution systems into the network, which achieved some performance improvement, but the speedup factor might not justify the use of power. In other formulations, the gas and thermal systems are modeled and coupled into the electrical system to calculate the energy flow using the Inexact Newtons method and \ac{GMRES} pre-conditioning \cite{Qi2019}. Contrary to previous studies on \ac{GPU}s, this study showed that the larger the system, the more time required. This is a testimony that care must be taken in routines and schemes that would scale with the hardware. Finally, the last notable work is in the AC TSC-\ac{OPF} using \ac{GPU} acceleration \cite{Geng2017}. The reduced-space \ac{IPM} was used in this study, and it was the portion that was decomposed and parallelized on \ac{GPU} using Schur's complement. They used a 12951 bus system in the largest case and compared their \ac{GPU} accelerated algorithm to a single sequential core and parallel 16-core \ac{CPU} implementations. Their algorithm achieved a speedup of x24 and x6.5, respectively.

\section{Grid and Cloud Computing Based Studies}

\subsection{Development} 

The use of \ac{HPC} facilities in the electrical power industry is not uncommon in various offline and some online applications, especially ones related to smart grids and microgrid planning \cite{Falcao1996}. For example, California Independent System Operator (CAISO) uses \ac{HPC} to perform various real-time assessments of the network, such as reliability assessments \cite{Alam2018}. Facing the huge computational load, ISO-NE installed an on-premise computer cluster in 2007 using EnFuzion as a job manager \cite{Luo2019}. They later faced challenges in choosing the optimal size for clusters and investment in computational power since the peak computing jobs and average ones were very different. Hence it made sense to move some non-emergent applications to Cloud. In fact, ISO-NE had already initiated a project to adopt cloud computing with emphasis on achieving privacy and security \cite{Feng2015} \cite{england_2021}. When facilities begin to struggle to meet the increasing requirement of deployed power system applications, it makes sense to resort to cloud services. Cloud computing really expands the realm in which algorithms and systems can be parallelized and exhausted. Regulators and players won't have to worry about the availability of resources anymore but only squeeze out every inch of performance and manage the ``rented" resources. Un-needed capital investment can be avoided, and real-time data can be shared with third parties.  

When it comes to Grid and Cloud computing studies, performance enhancement is often sought through scalability and resource availability rather than optimizing for specific hardware. While this works very well, the combination of fine optimization would be much more powerful. But this might be only possible through the Grid rather than the Cloud model since it is more controllable. Most of the work in this area is very recent, but it starts with a few studies on Grid Computing. In an application that is very similar to the Cloud computing paradigm, the work by Morante et al. \cite{Morante2006} might have been the first modular and hardware scalable implementation of parallel contingency analysis on a grid of 8 heterogeneous computers. A middleware called Hierarchical Metacomputers (HiMM) was used to allocate resources economically based on resource adequacy and a given budget value. By increasing the budget value, their middleware managed to lower the execution time by exploiting more expensive, more powerful resources. Other papers from the time explored the idea of monitoring and control of the power system using decentralized schemes on grid computing \cite{Taylor2006}. A few more studies explored Grid-based frameworks and applications, such as Huang et al. \cite{Huang2006}\cite{Huang2008} and Ali M. \cite{Ali2006}, load flow on Grid by Al-Khannak et al. \cite{Al-Khannak2007}, and dynamic security assessment by Xingzhi Wang et al. \cite{WANG2010}. In 2010, the literature almost completely shifted towards cloud computing and particularly an integrated framework combining smart Grid with Cloud, given the advent of AMI and big data at the dawn of that year. Several frameworks and models for smart grid co-ordination \cite{Rusitschka2010} \cite{Mohsenian2010} and power system assessment \cite{Huang2010c} using cloud appeared that year and later \cite{Markovic2013}. Ideas such as cloud-based demand response were being explored \cite{Hongseok2011}, and many papers suggested network architecture and control topologies that are realizable with cloud \cite{Bo2014}.

\subsection{State of the Art} 

In \cite{Yoon2018} a finer scope was taken on task management of massive parallel contingency analysis using Hadoop Distributed File System on the Cloud. They applied an N-1 transmission line contingency analysis and used the \ac{NR} method to solve the power flows. First, the system distributes the contingency and other parameters to separate nodes such that each node solves a contingency case. What's unique about their job management scheme is that when the number of cores increases, the network bandwidth automatically increases as well, further increasing the performance. With this approach, they could perform a full AC contingency analysis for a real network in less than  40s. 

For certain applications, such as Demand Side Management (DSM), fixed resources become an even greater issue as the amount of information processing and the computational requirement fluctuates based on the availability/flexibility of demand-side resources, which dictates the complexity of the problem at every instance. One of the options that are becoming more attractive is using cloud computing services, which could be much cheaper than expanding existing facilities. Using such services means an optimal allocation of the computational resource becomes much more crucial as cloud services are often billed based on the consumed resources and pay-as-you-go terms \cite{Amazon}. In 2016 Z. Cao et al.\cite{Cao2017} handled this issue with a source allocation algorithm that finds the optimal cloud computing resources for \ac{DSM} instances. Commercial cloud computing resources differ from regular \ac{HPC} clusters in the sense that there exists a greater variety in architecture, and the performance compactness might be lower than that of specialized \ac{HPC}s used in research.

Interesting network paradigms could be created given that computational resources can be flexible and scaled as cloud services provide. Sheikhi\cite{Sheikhi2015} explores the idea of an ``Energy Hub" where customers can be active in Demand Response Management by reducing their direct electricity consumption and using the output of the combined heat and power from the energy hub that the gas supplier supplies. This does not change customers' electricity consumption level, but the demand has reduced from the electrical supplier's point of view. This Energy Hub + Smart meter is now a Smart Energy Hub, and single or multiple customers could share it. The problem was formulated in a game theory approach where the Smart Energy hub is a price anticipator, which tries to predict the consequences of its own action on the price and chooses the optimal load shifting schedule to reduce the cost on customers on those bases. In a similar fashion to \cite{Rajeev2015}, the smart energy hubs read and control the outputs and send data to the Cloud to be aggregated and computed for decision-making, solving the game according to the cost function. They simulated their approach and showed that it resulted in a decrease in energy price compared to no \ac{DSM} game. Also, they compared the communication cost of direct messaging configuration vs. cloud configuration, where Cloud showed lower cost making the platform more suitable for such applications. 

On a much more refined and more local scale, Wang et al. \cite{Wang2020} attempt to decentralize the problem of Dynamic Economic Dispatch (DED), i.e., energy management in real-time, by using inverter Digital Signal Processor (DSP) chips and cloud computing. The paper solves a multi-parametric quadratic programming optimization problem, which has been highly applied in the area of coordinated power system \ac{ED} and TSO-DSO network coordinated dispatch. The solution involves two parts and is decomposed into two subproblems: 1- An offline calculation that Cloud carries out. 2- Real-time decision-making that the DSP carries out. In the cloud computing part, distributed renewable generation and loads are forecasted to create piecewise expressions. Every 4 hours, the expressions and information are sent to inverters such that the DSP chip can solve and optimize the output based on the real-time input of load and RE. The Cloud provides flexibility and handles the highest computation burden while simplifying the subproblem solved by the inverter. 

On a 14 node test case with PV, wind, Grid, diesel, and battery system, with they drew a comparison between their approach and a traditional implementation on an i7 regular laptop. \ac{AWS}instances were created, and a real DSP chip was used. Their test showed that by moving offline computations to Cloud, it is solved within 34us compared to the traditional algorithm (372ms). This is a colossal speedup but might not be fair since the traditional algorithm creates a whole new deterministic problem every time it collects new values. The main gist is that it achieves the needed solution time for using DSP since the calculations on the inverter must be lower than 100us not to cause issues and interruptions. Sharing the inverter's chip rather than adding local controllers lowers investment and maintenance costs, and the distributed nature makes it robust against single-point failure. However, care needs to be taken in the job timing such that the control functionality of the inverter is not interrupted.

Addressing the security concerns of Cloud, F. Ma et al. also proposed a cost-oriented model to optimize computing resource allocation, specifically for demand-side management problems using simulated annealing and modified priority list algorithms \cite{Ma2016}. The objective function parameters are based on actual Amazon cloud service pricing. This cost-oriented model was compared to a traditional O2O model, which allocates resources based on the peak computational load for the renting period. The proposed optimization method showed a significant cost reduction over the traditional source allocation method. There is a security concern that comes with outsourcing sensitive processes. For other players in a free market, there is an economic benefit to engaging in cyberattacks and accessing information from competitors' processes such as \ac{ED}, as it could help with their bidding strategies. In their study, they explain the use of the Virtual Private Cloud (VPC) scheme, which isolates their portion of the Cloud such that their resources are not shared with other organizations or applications even if idle. It is supposed to increase the security of the outsourcing process. Yet, one can see how the spread of such a strategy would create an impediment to the scalability and efficiency of the Cloud. 

The previous study was part of a diverse paper showcasing the challenges and experiences gained by ISO-Newengland in moving to cloud services. In their move their applications, they used Axceleon CloudFuzion \cite{cloudfuzion2020} job balancer, which provides high failure tolerance and job monitoring. The work involves heuristics and operational decisions, providing a great insight into the methodologies and equations used to choose the number of instances and squeeze every bit out of the rented hour. An N-2 contingency analysis was performed on a test case that takes 470h on a regular workstation; the case jobs were done in less than an hour with their scheme. CloudFuzaion wasn't flawless, however, as its workflow was often interrupted by manual steps (which meant it had to be monitored). Thus  ISO-NE  started a project with Axceleon to develop an independent power system simulation platform for cloud computing that addresses that issue, fully automating processes after receiving the user input. In a 2019 study \cite{Luo2019} they demonstrated that their platform managed to run multiple instances reaching near 100\% \ac{CPU} utilization of the instances launched for certain jobs and capable of many task computing and co-simulations. 

While the work above used a service-level security mechanism, Sarker and Wang \cite{Sarker2018} wanted to ensure security, assuming that the cloud in-house security infrastructure is compromised. They transform the \ac{ED} problem into a \ac{CPLP} formulation  \cite{mangasarian_2010}\cite{li_li_deng_2011}  to achieve holistic security, such that all sensitive information remains unknown by competitors. The approach protects against attacks from passive and active entities on the Cloud (administrators and customers). It works by converting inequality to equality constraints and multiplying the coefficients by randomly generated positive real numbers twice (a mononomial matrix U then H), which are held privately. The resultant constraint matrix is sent to the Cloud, and the equipment information implied in the constraint coefficients remains obscure to any attackers. This work enhances the security matrix reduction of \ac{CPLP}. Since the feasible region of the \ac{CPLP} that is produced after those operations is the same as the original LP problem, solving for those new constraints (the \ac{CPLP}) yields the same solution as the LP. A test of the algorithm was performed on a 2383-bus Polish system, including 327 generators solved using CPLEX, comparing its performance on four different cloud instances. The method showed scalability, but it wasn't tested against a regular \ac{ED} algorithm. 

Another paradigm that Cloud facilitates is the Many Task Computing paradigm. It facilitates Co-Simulations which involve solving many optimization problems and performing many studies apart then connecting them. In \cite{Overlin2018} they perform a large co-simulation by decomposing a network into heterogeneous partitions that are unique to each other, creating different problems for each partition (e.g., generators, passive components, loads, etc.). The dynamic resource allocation ability fits well with large-scale co-simulations because 1- different components have different transient reactions. 2- They might require different timesteps depending on transient status. 3- Each problem could have a different formulation (\ac{NLP}, \ac{MILP}, etc.) and require different solution times. The paper demonstrates the achievable co-simulation performance and interfacing on the Cloud using existing commercial tools. For example, in one instance, the network was divided into multiple Simulink models, launching Matlab script simulations in different processes. In another trial, a compiled \ac{MPI} C code was used and Simulink executables to run the simulation.

\section{Discussion}

Parallel Applications for power systems started showing up around the late 60s and early 70s, around the same time when a commercial market for supercomputers and clusters was sprouting. At that stage, parallel computers were still experimental in nature, and oftentimes their design targeted a specific problem type or structure. Very few computers were suitable for power system studies as most had low arithmetic precision that's equal to or less than 32-bit, which has been shown to be inadequate for direct solution methods \cite{Happ1979}.  

At an abstract level, computer hardware architecture and its uses in power system studies are still the same. What used to be a ``computer" or ``host" is today's \ac{CPU}, and \ac{SIMD}s like array processors were used just like today's \ac{GPU}s would be used for power system studies. Algorithms which include diakoptics/tearing and tree graphs, used to be a common theme at the start of vectorization and fine-grained parallelism, and it is still used in current \ac{GPU} power system studies. Another example of similarity is that one of the issues faced at the time was that transient simulation timestep iterations sometimes required substantial logic and data to model for each node \cite{Jose1982}. This means it would create a burden on the computer that hosts the array processes and cause communication bottlenecks. This is very analogous to what happens today in \ac{GPU}-\ac{CPU} optimization algorithms. Ironically, S. Jose argued in 1982 \cite{Jose1982} for the need for a general-purpose processor to tackle the previous issues and since vector/array computers pose software hurdles and challenges that are too great to justify the enhancements achieved. Yet the same challenges are faced today, just at a different scale and magnitude (i.e., \ac{GPU}-\ac{CPU} interfacing / Cloud Implementations). 
A major shift in the field occurred around the 90s; around the same time, general-purpose processors experienced significant innovation and cost reduction, and more parallel optimization algorithms started appearing. Studies in power system stability became abundant and \ac{UC} algorithms debuted with most papers using metaheuristics to solve the problem. While implementation would have been arguably doable, simulations of parallel hardware still existed because more care was placed on implementation optimization and ensuring the practicality/portability of the parallel algorithms. The meaning or extent of what is considered coarse-grained and fine-grained algorithms shifted over time. 

The main direction for \ac{HPC} incorporation in power system studies application is moving towards real-time applications, much more so than offline applications. From the literature, it seems that renewable energy generation is the urgent driver for resorting to using AC formulations in real-time applications, followed by annual cost savings of replacing DC formulations. Benders decomposition and Lagrangian relaxation seem to be the most common combination in decomposing stochastic full AC problems. In larger systems, the application of parallel computation is clearly more advantageous, while in smaller systems, serial programming performs better or at least matches parallel computational approaches, mainly due to the communication overhead as an increased number of processes means longer and more communication time between them. The extent of this effect depends highly on the strategies used in parallelization as well as the cluster architecture and hardware used. \ac{GPU}s, for example, exhibit extreme parallelism in processing architecture yet had superior performances over a more coarse \ac{CPU} implementation shown previously \cite{Roberge2017}. Many organizations and research teams are developing public tools and frameworks to help incorporate \ac{HPC} into power system studies. \ac{PNNL} developed HIPPO \cite{Pan} a tool to help grid operators tackle  \ac{SCUC} by leveraging optimization algorithms for \ac{HPC} deployment. \ac{PNNL} also initialized the development of another framework for power system simulations called GridPack, which falls under a larger suite called GridOPTICS \cite{Palmer2016}. While such tools facilitate the use of multi-processor parallelization, others such as Nividia \ac{CUDA}  \cite{Nividia} evolved \ac{GPU}s - which have an immensely parallel architecture - to become easily programmable and spouted the trend of using \ac{GPU}s for scientific calculations showing a promising future. 

On-premise \ac{HPC} is not future-proof as the grid organism keeps on evolving. A power system with $n$ components with each component having $m$ states can have  $m^n$ over all possible states. The Grid is quickly adding more components in terms of quantity and variety, AMI, EVs, IoT, etc. All power studies will keep on growing, and control rooms and operators will also need immediate visualizations for easy information analysis. This means that power system operators will inevitably resort to Cloud services. However, cloud computing has many of its own challenges related to policy, security, and cooperation before any solid adaptation is made. The Optimal placement of data centers depends on various stochastic factors, and the lack of interoperability between providers of cloud services doesn't make this problem any easier. Regulatory compliance in terms of security and access is extremely hard to ensure. Data and process locations are unknown, and it becomes hard to investigate any dysfunction or intrusion. An Efficient recovery mechanism needs to always be in place, and even if the host company's structure or owner-ships change, long-term data viability must be in place. Moreover, the business case for moving to cloud computing needs to be established first, which is different for every entity, and it is difficult to predict the future costs of the services. Lots of preparation and tools need to be created locally to ensure stable operation and inseparability and security, such as handling software licensing issues and data coordination/processing.

Computation aggregation evolved from a single processor to a processor and accelerator to a multi-processor system, Beowulf clusters and multi-core processors then grid. And even at a small level, much like vector arrays and ALUs were added to processors, future \ac{CPU}s and \ac{GPU}s will be integrated into the same device, and the cycle continues. In the future, the Cloud will be an integral part of all operational entities, including the electrical industry. The future electrical Grid and Cloud will look very different from today, both will be dynamic and transactive and will have a reciprocal relationship where the Cloud acts as the brain of the electrical network, and both will probably be driven by similar forces. 

\subsection{Software and Solvers}

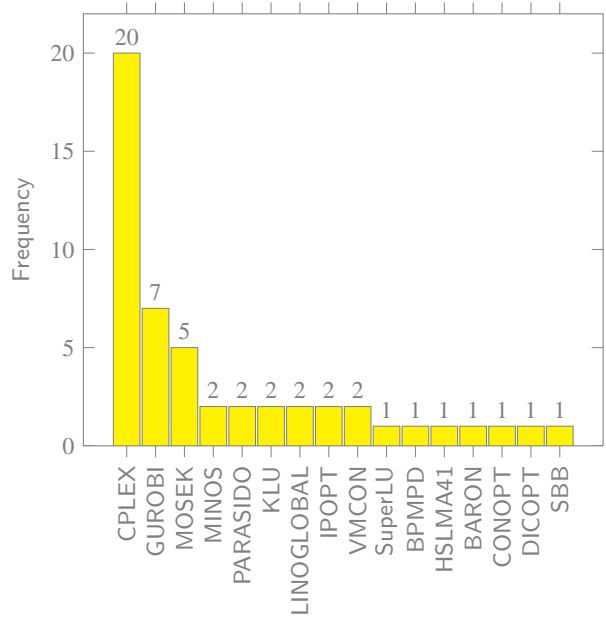
\begin{figure}
    \centering
\begin{tikzpicture}
\begin{axis}[
    ybar ,
    ymin=0,
    legend style={at={(0.5,-0.15)},
      anchor=north,legend columns=-1},
    ylabel={Frequency},
    ymin=0,
    symbolic x coords={CPLEX,GUROBI,MOSEK,MINOS,PARASIDO,KLU,LINOGLOBAL,IPOPT,VMCON,SuperLU,BPMPD,HSLMA41,BARON,CONOPT,DICOPT,SBB},
    xtick=data,
    nodes near coords,
    nodes near coords align={vertical},
     xticklabel style={rotate=90},
    ]
\addplot [fill=yellow, ultra thin]   coordinates{(CPLEX,20)(GUROBI,7)(MOSEK,5)(MINOS,2)(PARASIDO,2)(KLU,2)(VMCON,2)(LINOGLOBAL,2)(IPOPT,2)(SuperLU,1)(BPMPD,1)(HSLMA41,1)(BARON,1)(CONOPT,1)(DICOPT,1)(SBB,1)};

\end{axis}
\end{tikzpicture}
    \caption{The occurrences of solvers in the reviewed literature}
    \label{fig12}
\end{figure}

Commercial solver use can be traced back to the sixties with solvers such as the LP/90/94 \cite{MathStories} in conjunction with the development of the field of mathematical programming. Thus today, there is an abundance of open source and commercial solvers that race to employ the best techniques to solve standard problem formulations. This is evident in Fig. \ref{fig12} showing the variety of solvers used.

To a large degree, commercial solvers simplified optimization for engineers allowing them to focus on modeling, leading to the subfield of model decomposition. Nevertheless, a few challenges arise when using commercial solvers instead of employing a specific solution algorithm to the problem. The heuristics involved in solver design could create a vast disparity in performance even for solvers within the same caliber solving the same type of problem. Also, the ability of a solver to identify and exploit the structure of the model heavily determines whether the model can be solved within a reasonable time. If the solver fails to accomplish this step, it might exhibit exponential growth in running time as indicated by complexity analysis. Moreover, hidden bugs and issues with the source code of the solvers could exist, particularly true for commercial solvers. 

Established commercial solvers with full-time development teams such as CPLEX and Gurobi exhibit a more comprehensive dictionary of identifiable problem structures to accommodate the large user base. They are robust, scalable, and capable of handling large search spaces with multithreading and \ac{HPC} exploitation capabilities. Moreover, they are easy to install and interface with many programming languages. Fig \ref{fig12} shows the hierarchy of occurrences of different solvers in the surveyed literature, and it can be observed that the previously mentioned solvers dominate the literature for the previously mentioned reasons. But not to deter from experimenting with non-commercial solvers as they may be superior for specific problems. It is also worth noting that all the well-established general solvers in the tier of CPLEX and Gurobi are \ac{CPU}-based and none exploit \ac{GPU}s in their processes, an area worth exploring \cite{Gurobi1}.

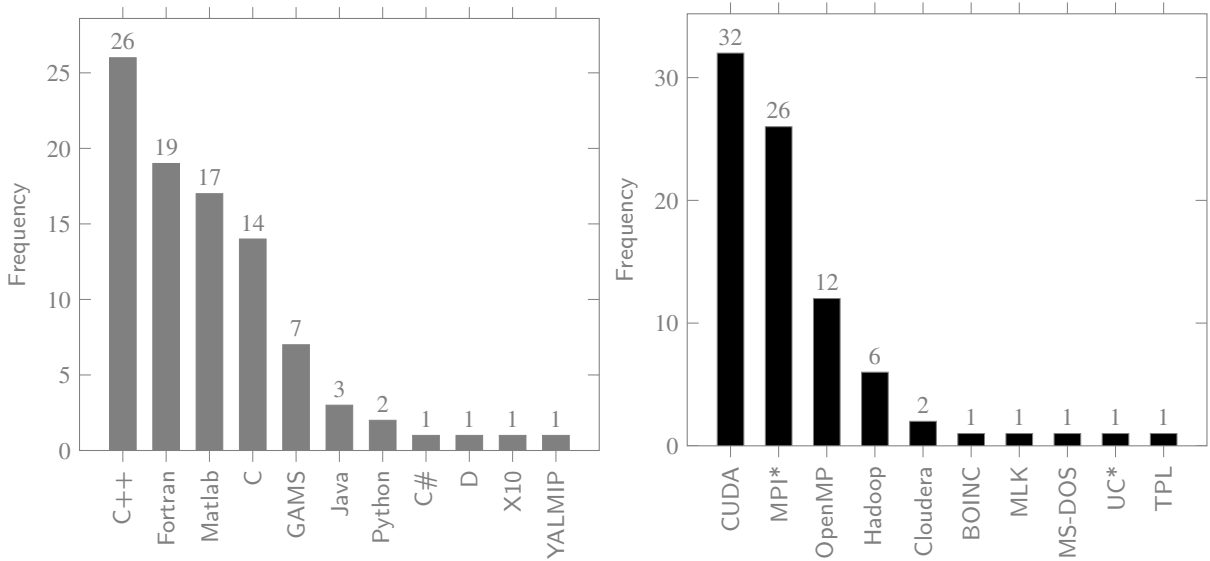
\begin{figure}
    \centering
\begin{tikzpicture}
\begin{axis}[
    ybar ,
    ymin =0,
    legend style={at={(0.5,-0.15)},
      anchor=north,legend columns=-1},
    ylabel={Frequency},
    ymin=0,
    symbolic x coords={C++,Fortran,Matlab,C,GAMS,Java,Python,C\#,D,X10,YALMIP},
    xtick=data,
    nodes near coords,
    nodes near coords align={vertical},
     xticklabel style={rotate=90}
    ]
\addplot[fill=gray, ultra thin]  coordinates{(C++,26)(Fortran,19)(Matlab,17)(C,14)(GAMS,7)(Java,3)(Python,2)(C\#,1) (X10,1)(D,1)(YALMIP,1)};

\end{axis}
\end{tikzpicture}
\begin{tikzpicture}
\begin{axis}[
    ybar ,
    ymin=0,
    legend style={at={(0.5,-0.15)},
      anchor=north,legend columns=-1},
    ylabel={Frequency},
    ymin=0,
    symbolic x coords={CUDA,MPI*,OpenMP, Hadoop,Cloudera,BOINC,MLK,MS-DOS,UC*,TPL},
    xtick=data,
    nodes near coords,
    nodes near coords align={vertical},
     xticklabel style={rotate=90},
    ]
\addplot[fill=black, ultra thin]  coordinates{(CUDA,32)(MPI*,26)(OpenMP,12)(Hadoop,6)(Cloudera,2)(BOINC,1)(MLK,1)(MS-DOS,1)(UC*,1)(TPL,1)};

\end{axis}
\end{tikzpicture}
    \caption{The occurrences of programming languages (left) and \ac{API}s (right) in the reviewed literature. *MPI includes MPICH, mpi4py, MultiMATLAB. UC: Unix Command }
    \label{fig13}
\end{figure}

Compiled and procedural languages such as C and Fortran dominate the literature due to their superior performance, as shown in Fig. \ref{fig13} (a). However, other multi-paradigm multi-paradigm and object-oriented languages (Matlab and python) started to infiltrate the literature due to their simplicity and convenient libraries. Other concurrent programming-oriented languages that might be of interest include Charm, Chapel, Cython, and Julia. Chapel has more advanced parallelism than Julia, while Julia has gained huge popularity since its release recently. Julia is expected to populate future literature due to its heavy emphasis on optimization and C-like performance. In terms of Parallel \ac{API}s, the fast adaptation of \ac{CUDA} as shown in Fig. \ref{fig13} 
(b) testifies the thirst for massive throughput and suggests that in terms of \ac{GPU}-based power system optimization studies, there is to come.  

There exist some integrated high-level frameworks designed to scale certain power system studies on \ac{HPC} such as BELTISTOS \cite{beltistos}, which solves multi-period, security-constrained, and stochastic \ac{OPF} problems incorporating multilevel solution strategy implemented in PARDISO. However, when compared to GridPack, this framework seems quite limited. As part of the HIPPO project mentioned earlier \cite{HIPPO}, \ac{PNNL} developed the software framework GridPack$^{TM}$ that lowers the barrier for power system research and analysis in creating parallel models for \ac{HPC} implementation. \cite{Jin2016}. 

Grid pack automates processes such as determining the Y-Bus of the network and solving \ac{PF} equations, integrating algebraic differential equations, coupling simulation components, distributing network and matrix representations of the network, and employing linear and non-linear solvers. GridPack has a partitioner that partitions the network module buses into several processors where it maximizes the interconnections between buses within the same processer and minimizes the ones between separate processors. It's based on ``Parmetis" partitioning software, which achieves graph mesh partitioning, matrix reordering, etc. The matrices of the distributed matrices of the partitioned network are then distributed by mappers, which determine the contribution buses and branches from each processor by getting the dimensions and locations of elements. The math module generates those matrices and supplies linear and non-linear solvers built on the PETSc library. GridPack also has libraries of already developed, ready-to-use parallel applications. This includes different types of contingency analysis, initialization of dynamic simulation, power flow, and voltage stability analysis.

\subsection{Challenges in the Literature}

This review did not delve into deep comparisons between the different approaches due to the lack of standardization in various aspects of the studies, making it hard to draw meaningful comparisons. These challenges start with network topologies, sizes, a difference in hardware, and a mere lack of information. This is further discussed later in this subsection and can be observed in Table \ref{table1} in the Appendix. 

First, the variety of test cases between studies causes a solution universality issue. Many solution approaches exploit the structure and properties of the problem, such as sparsity and asymmetry, which vary with different network topologies and the number of bus interconnections. The effect of that was evident in \cite{Li2017a} where the parallel \ac{FDPF} scheme performed better on the Pan-European system than the Polish system because it had a more orderly topology. Some studies boast remarkable results on massive synthetic bus systems that are an augmentation of the same small bus system connected with tie lines, such as in  \cite{Zhou2017c} where the IEEE-39 case was copied and connected over 6000 times. This creates a level of symmetry that does not exist in natural systems, one that certainly affects the performance rendition. Moreover, some \ac{SIMD}-based studies use made-up or modified power systems that are very dense, which suits what the hardware is designed for but presents a false or exaggerated sense of performance accomplishment since real systems are generally sparse. 

The second challenge is the lack of details in the experimentation setup essential for replication. Some papers provide the model of the hardware used without the number of threads or processes used and vice versa. Others claim a parallel application without mentioning the communication scheme used or the number of subproblems created. Furthermore, some papers use or compare iterative algorithms such as  ``Traditional ADMM/LR" or other generic algorithms without providing the user-adjusted parameters/heuristics involved in tweaking such algorithms, making it impossible to replicate and verify the results. Also, barely any of the studies explicitly mention the number and type of constraints and variables generated by the formulation and test cases used. 

There is also no standardization in the metrics used to evaluate performance. Some studies use absolute speed up; some use relative speedups. Some compare their parallel approach to a different parallel approach which is weak because the comparison loses its meaning if the proposed parallel approach is inferior to a sequential one. Some suffice by comparing their own parallelized approach to itself applied sequentially (scalability metric), which is problematic because decomposed task could perform worse than a coupled one when applied sequentially.  

The third challenge is the lack of parallel implementation of long-term grid planning models akin to Transmission System Planning or Generation System Planning or their combination. This type of study that helps us plan the transition to the future network struggle with a very small number of factors, accuracy, and uncertainty, and on small test cases that many studies started resorting to decomposition algorithms \cite{KOLTSAKLIS2018563} mainly benders decomposition to decouple investment variables in the models. Yet, it seems that almost non of the studies use or resort to parallel and high-performance computing, which is a huge lost opportunity as we need to add as many factors as possible to find the real optimal path of transitioning and investment given all the future policies technologies and scenarios that we can speculate at the moment. 

The final significant challenge is the lack of standardization in software and hardware used in the studies. The main issue with software is in the variety of solvers used in papers that employ model decomposition schemes. Commercial solvers operate as black boxes that use different techniques, some of which are trade secrets. They are coded with different efficacies and have their bugs and problems, amplifying the confusion in interpretation. 

The lack of hardware standardization in the literature has been highlighted since the 90s \cite{Ng1990}. Even very recently, within supposedly comparative work where four different parallel schemes were compared,  each scheme was performed on a different supercomputer and a different test case \cite{Jin2017}. The single study that provided a meaningful cross-hardware comparison was \cite{Blaskiewicz2015}. They experimented with different \ac{CUDA} routines on different but closely related NIVIDIA \ac{GPU} models, showing that their approach was not superior on every model proving the importance of hardware normalization. 

One way to help tackle the challenge of hardware standardization is by using cloud service instances, such as AWS, as a benchmark, as they are easily accessible globally. Especially since Virtual CPUs (vCPUs) handle the standardization of the heterogeneous hardware and usage (an Elastic Compute Unit is equivalent to the computing power of a 1.0-1.2GHz 2007 Opteron or 2007 Xeon Processor \cite{amazon2021}. Moreover, it fits the industry's trend of shifting computational power to the Cloud. Also, these services come with metric tools that allow the user to look into the actual hardware usage and CPU and memory efficiency of their algorithm. This leads us to the last point: most previous studies merely glance over memory and treat memory resources as a bottleneck rather than a shared and finite resource. More focus on data and memory efficiency is needed. Future studies need to mention the maximum amount of data that needs to be processed and the actual memory usage of their approaches. 

\subsection{Future and Recommendations}

At this point, the role cloud computation would play in power system \ac{HPC} applications and the future Grid is almost unquestionable due to its sheer scale and versatility. The Cloud acting as the brain of the Grid fits the notion of the living organism envisioned by many for the Future Grid. In a few studies, the Cloud has been recognized as the centerpiece for distributed computing paradigms such as fog computing and AMI resource leveraging. With that said, the increasing dependence on centralized cloud computing services is antithetical to the goal of energy decentralization/independence. Yet the trend couldn't be more natural,  manifesting a cycle, how it almost recreates the onset of electrical generation monopolies. The decentralization of computational resources for power systems over micro users, however, doesn't seem to be that far-fetched of an idea, especially with currently existing applications such as blockchains and volunteer computing. 

A lot of the earlier parallel computing studies for power systems modified or created the hardware around algorithms used \cite{Taoka1981}. This hardware manipulation to suit limited computational purposes is making a comeback due to moors law and other limitations. Analog computing hardware is making a comeback as it is way more efficient in matrix multiplications. Rather than turning on and off, analog transistors encode a range of numbers based on the conductance magnitude, which is dictated by the gate. Their level of precision, however, makes them mainly suitable for AI chips and algorithms. Provided that their future precision becomes comparable to digital computers, they might be a contender to \ac{GPU}s in mathematical optimization matrix operations. 

The advancement in Quantum Computing research is creating a creeping disruptor of classical computation and algorithms as we know them. It has been shown that current Quantum Computers can solve combinatorial optimization problems that resemble ones related to energy system problems. Namely on the IBMs D-WAVE solving facility location problem \cite{AJAGEKAR201976}. The potential of applying quantum computation for dynamic stability simulations, \ac{OPF} and \ac{UC} was discussed in the early 2000s \cite{Tylavsky2003}. In fact, the mixed-integer quadratic \ac{UC} problem can be transformed into a Quadratic Unconstrained Binary Optimization (QUBO) by discretizing the problem space, a form that can be turned into a quantum program. In a merely experimental effort, this was actually implemented by \cite{AJAGEKAR201976}, the test systems were very small, from 3 to 12 units, and the solutions of the D-Wave were accurate for a smaller number of units but quickly started deviating. The DC power flow was also implemented on the D-Wave with an HHL algorithm process on a 3-bus test case showing accuracy \cite{eskandarpour2020quantum}. These might be the first experimental efforts in employing quantum computation for operational power system studies. 

Looking back at the studies, one can observe that our current parallel studies don't come anywhere near covering the potential variables of the future Grid. The models are highly simplified and filled with assumptions. The amount of detail, planning factors, and uncertainties are not close to what needs to be considered in grid modernization and future transition. Yet the accuracy and computational performance of the solutions are sometimes not impressive. And even when decomposition techniques are used, and the created parallel structures are exploited with \ac{HPC}, often we are faced with not-so-impressive outcomes, probably due to the lack of understanding and ingenuity in employing the parallel and decomposition and parallel techniques. The hardware that is used in many of the studies is often limited to a multi-core processor limiting the potential throughput. A complicated brain is needed to operate the complicated organism that is the future Grid. In the face of the Grid transformational changes, the power system community needs to start heavily adapting \ac{HPC} techniques and utilization, incorporating them into future operational and planning studies. Moreover, a high level of transparency and collaboration is needed to accelerate the adaptation of parallel techniques making such knowledge the norm in power system studies for the Future Grid.  

The lack of standardization makes it very hard to replicate techniques from different works. Therefore, it is very important to have a standard framework and minimum information requirement in future power system study publications. This is especially important to ensure published models and techniques validity, given the scientific reproducibility crisis \cite{Baker20161500SL} \cite{ReplicationCrisis} \cite{Cockburn2020ThreatsOA}. The following points should serve as a guideline for future parallel studies in the field: 

\begin{enumerate}[label=\roman*.]

\item  A small validation test case, including any modifications, should be presented with all of the parameters and results.
\item All the model expressions must be fully indexed without brevity or detail omissions. This includes both the model pre and post-decomposition. If possible, the full extended model specific to the validation test case should be provided.

\item The pseudocode of the algorithm and flow chart demonstrating the parallel task splitting and synchronization should be included. The values of any tuning factors or heuristic parameters used should be provided.

\item All the platforms, software tools used, parallel strategies and metrics should be specified, this includes:
    \begin{enumerate}
    \item Operating system (e.g. Windows 10)
    \item Coding language (e.g. Python or Julia)
    \item Commercial solvers \& version (e.g. Gurobi 9.0.1)
    \item Parallelization \ac{API} or pacakge (e.g. mpi4py)
    \item Processes communication protocol (e.g. point-to-point or collective, etc.)
    \item Machine used (e.g. local university cluster, personal laptop)
    \item Type of worker allocated and all its model specs (e.g. 8-core 2.1 GHz 4mb intel i-7400) 
    \item Memory allocated and technology (e.g. 10GB DDR5 RAM)
    \item Number of processeses, threads \& allocation per worker (e.g. the 100 subproblems were divided on 5\ac{CPU}s (20 sub-problem/processes per \ac{CPU}, each subproblem was solved using 6 threads (12 hyper-threads) automatically allocated by the solver.
    \item Average \& Peak efficiency of memory and \ac{CPU} usage. (e.g. CPUS efficiency: 100\% peak and 91\% average. Memory utilization: 80\% peak and 40\% average. 
    \end{enumerate}
\item A test should be carried out on test cases incrementally increasing in size with a variety of network topologies to demonstrate the scalability and universality of the proposed method. And an effort should be placed to compare the speed up to the fastest known algorithm.

\end{enumerate}

\section{Conclusion}

In this article, the beginnings of parallel computation and its appearances in power system studies were recounted, and the recent research and literature were reviewed. Several past reviews conducted that were previously conducted were cited, distinguishing this work from them. The significance of hardware, paradigms, and the history of parallel computing was then discussed. Studies of parallel power systems were summarized later, starting by reciting the development of studies up until the 21st century, with emphasis on the most impactful papers from the last decade. Studies included analyses of the stability of the power systems, state estimation and operation of the power systems, and market optimization. The state-of-the-art was also discussed, highlighting the need for standardization in the literature and showcasing the future of computation in power system studies. Given the grid modernization and transition towards net-zero emission, power systems are increasingly becoming more complex and resorting to high-performance, and parallel computing and cloud computing has never been more important.  

\section*{Acknowledgement}
We thank Manuel Zamudio Lopez (University of Calgary) for his comments on the manuscript.

\printcredits

\bibliographystyle{vancouver1}

\bibliography{vancbib}

\begin{thebibliography}{100}

\bibitem{FACTSHEET_2021}
{The White House}. Building on past u.s. Leadership, including efforts by
  states, cities, tribes, and territories, the new target aims at 50-52 percent
  reduction in u.s.; 2021.
\newblock [Online; accessed 29-Nov-2021].
\newblock Available from: \url{https://bit.ly/3xEDGSn}.

\bibitem{CoalExitTimeline}
{Coal Exit}; 2022.
\newblock [Online; accessed 25-May-2022].
\newblock \url{https://beyond-coal.eu/coal-exit-timeline/}.

\bibitem{Mapped_2020}
{Carbon Brief}; 2020.
\newblock [Online; accessed 29-Nov-2021].
\newblock Available from: \url{https://bit.ly/3HFuvFR}.

\bibitem{Carboncapture}
IEA; 2022.
\newblock [Online; accessed 25-May-2022].
\newblock \url{https://bit.ly/3QB9fFh}.

\bibitem{Ritchie_Roser_Rosado_2020}
Ritchie H, Roser M, Rosado P.
\newblock Energy.
\newblock Our World in Data  2020 Nov.
\newblock Available from: \url{https://ourworldindata.org/renewable-energy}.

\bibitem{Renewableaccelerating}
IEA; 2022.
\newblock [Online; accessed 25-May-2022].
\newblock \url{https://bit.ly/3njRMEd}.

\bibitem{ltdelectrif}
ltd R, Markets. Industrial electrification growth opportunities - research and
  markets; 2022.
\newblock [Online; accessed 25-May-2022].
\newblock Available from: \url{https://bit.ly/3QKmIej}.

\bibitem{TrendsEV}
IEA; 2021.
\newblock Available from: \url{https://bit.ly/3becrGP}.

\bibitem{GONZALEZTORRES2022626}
González-Torres M, Pérez-Lombard L, Coronel JF, Maestre IR, Yan D.
\newblock A review on buildings energy information: trends, end-uses, fuels and
  drivers.
\newblock Energy Reports  2022;8:626-37.
\newblock Available from:
  \url{https://www.sciencedirect.com/science/article/pii/S235248472101427X}.

\bibitem{B2EIndustry}
Industry B; 2021.
\newblock Available from: \url{https://bit.ly/3HFyceH}.

\bibitem{ZHU2021125209}
Zhu X, Liu K.
\newblock A systematic review and future directions of the sharing economy:
  business models, operational insights and environment-based utilities.
\newblock J Cleaner Prod  2021;290:125209.
\newblock Available from:
  \url{https://www.sciencedirect.com/science/article/pii/S0959652620352537}.

\bibitem{Frequentlysmart}
EIA.
\newblock Annual electric power industry report; 2021.
\newblock Available from: \url{https://www.eia.gov/tools/faqs/faq.php}.

\bibitem{BORENIUS2021107367}
Borenius S, Hämmäinen H, Lehtonen M, Ahokangas P.
\newblock Smart grid evolution and mobile communications—scenarios on the
  finnish power grid.
\newblock Electr Pow Syst Res  2021;199:107367.
\newblock Available from:
  \url{https://www.sciencedirect.com/science/article/pii/S0378779621003485}.

\bibitem{9505699}
Hua H, Liu T, He C, Nan L, Zeng H, Hu X, et~al.
\newblock Day-ahead scheduling of power system with short-circuit current
  constraints considering transmission switching and wind generation.
\newblock IEEE Access  2021;9:110735-45.

\bibitem{8663395}
Daly P, Qazi HW, Flynn D.
\newblock Rocof-constrained scheduling incorporating non-synchronous
  residential demand response.
\newblock IEEE Trans Power Syst  2019;34(5):3372-83.

\bibitem{9215152}
Nawaz A, Wang H.
\newblock Distributed stochastic security constrained unit commitment for
  coordinated operation of transmission and distribution system.
\newblock CSEE Journal of Power and Energy Systems  2021;7(4):708-18.

\bibitem{9250454}
Luburić Z, Pandžić H, Carrión M.
\newblock Transmission expansion planning model considering battery energy
  storage, tcsc and lines using ac opf.
\newblock IEEE Access  2020;8:203429-39.

\bibitem{8931650}
Zhuo Z, Zhang N, Yang J, Kang C, Smith C, O’Malley MJ, et~al.
\newblock Transmission expansion planning test system for ac/dc hybrid grid
  with high variable renewable energy penetration.
\newblock IEEE Trans Power Syst  2020;35(4):2597-608.

\bibitem{GONZALEZROMERO2019154}
Gonzalez-Romero IC, Wogrin S, Gomez T.
\newblock Proactive transmission expansion planning with storage
  considerations.
\newblock Energy Strategy Reviews  2019;24:154-65.
\newblock Available from:
  \url{https://www.sciencedirect.com/science/article/pii/S2211467X19300227}.

\bibitem{CarbonTaxBasics}
Basics CT.
\newblock State and trends of carbon pricing 2021; 2021.
\newblock Available from:
  \url{https://www.c2es.org/content/carbon-tax-basics/}.

\bibitem{9011131}
Prahastono I, Sinisuka NI, Nurdin M, Nugraha H.
\newblock A review of feed-in tariff model (fit) for photovoltaic (pv).
\newblock In: 2019 2nd International Conference on High Voltage Engineering and
  Power Systems (ICHVEPS); 2019. p. 076-9.

\bibitem{8970543}
Oprea SV, Bâra A.
\newblock Setting the time-of-use tariff rates with nosql and machine learning
  to a sustainable environment.
\newblock IEEE Access  2020;8:25521-30.

\bibitem{SHAHZAD2020106739}
Shahzad Y, Javed H, Farman H, Ahmad J, Jan B, Zubair M.
\newblock Internet of energy: opportunities, applications, architectures and
  challenges in smart industries.
\newblock Computers \& Electrical Engineering  2020;86:106739.
\newblock Available from:
  \url{https://www.sciencedirect.com/science/article/pii/S0045790620305942}.

\bibitem{9248892}
Kaya O, van~der Roest E, Vries D, Keviczky T.
\newblock Hierarchical model predictive control for energy management of
  power-to-x systems.
\newblock In: 2020 IEEE PES Innovative Smart Grid Technologies Europe
  (ISGT-Europe); 2020. p. 1094-8.

\bibitem{9535388}
Bokkisam HR, Singh S, Acharya RM, Selvan MP.
\newblock Blockchain-based peer-to-peer transactive energy system for community
  microgrid with demand response management.
\newblock CSEE Journal of Power and Energy Systems  2022;8(1):198-211.

\bibitem{PowerledgerEnergyProjects}
Ledger P.
\newblock {Utility tokens and erc-20 icos: where are we now?}; 2022.
\newblock Available from:
  \url{https://gbbcouncil.org/wp-content/uploads/2020/05/GBBC-Open-Source-Ideas-Special-Report-Utility-Tokens-and-ERC-20-ICOS.pdf}.

\bibitem{Conejo2019}
Conejo AJ, Baringo L.
\newblock Power electronics and power systems power system operations.
\newblock Springer Cham; 2019.
\newblock Available from: \url{http://www.springer.com/series/6403}.

\bibitem{CooperWilliamW.2010}
{Cooper William W } SLM, Joe Z.
\newblock International series in operations research {\&} management science
  introduction. vol. 139.
\newblock Springe; 2010.

\bibitem{NERC2019}
NERC.
\newblock Reliability standards for the bulk electric systems of north america.
\newblock Standard  2019:1-1044.
\newblock Available from:
  \url{http://www.nerc.com/docs/standards/rs/Reliability{\_}Standards{\_}Complete{\_}Set.pdf}.

\bibitem{Ma2016}
Ma F, Luo X, Litvinov E.
\newblock Cloud computing for power system simulations at iso new england -
  experiences and challenges.
\newblock IEEE Trans Smart Grid  2016;7(6):2596-603.

\bibitem{Alam2018}
Alam A, Gopinathan G, Shrestha B, Zhao R, Wu J, Xu R.
\newblock High preformance computing for operations and trnasmission planning
  at caiso.
\newblock 2018 IEEE/PES Transmission and Distribution Conference and Exposition
   2018:3-7.

\bibitem{Chen2020}
Chen Y, Member S, Wang F, Ma Y, Yao Y.
\newblock A distributed framework for solving and benchmarking security
  constrained unit commitment with warm start.
\newblock IEEE Trans Power Syst  2020;35(1):711-20.

\bibitem{E.M.Constantinescu2011}
Constantinescu EM, Zavala VM, M~Rocklin SL, Anitescu M.
\newblock A computational framework for uncertainty quantification and
  stochastic optimization in unit commitment with wind power generation.
\newblock IEEE Trans Power Syst  2011;26(1):431-41.

\bibitem{FERC2013}
O'Neill R, Castillo A, Cain B.
\newblock Optimal power flow and formulation papers; 2013.

\bibitem{Happ1969}
Happ HH, Undrill JM.
\newblock Multicomputer configurations and diakoptics: real power flow in power
  pools.
\newblock IEEE Trans Power App Syst  1969;PAS-88(6):789-96.

\bibitem{Ewart1971}
Ewart DN, Kirchmayer LK.
\newblock Automation and utility system security.
\newblock IEEE Spectr  1971;8(7):37-42.

\bibitem{Narita1973}
Narita S, Hammam MSAA.
\newblock Multicomputer control of system voltage and reactive power on
  real-time basis.
\newblock IEEE Trans Power App Syst  1973;PAS-92(1):278-86.

\bibitem{Wu1976}
Wu F.
\newblock Solution of large-scale networks by tearing.
\newblock IEEE Trans Circuits Syst  1976;23(12):706-13.

\bibitem{P.M.Anderson}
Anderson PM.
\newblock Exploring applications of parallel processing to power system
  analysis problems.
\newblock Seminar Proceeding Special Report. Electric Power Research Institute;
  1977.

\bibitem{Podmore1982}
Podmore R.
\newblock Application of an array processor for power system network
  computations.
\newblock Build Environ  1982;17(2):95-105.

\bibitem{Tylavsky1992}
Tylavsky CcDJ, Bose A, Alvarado MF, Betancourt R, Clements K, Heydt GT, et~al.
\newblock Parallel processing in power systems computation.
\newblock IEEE Trans Power Syst  1992;7(2):629-38.

\bibitem{Bose1993}
Bose A.
\newblock Parallel processing in dynamic simulation of power systems.
\newblock Sadhana  1993;18(5):815-41.

\bibitem{Falcao1996}
Falcao DM.
\newblock High perfromance computing in power system applications.
\newblock Lecture Notes in Computer Science (including subseries Lecture Notes
  in Artificial Intelligence and Lecture Notes in Bioinformatics)
  1996;1215(June):v.

\bibitem{Ramesh1996}
Ramesh VC.
\newblock On distributed computing for on-line power system applications.
\newblock Electrical Power \& Energy Systems  1996;18(8):527-33.

\bibitem{Ii2011}
Ii RCG, Wang L, Alam M.
\newblock High performance computing for electric power systems : applications
  and trends.
\newblock 2011 IEEE Power and Energy Society General Meeting  2011:1-8.

\bibitem{Green2013}
Green RC, Wang L, Alam M.
\newblock Applications and trends of high performance computing for electric
  power systems: focusing on smart grid.
\newblock IEEE Trans Smart Grid  2013;4(2):922-31.

\bibitem{TAN2019727}
Tan L, Jiang J.
\newblock Chapter 14 - hardware and software for digital signal processors.
\newblock In: Tan L, Jiang J, editors. Digital Signal Processing (Third
  Edition). third edition ed.  Academic Press; 2019. p. 727-84.
\newblock Available from:
  \url{https://www.sciencedirect.com/science/article/pii/B9780128150719000142}.

\bibitem{ChipFame_2018}
Aspray W.
\newblock The intel 4004 microprocessor: what constituted invention?
\newblock IEEE Annals Hist Comput  1997 Jul;19(3).
\newblock Arxiv article.

\bibitem{Stringer_2016}
Stringer L. Stringer L, editor. Vectors: how the old became new again in
  supercomputing. hpcwire; 2016.
\newblock Available from: \url{https://bit.ly/3HEa0Jz}.

\bibitem{Transputers1990}
Hey AJG.
\newblock Supercomputing with transputers—past, present and future.
\newblock In: Proceedings of the 4th International Conference on Supercomputing
   ICS '90. New York, NY, USA: Association for Computing Machinery; 1990. p.
  479-89.

\bibitem{Bose2011}
Bose P.
\newblock Power wall.
\newblock In: Padua D, editor. encyclopedia of parallel computing  Boston, MA:
  Springer US; 2011. p. 1593-608.

\bibitem{IntelProc}
Intel.
\newblock 8th and 9th generation intel core processor families and intel xeon e
  processor families; 2020.

\bibitem{vectorization2021}
Intel. Intel, editor. {Intel oneAPI DPC++/C++ Compiler developer guide and
  reference}. Intel; 2022.
\newblock [Online; accessed 29-Nov-2021].
\newblock
  \url{https://software.intel.com/content/www/us/en/develop/documentation/cpp-compiler-developer-guide-and-reference/top/optimization-and-programming-guide/vectorization/automatic-vectorization/programming-guidelines-for-vectorization.html}.

\bibitem{intel512}
Intel.
\newblock Intel {AVX}-512 - instruction set for packet processing.
\newblock Intel Corporation; 2021.
\newblock [Online; accessed 29-Nov-2021].

\bibitem{Productxeon}
Intel.
\newblock Intel xeon silver 4110 processor.
\newblock Inter Coorperation; 2018.
\newblock Available from:
  \url{https://www.intel.com/content/www/us/en/products/sku/123547/intel-xeon-silver-4110-processor-11m-cache-2-10-ghz.html}.

\bibitem{altra2021}
Computing A.
\newblock Ampere altra max 64-bit multi-core processor features.
\newblock Ampere Computing; 2021.
\newblock [Online; accessed 29-Nov-2021].

\bibitem{Forbes2021}
Teich P. Teich P, editor. Nvidia dominates the market for cloud ai accelerators
  more than you think. Forbes; 2021.
\newblock [Online; accessed 29-Nov-2021].
\newblock \url{https://bit.ly/3QIvdqa}.

\bibitem{navarro2014}
Navarro CA, Hitschfeld-Kahler N, Mateu L.
\newblock A survey on parallel computing and its applications in data-parallel
  problems using gpu architectures.
\newblock Comm Comput Phys  2014;15(2):285-329.

\bibitem{ARMLtd}
ARM. ARM, editor. Defining the future of computing. Arm | The Architecture for
  the Digital World; 2022.
\newblock Available from: \url{https://www.arm.com/}.

\bibitem{ARPANET}
Leiner BM, Cerf VG, Clark DD, Kahn RE, Kleinrock L, Lynch DC, et~al.
\newblock Brief history of the internet.
\newblock Internet Society; 1997.
\newblock Available from:
  \url{https://online.jefferson.edu/business/internet-history-timeline/}.

\bibitem{UNIX}
Spinellis D.
\newblock A repository of unix history and evolution.
\newblock Empirical Softw Engg  2017 Jun;22(3):1372-404.

\bibitem{InternetTimeline_2016}
Stott M.
\newblock {ARCNETworks}.
\newblock Arcnet Trade Association; 1998.
\newblock Available from:
  \url{https://www.darpa.mil/about-us/timeline/arpanet}.

\bibitem{9388639}
Majidha~Fathima KM, Santhiyakumari N.
\newblock A survey on evolution of cloud technology and virtualization.
\newblock In: 2021 Third International Conference on Intelligent Communication
  Technologies and Virtual Mobile Networks (ICICV); 2021. p. 428-33.

\bibitem{AlanBeck1997}
Beck A. Beck A, editor. {High throughput computing: an interview with miron
  livny}. Research.cs.wisc.edu; 2021.
\newblock [Online; accessed 29-Nov-2021].
\newblock \url{https://bit.ly/3y2Tuje}.

\bibitem{4777912}
Raicu I, Foster IT, Yong Z.
\newblock Many-task computing for grids and supercomputers.
\newblock In: 2008 Workshop on Many-Task Computing on Grids and Supercomputers;
  2008. p. 1-11.

\bibitem{globus_2021}
Globus. Globus, editor. Globus toolkit. Toolkit.globus.org; 2021.
\newblock [Online; accessed 29-Nov-2021].
\newblock \url{https://toolkit.globus.org/}.

\bibitem{EC2amazon}
Amazon.
\newblock Overview of amazon web services: aws whitepaper.
\newblock Amazon Web Services; 2022.
\newblock [Online; accessed 29-Nov-2021].

\bibitem{Geist1992}
Geist GA, Sunderam VS.
\newblock Network-based concurrent computing on the pvm system.
\newblock Concurrency: Pract Exper  1992 Jun;4(4):293-311.

\bibitem{UTM}
UTM. UTM, editor. Run virtual machines on iOS. UTM; 2022.
\newblock Available from: \url{https://getutm.app/}.

\bibitem{apptainer}
Apptainer. Apptainer, editor. Apptainer container engine. Apptainer; 2022.
\newblock Available from: \url{https://bit.ly/3bcICXb}.

\bibitem{Docker_2022}
Docker. Docker, editor. Docker container engine. Docker; 2022.
\newblock Available from: \url{https://www.docker.com/}.

\bibitem{gimps}
Mersenne. Mersenne, editor. Great internet mersenne prime search - primenet.
  GIMPS; 2022.
\newblock [Online; accessed 15-Mar-2022].
\newblock \url{https://www.mersenne.org/}.

\bibitem{boinc}
BOINC. BOINC, editor. News from boinc projects. BOINC; 2022.
\newblock [Online; accessed 15-Mar-2022].
\newblock \url{https://boinc.berkeley.edu/}.

\bibitem{appleserver}
Apple. Apple, editor. Macos server. Apple; 2022.
\newblock [Online; accessed 15-Mar-2022].
\newblock \url{https://www.apple.com/macos/server/}.

\bibitem{DASKgraph}
DASK. DASK, editor. Task graph optimization. DASK; 2022.
\newblock [Online; accessed 15-Mar-2022].
\newblock \url{https://docs.dask.org/en/stable/optimize.html}.

\bibitem{Wu1995}
Wu JQ, Bose A, Huang JA, Valette A, Lafrance F.
\newblock Parallel implementation of power system transient stability analysis.
\newblock IEEE Trans Power Syst  1995;10(3):1226-33.

\bibitem{parallel02}
Mccool MD.
\newblock Parallel programming, chapter 2.
\newblock No. March in 1. Morgan Kaufmann; 2012.

\bibitem{ADAMS198517}
Adams L.
\newblock Cosmic ray effects in microelectronics.
\newblock Microelectron J  1985;16(2):17-29.
\newblock Available from:
  \url{https://www.sciencedirect.com/science/article/pii/S0026269285802135}.

\bibitem{Flight72}
Government A.
\newblock {Aviation safety investigations \& reports: In-flight upset - Airbus
  A330-303, VH-QPA, 154 km west of learmonth, WA, 7 October 2008}.
\newblock Ausstralian Transport Safety Bureau; 2021.
\newblock [Online; accessed 29-Nov-2021].

\bibitem{Taoka1981}
Taoka H, Abe S, Takeda S.
\newblock Multiprocessor system for power system analysis.
\newblock Ifac Work S  1981:101-6.

\bibitem{Ward1956}
Ward JB, Hale HW.
\newblock Digital computer solution of power-flow problems [includes
  discussion].
\newblock Transactions of the American Institute of Electrical Engineers Part
  III: Power Apparatus and Systems  1956;75(3):398-404.

\bibitem{Chan1982}
Hulskamp JP, Chan Sm, Fazio JF.
\newblock Power flow outage studies using an array processor.
\newblock IEEE Trans Power App Syst  1982;PAS-101(1):254-61.

\bibitem{Foertsch2005}
Foertsch J, Johnson J, Nagvajara P.
\newblock Jacobi load flow accelerator using fpga.
\newblock In: Proceedings of the 37th Annual North American Power Symposium,
  2005.; 2005. p. 448-54.

\bibitem{Wang2007}
Wang X, Ziavras SG, Nwankpa C, Johnson J, Nagvajara P.
\newblock Parallel solution of newton’s power flow equations on configurable
  chips.
\newblock International Journal of Electrical Power \& Energy Systems
  2007;29(5):422-31.
\newblock Available from:
  \url{https://www.sciencedirect.com/science/article/pii/S014206150600192X}.

\bibitem{Happ1966}
Happ HH.
\newblock Special cases of orthogonal networks - tree and link.
\newblock IEEE Trans Power App Syst  1966;PAS-85(8):880-91.

\bibitem{Happ1967}
Happ HH.
\newblock Z diakoptics - torn subdivisions radially attached.
\newblock IEEE Trans Power App Syst  1967;PAS-86(6):751-69.

\bibitem{Carre1968}
Carre BA.
\newblock Solution of load-flow problems by partitioning systems into trees.
\newblock IEEE Trans Power App Syst  1968;PAS-87(11):1931-8.

\bibitem{Andretich1968}
Andretich RG, Brown HE, Happ HH, Person CE.
\newblock The piecewise solution of the impedance matrix load flow.
\newblock IEEE Trans Power App Syst  1968;PAS-87(10):1877-82.

\bibitem{Takatoo1985}
Takatoo M, Abe S, Bando T, Hirasawa K, Goto M, Kato T, et~al.
\newblock Floating vector processor for power system simulation.
\newblock IEEE Power Eng Rev  1985;PER-5(12):29-30.

\bibitem{Lau1991}
Lau K, Tylavsky DJ, Bose A.
\newblock Coarse grain scheduling in parallel triangular factorization and
  solution of power system matrices.
\newblock IEEE Trans Power Syst  1991;6(2):708-14.

\bibitem{Huang1994}
Huang G, Ongsakul W.
\newblock An adaptive sor algorithm and its parallel implementation for power
  system applications.
\newblock In: Proceedings of 1994 6th IEEE Symposium on Parallel and
  Distributed Processing; 1994. p. 84-91.

\bibitem{Greenberg1990}
Gomez A, Betancourt R.
\newblock Implementation of the fast decoupled load flow on a vector computer.
\newblock IEEE Trans Power Syst  1990;5(3):977-83.

\bibitem{Gomez1990}
Gomez A, Betancourt R.
\newblock Implementation of the fast decoupled load flow on a vector computer.
\newblock IEEE Trans Power Syst  1990;5(3):977-83.

\bibitem{Housos1982}
Housos EC, Wing O.
\newblock Parallel optimization with applications to power systems.
\newblock IEEE Trans Power App Syst  1982;PAS-101(1):244-8.

\bibitem{Chen2000}
Chen SD, Chen JF.
\newblock Fast load flow using multiprocessors.
\newblock International Journal of Electrical Power \& Energy Systems
  2000;22(4):231-6.
\newblock Available from:
  \url{https://www.sciencedirect.com/science/article/pii/S0142061599000538}.

\bibitem{Chen2005}
Chen SD, Chen JF.
\newblock A novel approach based on global positioning system for parallel load
  flow analysis.
\newblock International Journal of Electrical Power \& Energy Systems
  2005;27(1):53-9.
\newblock Available from:
  \url{https://www.sciencedirect.com/science/article/pii/S0142061504001048}.

\bibitem{Feng2002}
Feng T, Flueck AJ.
\newblock A message-passing distributed-memory newton-gmres parallel power flow
  algorithm.
\newblock In: IEEE Power Engineering Society Summer Meeting,. vol.~3; 2002. p.
  1477-82 vol.3.

\bibitem{Yalou2010}
Li Y, Li F, Li W.
\newblock Parallel power flow calculation based on multi-port inversed matrix
  method.
\newblock In: 2010 International Conference on Power System Technology; 2010.
  p. 1-6.

\bibitem{Sun2015}
Sun H, Guo Q, Zhang B, Guo Y, Li Z, Wang J.
\newblock Master–slave-splitting based distributed global power flow method
  for integrated transmission and distribution analysis.
\newblock IEEE Trans Smart Grid  2015;6(3):1484-92.

\bibitem{Su2018}
Su X, Liu T, Wu L.
\newblock Fine-grained fully parallel power flow calculation by incorporating
  bbdf method into a multistep nr algorithm.
\newblock IEEE Trans Power Syst  2018;33(6):7204-14.

\bibitem{Garcia2010}
Garcia N.
\newblock Parallel power flow solutions using a biconjugate gradient algorithm
  and a newton method: a gpu-based approach.
\newblock In: IEEE PES General Meeting; 2010. p. 1-4.

\bibitem{Singh2010}
Singh J, Aruni I.
\newblock Accelerating power flow studies on graphics processing unit.
\newblock In: 2010 Annual IEEE India Conference (INDICON); 2010. p. 1-5.

\bibitem{Dag2011}
Dağ H, Soykan G.
\newblock Power flow using thread programming.
\newblock In: 2011 IEEE Trondheim PowerTech; 2011. p. 1-5.

\bibitem{Vilacha2011}
Vilachá C, Moreira JC, Míguez E, Otero AF.
\newblock Massive jacobi power flow based on simd-processor.
\newblock In: 2011 10th International Conference on Environment and Electrical
  Engineering; 2011. p. 1-4.

\bibitem{Mei2012}
Yang M, Sun C, Li Z, Cao D.
\newblock An improved sparse matrix-vector multiplication kernel for solving
  modified equation in large scale power flow calculation on cuda.
\newblock In: Proceedings of The 7th International Power Electronics and Motion
  Control Conference. vol.~3; 2012. p. 2028-31.

\bibitem{XueLi2013}
Xue L, Fangxing L, Clark JM.
\newblock Exploration of multi-frontal method with gpu in power flow
  computation.
\newblock In: 2013 IEEE Power Energy Society General Meeting; 2013. p. 1-5.

\bibitem{Ablakovic2012}
Ablakovic D, Dzafic I, Kecici S.
\newblock Parallelization of radial three-phase distribution power flow using
  gpu.
\newblock In: 2012 3rd IEEE PES Innovative Smart Grid Technologies Europe (ISGT
  Europe); 2012. p. 1-7.

\bibitem{Blaskiewicz2015}
Blaskiewicz P, Zawada M, Balcerek P, Dawidowski P.
\newblock An application of gpu parallel computing to power flow calculation in
  hvdc networks.
\newblock In: 2015 23rd Euromicro International Conference on Parallel,
  Distributed, and Network-Based Processing; 2015. p. 635-41.

\bibitem{Huang2017}
Huang RH, Jin S, Chen Y, Diao R, Palmer B, Qiuhua.
\newblock Faster than real-time dynamic simulation for large-size power system
  with detailed dynamic models using high-performance computing platform.
\newblock IEEE Power \& Energy Society General Meeting  2017:0-4.

\bibitem{Guo2012}
Guo C, Jiang B, Yuan H, Yang Z, Wang L, Ren S.
\newblock Performance comparisons of parallel power flow solvers on gpu system.
\newblock 2012 IEEE International Conference on Embedded and Real-Time
  Computing Systems and Applications Performance  2012:232-9.

\bibitem{Wang2017}
Wang Y, Member S, Wu L, Member S, Wang S, Member S.
\newblock A fully-decentralized consensus-based admm approach for dc-opf with
  demand response.
\newblock IEEE Trans Smart Grid  2017;8(6):2637-47.

\bibitem{Marin2017}
Marin M, Defour D, Milano F.
\newblock Asynchronous power flow on graphic processing units.
\newblock In: 2017 25th Euromicro International Conference on Parallel,
  Distributed and Network-based Processing (PDP); 2017. p. 255-61.

\bibitem{Gnanvignesh2019}
Gnanavignesh R, Shenoy UJ.
\newblock Gpu-accelerated sparse lu factorization for power system simulation.
\newblock In: 2019 IEEE PES Innovative Smart Grid Technologies Europe
  (ISGT-Europe); 2019. p. 1-5.

\bibitem{Tang2019}
Tang K, Fang R, Wang X, Dong S, Song Y.
\newblock Mass expression evaluation parallel algorithm based on ‘expression
  forest’ and its application in power system calculation.
\newblock In: 2019 IEEE Power Energy Society General Meeting (PESGM); 2019. p.
  1-5.

\bibitem{ARAUJO2019}
Araújo I, Tadaiesky V, Cardoso D, Fukuyama Y, Ádamo Santana.
\newblock Simultaneous parallel power flow calculations using hybrid cpu-gpu
  approach.
\newblock International Journal of Electrical Power \& Energy Systems
  2019;105:229-36.
\newblock Available from:
  \url{https://www.sciencedirect.com/science/article/pii/S0142061518304320}.

\bibitem{DongHee2020}
Yoon DH, Han Y.
\newblock Parallel power flow computation trends and applications: a review
  focusing on gpu.
\newblock Energies  2020;13(9).
\newblock Available from: \url{https://www.mdpi.com/1996-1073/13/9/2147}.

\bibitem{Daher2022}
Daher~Daibes JV, Brown Do Coutto~Filho M, Stacchini~de Souza JC, Gonzalez~Clua
  EW, Zanghi R.
\newblock Experience of using graphical processing unit in power flow
  computation.
\newblock Concurrency and Computation: Practice and Experience
  2022;34(6):e6762.

\bibitem{Abhyankar2021}
Abhyankar S, Peles S, Rutherford R, Mancinelli A.
\newblock Evaluation of ac optimal power flow on graphical processing units.
\newblock In: 2021 IEEE Power Energy Society General Meeting (PESGM); 2021. p.
  01-5.

\bibitem{Dag1999}
Dag H, Alvarado FL.
\newblock Computation-free preconditioners for the parallel solution of power
  system problems.
\newblock IEEE Trans Power Syst  1997;12(2):585-91.

\bibitem{Li2014}
Li X, Li F.
\newblock Gpu-based power flow analysis with chebyshev preconditioner and
  conjugate gradient method.
\newblock Electr Pow Syst Res  2014;116:87-93.

\bibitem{Li2015}
Li X, Li F.
\newblock Gpu-based two-step preconditioning for conjugate gradient method in
  power flow.
\newblock In: 2015 IEEE Power Energy Society General Meeting; 2015. p. 1-5.

\bibitem{Li2017a}
Li X, Li F, Yuan H, Member S.
\newblock Gpu-based fast decoupled power flow with preconditioned iterative
  solver and inexact newton method.
\newblock IEEE Trans Power Syst  2017;32(4):2695-703.

\bibitem{Wang2018}
Wang M, Chen Y, Huang S.
\newblock Gpu-based power flow analysis with continuous newton ' s method.
\newblock IEEE Conference on Energy Internet and Energy System Integration
  (EI2)  2018.

\bibitem{Su2020}
Su X, He C, Liu T, Wu L.
\newblock Full parallel power flow solution: a gpu-cpu-based vectorization
  parallelization and sparse techniques for newton-raphson implementation.
\newblock IEEE Trans Smart Grid  2020;11(3):1833-44.

\bibitem{ZHOU2019}
Zhou G, Feng Y, Bo R, Zhang T.
\newblock Gpu-accelerated sparse matrices parallel inversion algorithm for
  large-scale power systems.
\newblock International Journal of Electrical Power\& Energy Systems
  2019;111:34-43.
\newblock Available from:
  \url{https://www.sciencedirect.com/science/article/pii/S0142061518325109}.

\bibitem{Zhou2018}
Zhou G, Bo R, Chien L, Zhang X, Yang S, Su D.
\newblock Gpu-accelerated algorithm for online probabilistic power flow.
\newblock IEEE Trans Power Syst  2018;33(1):1132-5.

\bibitem{Zhou2017}
Zhou G, Bo R.
\newblock Gpu-based batch lu-factorization solver for concurrent analysis of
  massive power flows.
\newblock IEEE Trans Smart Grid  2017;32(6):4975-7.

\bibitem{Kardos2020}
Kardo{\v{s}} J, Kourounis D, Schenk O, Member S.
\newblock Two-level parallel augmented schur complement interior-point
  algorithms for the solution of security constrained optimal power flow
  problems.
\newblock 1340 IEEE TRANSACTIONS ON POWER SYSTEMS  2020;35(2):1340-50.

\bibitem{beltistos}
Beltistos. Beltistos, editor. Beltistos. Beltistos; 2022.
\newblock [Online; accessed 29-Nov-2021].
\newblock \url{http://www.beltistos.com/}.

\bibitem{CARPENTIER19793}
Carpentier J.
\newblock Optimal power flows.
\newblock International Journal of Electrical Power \& Energy Systems
  1979;1(1):3-15.
\newblock Available from:
  \url{https://www.sciencedirect.com/science/article/pii/0142061579900267}.

\bibitem{Huneault1991}
Huneault M, Galiana FD.
\newblock A survey of the optimal power flow literature.
\newblock IEEE Trans Power Syst  1991;6(2):762-70.

\bibitem{Shoults1982}
Shoults RR, Sun DT.
\newblock Optimal power flow based upon p-q decomposition.
\newblock IEEE Trans Power App Syst  1982;PAS-101(2):397-405.

\bibitem{Talukdar1983}
Talukdar SN, Giras TC, Kalyan VK.
\newblock Decompositions for optimal power flows.
\newblock IEEE Trans Power App Syst  1983;PAS-102(12):3877-84.

\bibitem{Pereira1987}
Monticelli A, Pereira MVF, Granville S.
\newblock Security-constrained optimal power flow with post-contingency
  corrective rescheduling.
\newblock IEEE Power Eng Rev  1987;PER-7(2):43-4.

\bibitem{Huang1992}
Huang GM, Hsieh SC.
\newblock Exact convergence of a parallel textured algorithm for constrained
  economic dispatch control problems.
\newblock Proceedings of the 31st IEEE Conference on Decision and Control
  1992:570-5.

\bibitem{Huang1995}
Huang G, Hsieh SC.
\newblock A parallel had-textured algorithm for constrained economic dispatch
  control problems.
\newblock IEEE Trans Power Syst  1995;10(3):1553-8.

\bibitem{Teixeira1990}
Teixeira MJ, Pinto HJCP, Pereira MVF, McCoy MF.
\newblock Developing concurrent processing applications to power system
  planning and operations.
\newblock IEEE Trans Power Syst  1990;5(2):659-64.

\bibitem{Kim1997}
Kim BH, Baldick R.
\newblock Coarse-grained distributed optimal power flow.
\newblock IEEE Trans Power Syst  1997;12(2):932-9.

\bibitem{Kim1999}
Baldick R, Kim BH, Chase C, Luo Y.
\newblock A fast distributed implementation of optimal power flow.
\newblock IEEE Trans Power Syst  1999;14(3):858-64.

\bibitem{Baldick1999}
Baldick R, Kim BH, Chase C, Yufeng L.
\newblock A fast distributed implementation of optimal power flow.
\newblock IEEE Trans Power Syst  1999;14(3):858-64.

\bibitem{Anese2013}
Anese ED, Zhu H, Giannakis GB.
\newblock Distributed optimal power flow for smart microgrids.
\newblock IEEE Trans Smart Grid  2013;4(3):1464-75.

\bibitem{Liu2011}
Liu K, Li Y, Sheng W.
\newblock The decomposition and computation method for distributed optimal
  power flow based on message passing interface (mpi).
\newblock International Journal of Electrical Power \& Energy Systems
  2011;33(5):1185-93.
\newblock Available from:
  \url{https://www.sciencedirect.com/science/article/pii/S0142061511000718}.

\bibitem{Kim2000}
Kim BH, Baldick R.
\newblock A comparison of distributed optimal power flow algorithms.
\newblock IEEE Trans Power Syst  2000;15(2):599-604.

\bibitem{Talukdar1994}
Talukdar S, Ramesh VC.
\newblock A multi-agent technique for contingency constrained optimal power
  flows.
\newblock IEEE Trans Power Syst  1994;9(2):855-61.

\bibitem{Rodrigues1994}
Rodrigues M, Saavedra OR, Monticelli A.
\newblock Asynchronous programming model for the concurrent solution of the
  security constrained optimal power flow problem.
\newblock IEEE Trans Power Syst  1994;9(4):2021-7.

\bibitem{Wei2005}
Wei Q, Flueck AJ, Feng T.
\newblock A new parallel algorithm for security constrained optimal power flow
  with a nonlinear interior point method.
\newblock In: IEEE Power Engineering Society General Meeting, 2005; 2005. p.
  447-53 Vol. 1.

\bibitem{Borges2007}
Borges CLT, Alves JMT.
\newblock Power system real time operation based on security constrained
  optimal power flow and distributed processing.
\newblock In: 2007 IEEE Lausanne Power Tech; 2007. p. 960-5.

\bibitem{Yuan2016}
Yuan Z, Hesamzadeh MR, Cui Y, {Bertling Tjernberg} L.
\newblock Applying high performance computing to probabilistic convex optimal
  power flow.
\newblock 2016 International Conference on Probabilistic Methods Applied to
  Power Systems, PMAPS 2016 - Proceedings  2016.

\bibitem{Farivar2013}
Farivar M, Low SH.
\newblock Branch flow model: relaxations and convexification—part i.
\newblock IEEE Trans Power Syst  2013;28(3):2554-64.

\bibitem{Yuan2019}
Yuan Z, Member S, Hesamzadeh MR, Member S.
\newblock A modified benders decomposition algorithm to solve second-order cone
  ac optimal power flow.
\newblock IEEE TRANSACTIONS ON SMART GRID,  2019;10(2):1713-24.

\bibitem{Lan2017}
Lan T, Huang GM.
\newblock An intelligent parallel scheduling method for optimal transmission
  switching in power systems with batteries.
\newblock 19th International Conference on Intelligent System Application to
  Power Systems (ISAP)  2017.

\bibitem{Zhang2018}
Zhang Q, Sahraei-ardakani M.
\newblock Distributed dcopf with flexible transmission.
\newblock Electr Power Syst Res  2018;154:37-47.

\bibitem{Mohammadi2015}
Mohammadi J, Zhang J, Kar S, Hug G, Moura JMF.
\newblock Multilevel distributed approach for dc optimal power flow.
\newblock In: 2015 IEEE Global Conference on Signal and Information Processing
  (GlobalSIP); 2015. p. 1121-5.

\bibitem{Sadnan2020}
Sadnan R, Member S, Dubey A.
\newblock Distributed optimization using reduced network equivalents for radial
  power distribution systems.
\newblock IEEE Transactions on Power Systems 1  2020;8950(c):1-12.

\bibitem{Molzahn2017}
Molzahn DK, Dorfler F, Sandberg H, Low SH, Chakrabarti S, Baldick R, et~al.
\newblock A survey of distributed optimization and control algorithms for
  electric power systems.
\newblock IEEE Trans Smart Grid  2017;8(6):2941-62.

\bibitem{Tu2020}
Tu S, W{\"{a}}chter A, Wei E.
\newblock A two-stage decomposition approach for ac optimal power flow.
\newblock arXiv  2020;8950(c):1-10.

\bibitem{DeMiguel2006}
DeMiguel V, Murray W.
\newblock A local convergence analysis of bilevel decomposition algorithms.
\newblock Optim Eng  2006;7(2):99-133.

\bibitem{Kerr1966}
Kerr RH, Scheidt JL, Fontanna AJ, Wiley JK.
\newblock Unit commitment.
\newblock IEEE Trans Power App Syst  1966;PAS-85(5):417-21.

\bibitem{Midwest2009}
Ma X, Song H, Hong M, Wan J, Chen Y, Zak E.
\newblock The security-constrained commitment and dispatch for midwest iso
  day-ahead co-optimized energy and ancillary service market.
\newblock In: 2009 IEEE Power Energy Society General Meeting; 2009. p. 1-8.

\bibitem{Shiina2003}
Shiina T, Birge JR.
\newblock Stochastic unit commitment problem.
\newblock International Transactions In Operational Research  2003;11(1).

\bibitem{Papavasiliou2013}
Papavasiliou A, Oren SS.
\newblock A comparative study of stochastic unit commitment and
  security-constrained unit commitment using high performance computing.
\newblock 2013 European Control Conference, ECC 2013  2013:2507-12.

\bibitem{Ji2021}
Ji X, Zhang Y, Han X, Ye P, Xu B, Yu Y.
\newblock Electrical power and energy systems multi-level interactive unit
  commitment of regional power system.
\newblock International Journal of Electrical Power \& Energy Systems
  2021;125(July 2020).

\bibitem{L.Wu2007}
Wu L, Shahidehpour M, Li T.
\newblock Stochastic security-constrained unit commitment.
\newblock IEEE Trans Power Syst  2007;2(2):800-11.

\bibitem{Liu2019AC}
Liu J, Laird CD, Scott JK, Watson JP, Castillo A.
\newblock Global solution strategies for the network-constrained unit
  commitment problem with ac transmission constraints.
\newblock IEEE Trans Power Syst  2019;34(2):1139-50.

\bibitem{Wong1993}
Wong KP, Wong YW.
\newblock Short-term hydrothermal scheduling with reservoir volume constraints.
  Ii. Parallel simulated annealing approach.
\newblock In: 1993 2nd International Conference on Advances in Power System
  Control, Operation and Management, APSCOM-93.; 1993. p. 565-70 vol.2.

\bibitem{Numnoda1996}
Numnonda T, Annakkage UD, Pahalawaththa NC.
\newblock Unit commitment using stochastic optimization.
\newblock In: Proceedings of International Conference on Intelligent System
  Application to Power Systems; 1996. p. 428-33.

\bibitem{Misra1994}
Misra N, Baghzouz Y.
\newblock Implementation of the unit commitment problem on supercomputers.
\newblock IEEE Trans Power Syst  1994;9(1):305-10.

\bibitem{Lau1997}
Lau KK, Kumar MJ.
\newblock Parallel implementation of the unit commitment problem on nows.
\newblock In: Proceedings High Performance Computing on the Information
  Superhighway. HPC Asia '97; 1997. p. 128-33.

\bibitem{Numnonda1996}
Numnonda T, Annakkage UD, Pahalawaththa NC.
\newblock Unit commitment using stochastic optimization.
\newblock Proceedings of the International Conference on Intelligent Systems
  Applications to Power Systems, ISAP  1996;1(2):428-33.

\bibitem{Yang1995}
Yang HT, Yang PC, Huang CL.
\newblock Optimization of unit commitment using parallel structures of genetic
  algorithm.
\newblock In: Proceedings 1995 International Conference on Energy Management
  and Power Delivery EMPD '95. vol.~2; 1995. p. 577-82 vol.2.

\bibitem{Yang1997}
Yang HT, Yang PC, Huang CL.
\newblock A parallel genetic algorithm approach to solving the unit commitment
  problem: implementation on the transputer networks.
\newblock IEEE Power Eng Rev  1997;17(5):58-9.

\bibitem{Murillo-s2000}
Murillo-s CE, Thomas RJ.
\newblock Parallel processing implementation of the unit commitment problem
  with full ac power flow constraints.
\newblock Proceedings of the 33rd Annual Hawaii International Conference on
  System Sciences  2000;00(c):1-9.

\bibitem{Baslis2009}
Baslis CG, Papadakis SE, Bakirtzis AG.
\newblock Simulation of optimal medium-term hydro-thermal system operation by
  grid computing.
\newblock IEEE Trans Power Syst  2009;24(3):1208-17.

\bibitem{PapavasiliouA;OrenS.S;ONeill2011}
Papavasiliou A, Oren SS, O'Neill RP.
\newblock No titlereserve requirements for wind power integration: a
  scenario-based stochastic programming framework.
\newblock IEEE Trans Power Syst  2011;26:2197-206.

\bibitem{Papavasiliou2015}
Papavasiliou A, Oren SS, Rountree B.
\newblock Applying high performance computing to transmission-constrained
  stochastic unit commitment for renewable energy integration.
\newblock IEEE Trans Power Syst  2015;30(3):1109-20.

\bibitem{aravena2015}
Aravena I, Papavasiliou A.
\newblock A distributed asynchronous algorithm for the two-stage stochastic
  unit commitment problem.
\newblock In: 2015 IEEE Power Energy Society General Meeting; 2015. p. 1-5.

\bibitem{Bai2015}
Bai Y, Zhong H, Xia Q, Kang C, Xie L.
\newblock A decomposition method for network-constrained unit commitment with
  ac power fl ow constraints.
\newblock Energy  2015;88:595-603.

\bibitem{BAI2008}
Bai X, Wei H, Fujisawa K, Wang Y.
\newblock Semidefinite programming for optimal power flow problems.
\newblock International Journal of Electrical Power \& Energy Systems
  2008;30(6):383-92.
\newblock Available from:
  \url{https://www.sciencedirect.com/science/article/pii/S0142061507001378}.

\bibitem{Khanabadi2016}
Khanabadi M, Wang C.
\newblock Security-constrained unit commitment considering voltage stability :
  a parallel solution.
\newblock 2016 North American Power Symposium (NAPS)  2016.

\bibitem{KARGARIAN2011}
Kargarian A, Raoofat M, Mohammadi M.
\newblock Reactive power market management considering voltage control area
  reserve and system security.
\newblock Appl Energy  2011;88(11):3832-40.
\newblock Available from:
  \url{https://www.sciencedirect.com/science/article/pii/S0306261911002467}.

\bibitem{SIOSHANSI2008245}
Sioshansi R, Oren S, O'neill R.
\newblock Chapter 6 - the cost of anarchy in self-commitment-based electricity
  markets.
\newblock In: Sioshansi FP, editor. Competitive Electricity Markets  Elsevier
  Global Energy Policy and Economics Series. Oxford: Elsevier; 2008. p. 245-66.
\newblock Available from:
  \url{https://www.sciencedirect.com/science/article/pii/B9780080471723500106}.

\bibitem{Feizollahi2015}
Feizollahi MJ, Costley M, Ahmed S, Grijalva S.
\newblock Large-scale decentralized unit commitment.
\newblock International Journal of Electrical Power and Energy Systems
  2015;73:97-106.

\bibitem{Doostizadeh2016}
Doostizadeh M, Aminifar F, Lesani H, Ghasemi H.
\newblock Multi-area market clearing in wind-integrated interconnected power
  systems : a fast parallel decentralized method.
\newblock Energy Convers Manage  2016;113:131-42.

\bibitem{Ramanan2017}
Ramanan P, Yildirim M, Gebraeel N, Ramanan P, Gebraeel N.
\newblock Asynchronous decentralized framework for unit commitment in power
  systems.
\newblock IEEE Trans Power Syst  2017:665-74.

\bibitem{Shi2014}
Shi W, Ling Q, Yuan K, Wu G, Yin W.
\newblock On the linear convergence of the admm in decentralized consensus
  optimization.
\newblock IEEE Trans Signal Processing  2014;62(7):1750-61.

\bibitem{Ramanan2019}
Ramanan P, Yildirim M, Chow E, Gebraeel N.
\newblock An asynchronous , decentralized solution framework for the large
  scale unit.
\newblock IEEE Trans Power Syst  2019;34(5):3677-86.

\bibitem{Boyd2011}
Boyd S, Parikh N, Chu E, Peleato B, Eckstein J.
\newblock Distributed optimization and statistical learning via the alternating
  direction method of multipliers.
\newblock Found Trends Mach Learn  2011 Jan;3(1):1-122.

\bibitem{Bragin2017}
Bragin MAS, Luh PB.
\newblock Distributed and asynchronous unit commitment and economic dispatch.
\newblock 2017 IEEE Power\& Energy Society General Meeting  2017.

\bibitem{Bragin2021}
Bragin MA, Yan B, Luh PB.
\newblock Distributed and asynchronous coordination of a mixed-integer linear
  system via surrogate lagrangian relaxation.
\newblock IEEE Trans Automat Sci Eng  2021;18(3):1191-205.

\bibitem{Kargarian2018}
Kargarian A, Mehrtash M, Falahati B.
\newblock Decentralized implementation of unit commitment with analytical
  target cascading.
\newblock IEEE Trans Power Syst  2018;33(4):3981-93.

\bibitem{Fu2013}
Fu Y, Li Z, Wu L.
\newblock Modeling and solution of the large-scale security-constrained unit
  commitment.
\newblock IEEE Trans Power Syst  2013;28(4):3524-33.

\bibitem{Kargarian2015}
Kargarian A, Fu Y, Li Z.
\newblock Distributed security-constrained unit commitment for large-scale
  power systems.
\newblock IEEE Trans Power Syst  2015;30(4):1925-36.

\bibitem{MingZhou2018}
Ming Z, Junyi Z, Gengyin L, Jianwen R.
\newblock Distributed dispatch approach for bulk ac / dc hybrid systems with
  high wind power penetration.
\newblock IEEE Trans Power Syst  2018;33(3):3325-36.

\bibitem{Wei2020}
Wei L, Liu G, Member S, Yan S, Member S.
\newblock Graph computing based security constrained unit commitment in
  hydro-thermal power systems incorporating pumped hydro storage.
\newblock CSEE Journal of Power and Energy Systems  2020:1-11.

\bibitem{YongFu2007}
Fu Y, Shahidehpour M.
\newblock Fast scuc for large-scale power systems.
\newblock IEEE Trans Power Syst  2007;22(4):2144-51.

\bibitem{FengWei2018}
Feng W, Yuan C, Dai R, Liu G, Li F.
\newblock Graph computation based power flow for large-scale ac/dc system.
\newblock In: 2018 International Conference on Power System Technology
  (POWERCON); 2018. p. 468-73.

\bibitem{Yuan2020}
Yuan C, Zhou Y, Liu G, Dai R, Lu Y, Wang Z.
\newblock Graph computing-based wls fast decoupled state estimation.
\newblock IEEE Trans Smart Grid  2020;11(3):2440-51.

\bibitem{ShiQingxin2019}
Shi Q, Yuan C, Feng W, Liu G, Dai R, Wang Z, et~al.
\newblock Enabling model-based lti for large-scale power system security
  monitoring and enhancement with graph-computing-based power flow calculation.
\newblock IEEE Access  2019;7:167010-8.

\bibitem{Chen2016}
Chen Y, Casto A, Wang F, Wang Q, Wang X, Wan J, et~al.
\newblock Improving large scale day-ahead security constrained unit commitment
  performance.
\newblock IEEE Trans Power Syst  2016;31(6):4732-43.

\bibitem{Chen2021}
Chen Y, Member S, Pan F, Holzer J, Rothberg E.
\newblock A high performance computing based market economics driven
  neighborhood search and polishing algorithm for security constrained unit
  commitment.
\newblock IEEE Trans Power Syst  2021;36(1):292-302.

\bibitem{HIPPO}
PNNL. PNNL, editor. High-performance computing helps grid operators manage
  increasing complexity. High-Performance Computing Helps Grid Operators Manage
  Increasing Complexity; 2020.
\newblock [Online; accessed 29-Nov-2021].
\newblock \url{https://bit.ly/3tLmUA6}.

\bibitem{stott1987}
Stott B, Alsac O, Monticelli AJ.
\newblock Security analysis and optimization.
\newblock Proc IEEE  1987;75(12):1623-44.

\bibitem{Balu1992}
Balu N, Bertram T, Bose A, Brandwajn V, Cauley G, Curtice D, et~al.
\newblock On-line power system security analysis.
\newblock Proc IEEE  1992;80(2):262-82.

\bibitem{Hao1995}
Hao S, Papalexopoulos A, Peng T.
\newblock Distributed processing for contingency screening applications.
\newblock IEEE Trans Power Syst  1995;10(2):838-44.

\bibitem{Mendes2000}
Mendes JC, Saavedra OR, Feitosa SA.
\newblock A parallel complete method for real-time security analysis in power
  systems.
\newblock IEEE Trans Power Syst  2000;56:27-34.

\bibitem{Balduino2004}
Balduino L, Alves ACB.
\newblock Parallel processing in a cluster of microcomputers with application
  in contingency analysis.
\newblock In: 2004 IEEE/PES Transmision and Distribution Conference and
  Exposition: Latin America (IEEE Cat. No. 04EX956); 2004. p. 285-90.

\bibitem{Huang2009}
Huang Z, Chen Y, Nieplocha J.
\newblock Massive contingency analysis with high performance computing.
\newblock IEEE Power \& Energy Society General Meeting  2009:1-8.

\bibitem{Huang2010}
Huang Q, Zhou M, Zhang Y, Wu Z.
\newblock Exploiting cloud computing for power system.
\newblock International Conference on Power System Technology  2010:1-6.

\bibitem{Khaitan2013}
Khaitan SK, McCalley JD.
\newblock Parallelizing power system contingency analysis using d programming
  language.
\newblock In: 2013 IEEE Power Energy Society General Meeting; 2013. p. 1-5.

\bibitem{Khaitan2014}
Khaitan SK, McCalley JD.
\newblock Scale: a hybrid mpi and multithreading based work stealing approach
  for massive contingency analysis in power systems.
\newblock Electr Pow Syst Res  2014;114:118-25.
\newblock Available from:
  \url{https://www.sciencedirect.com/science/article/pii/S0378779614001576}.

\bibitem{SEKINE1994}
Sekine Y, Takahashi K, Sakaguchi T.
\newblock Real-time simulation of power system dynamics.
\newblock International Journal of Electrical Power \& Energy Systems
  1994;16(3):145-56.
\newblock Special Issue 11th Power Systems Computation Conference.
\newblock Available from:
  \url{https://www.sciencedirect.com/science/article/pii/0142061594900043}.

\bibitem{Pai1992}
Pai MA, Sauer PW, Kulkarni AY.
\newblock Conjugate gradient approach to parallel processing in dynamic
  simulation of power systems.
\newblock In: 1992 American Control Conference; 1992. p. 1644-7.

\bibitem{Decker1996}
Decker IC, Falcao DM, Kaszkurewicz E.
\newblock Conjugate gradient methods for power system dynamic simulation on
  parallel computers.
\newblock IEEE Trans Power Syst  1996;11(3):1218-27.

\bibitem{Xue2005}
Shu J, Xue W, Zheng W.
\newblock A parallel transient stability simulation for power systems.
\newblock IEEE Trans Power Syst  2005;20(4):1709-17.

\bibitem{Jin2013}
Jin S, Huang Z, Diao R, Wu D, Chen Y.
\newblock Parallel implementation of power system dynamic simulation.
\newblock IEEE Power and Energy Society General Meeting  2013.

\bibitem{Alvarado1979}
Alvarado FL.
\newblock Parallel solution of transient problems by trapezoidal integration.
\newblock IEEE Trans Power App Syst  1979;PAS-98(3):1080-90.

\bibitem{Hatcher1977}
Hatcher WL, Brasch FM, Van~Ness JE.
\newblock A feasibility study for the solution of transient stability problems
  by multiprocessor structures.
\newblock IEEE Trans Power App Syst  1977;96(6):1789-97.

\bibitem{Hatcher1976}
Hatcher L William. Lloyd H, editor. A special purpose multiprocessor for the
  simulation of dynamic systems.. Naval Postgraduate School; 1976.
\newblock Available from: \url{https://calhoun.nps.edu/handle/10945/17950}.

\bibitem{Brasch1979}
Brasch FM, Van~Ness JE, Sang-Chul K.
\newblock The use of a multiprocessor network for the transient stability
  problem.
\newblock IEEE Conference Proceedings Power Industry Computer Applications
  Conference  1979:337-44.

\bibitem{Brasch1982}
Brasch FM, {Van Ness} JE, Kang SC.
\newblock Simulation of a multiprocessor network for power system problems.
\newblock IEEE Trans Power App Syst  1982;PAS-101(2):295-301.

\bibitem{Taoka1983}
Taoka H, Abe S, Takeda S.
\newblock Fast transient stability solution using an array processor.
\newblock IEEE Trans Power App Syst  1983;PAS-102(12):3835-41.

\bibitem{Taoka1984}
Taoka H, Abe S.
\newblock Fast transient stability solution adapted for an array processor.
\newblock IEEJ Transactions on Power and Energy  1984;104(5):297-304.

\bibitem{LaScala1990}
{La Scala} M, Bose A, Tylavsky DJ, Chai JS.
\newblock A highly parallel method for transient stability analysis.
\newblock IEEE Trans Power Syst  1990;5(4):1439-46.

\bibitem{Scala1990gaussi}
Scala ML, Brucoli M, Torelli F, Trovato M.
\newblock A gauss-jacobi-block-newton method for parallel transient stability
  analysis.
\newblock IEEE Trans Power Syst  1990;5(4):1168-77.

\bibitem{Scala1991}
Scala ML, Sbrizzai R, Torclli F.
\newblock A pipelined-in-time parallel algoiiitiim for transient stability
  analysis.
\newblock Test  1991;6(2).

\bibitem{Zhu1992}
Zhu N, Bose A.
\newblock A dynamic partitioning scheme for parallel transient stability
  analysis.
\newblock IEEE Trans Power Syst  1992;7(2):940-6.

\bibitem{Chat1993}
Chat JS, Bose A.
\newblock Bottlenecks in parallel algorithms for power system stability
  analysis.
\newblock IEEE Trans Power Syst  1993;8(1):9-15.

\bibitem{Spong1987}
Ilic'-Spong M, Crow ML, Pai MA.
\newblock Transient stability simulation by waveform relaxation methods.
\newblock IEEE Trans Power Syst  1987;2(4):943-9.

\bibitem{Scala1990}
La~Scala M, Bose A, Tylavsky DJ, Chai JS.
\newblock A highly parallel method for transient stability analysis.
\newblock IEEE Trans Power Syst  1990;5(4):1439-46.

\bibitem{Crow1990}
Crow ML, Ilic M.
\newblock The parallel implementation of the waveform relaxation method for the
  simulation of structure-preserved power systems.
\newblock In: 1990 IEEE International Symposium on Circuits and Systems
  (ISCAS); 1990. p. 1285-8 vol.2.

\bibitem{Tylavsky1993}
Tylavsky DJ, Nagaraj S, Crouch PE, Gaye AA, Jarriel LF.
\newblock Parallel-vector processing synergy and frequency domain transient
  stability simulations.
\newblock Electr Pow Syst Res  1993;28(2):89-97.

\bibitem{Granelli1994}
Granelli GP, Montagna M, {La Scala} M, Torelli F.
\newblock Relaxation-newton methods for transient stability analysis on a
  vector/parallel computer.
\newblock IEEE Trans Power Syst  1994;9(2):637-43.

\bibitem{Taoka1992}
Taoka H, Iyoda I, Noguchi H, Sato N, Nakazawa T.
\newblock Real-time digital simulator for power system analysis on a hypercube
  computer.
\newblock IEEE Trans Power Syst  1992;7(1):1-10.

\bibitem{Chai1991}
Chai JS, Zhu N, Bose A, Tylavsky DJ.
\newblock Parallel newton type methods for power system stability analysis
  using local and shared memory multiprocessors.
\newblock IEEE Trans Power Syst  1991;6(4):1539-45.

\bibitem{Lee1991}
Lee SY, Chiang HD, Lee KG, Ku BY.
\newblock Parallel power system transient stability analysis on hypercube
  multiprocessors.
\newblock IEEE Trans Power Syst  1991;6(3):1337-43.

\bibitem{Aloisio1997}
Aloisio G, Bochicchio MA, La~Scala M, Sbrizzai R.
\newblock A distributed computing approach for real-time transient stability
  analysis.
\newblock IEEE Trans Power Syst  1997;12(2):981-7.

\bibitem{Hong1997}
C~Hong XM Shen.
\newblock Parallel transient stability analysis on distributed memory message
  passing multiprocessors.
\newblock IET Conference Proceedings  1997 Jan:304-9(5).

\bibitem{Hong2000}
Hong C.
\newblock Implementation of parallel algorithms for transient stability
  analysis on a message passing multicomputer.
\newblock 2000 IEEE Power Engineering Society Winter Meeting Conference
  Proceedings  2000:1410-5.

\bibitem{Xue2004}
Xue W, Shu J, Zheng W.
\newblock Parallel transient stability simulation for national power grid of
  china.
\newblock In: Proceedings of the Second International Conference on Parallel
  and Distributed Processing and Applications  ISPA'04. Berlin, Heidelberg:
  Springer-Verlag; 2004. p. 765-76.

\bibitem{Jikeng2008}
Jikeng L, Xinyu T, Xudong W, Weicheng W.
\newblock Parallel simulation for the transient stability of power system.
\newblock In: 2008 Third International Conference on Electric Utility
  Deregulation and Restructuring and Power Technologies; 2008. p. 1325-9.

\bibitem{Jikeng2009}
Lin~Jikeng L, Xudong W, Xinyu T.
\newblock Asynchronous parallel simulation of transient stability based on
  equivalence.
\newblock In: 2009 International Conference on Sustainable Power Generation and
  Supply; 2009. p. 1-5.

\bibitem{Jalili2010}
Jalili-Marandi V.
\newblock Acceleration of transient stability simulation for large-scale power
  systems on parallel and distributed hardware.
\newblock University of Alberta; 2010.

\bibitem{Werlen1993}
Werlen K, Glavitsch H.
\newblock Computation of transients by parallel processing.
\newblock IEEE Trans Power Delivery  1993;8(3):1579-85.

\bibitem{Falcao1993}
Falcao DM, Kaszkurewicz E, Almeida HLS.
\newblock Application of parallel processing techniques to the simulation of
  power system electromagnetic transients.
\newblock IEEE Trans Power Syst  1993;8(1):90-6.

\bibitem{Morales2001}
Morales F, Rudnick H, Cipriano A.
\newblock Electromechanical transients simulation on a multicomputer via the
  vdhn – maclaurin method.
\newblock IEEE Trans Power Syst  2001;16(3):418-26.

\bibitem{Dufour2012}
Dufour C, Jalili-Marandi V, Bélanger J.
\newblock Real-time simulation using transient stability, electromagnetic
  transient and fpga-based high-resolution solvers.
\newblock In: 2012 SC Companion: High Performance Computing, Networking Storage
  and Analysis; 2012. p. 283-8.

\bibitem{Peng2017}
Peng Z, Shu-jun Y, Hai-lin Z, Shuo Z.
\newblock Multi-rate electromagnetic transient simulation of large-scale power
  system based on multi-core.
\newblock J Eng  2017;2017(October):1106-12.

\bibitem{Beaudin2003}
Beaudin S, Marceau RJ, Bois G, Savaria Y, Kandil N.
\newblock An economic parallel processing technology for faster than real-time
  transient stability simulation.
\newblock Eur Trans Electr Power  2003;13:105-12.

\bibitem{Duan2018}
Duan P, Xu S, Chen H, Yang X, Wang S, Hu E.
\newblock High performance computing (hpc)for advanced power system studies.
\newblock 2nd IEEE Conference on Energy Internet and Energy System Integration,
  EI2 2018 - Proceedings  2018.

\bibitem{Li2017}
Li X, Li F, Yuan H, Cui H, Hu Q.
\newblock Gpu-based fast decoupled power flow with preconditioned iterative
  solver and inexact newton method.
\newblock IEEE Trans Power Syst  2017;32(4):2695-703.

\bibitem{Jin2017}
Jin S, Huang Z, Diao R, Wu D, Chen Y.
\newblock Comparative implementation of high performance computing for power
  system dynamic simulations.
\newblock IEEE Trans Smart Grid  2017;8(3):1387-95.

\bibitem{Aristidou2014}
Aristidou P, Fabozzi D, Van~Cutsem T.
\newblock Dynamic simulation of large-scale power systems using a parallel
  schur-complement-based decomposition method.
\newblock IEEE Trans Parallel Distrib Syst  2014;25(10):2561-70.

\bibitem{ARISTIDOU2015}
Aristidou P, {Van Cutsem} T.
\newblock A parallel processing approach to dynamic simulations of combined
  transmission and distribution systems.
\newblock International Journal of Electrical Power and Energy Systems
  2015;72:58-65.

\bibitem{Aristidou2016}
Aristidou P, Lebeau S, {Van Cutsem} T.
\newblock Power system dynamic simulations using a parallel two-level
  schur-complement decomposition.
\newblock IEEE Trans Power Syst  2016;31(5):3984-95.

\bibitem{Gopal2007}
Gopal A, Niebur D, Venkatasubramanian S.
\newblock Dc power flow based contingency analysis using graphics processing
  units.
\newblock In: 2007 IEEE Lausanne Power Tech; 2007. p. 731-6.

\bibitem{Tang2018}
Tang K, Dong S, Zhu B, Ni Q, Song Y.
\newblock Gpu-based real-time n-1 ac power flow algorithm with preconditioned
  iterative method.
\newblock In: 2018 IEEE Power Energy Society General Meeting (PESGM); 2018. p.
  1-5.

\bibitem{Fu2020}
Fu M, Zhou G, Zhao J, Feng Y, He H, Liang K.
\newblock Gpu-based n-1 static security analysis algorithm with preconditioned
  conjugate gradient method.
\newblock IEEE Access  2020;8:124066-75.

\bibitem{Huang2018a}
Huang S, Dinavahi V.
\newblock Real-time contingency analysis on massively parallel architectures
  with compensation method.
\newblock IEEE Access  2018;6:44519-30.

\bibitem{Hou1997}
Hou L, Bose A.
\newblock Implementation of the waveform relaxation algorithm on a shared
  memory computer for the transient stability problem.
\newblock IEEE Trans Power Syst  1997;12(3):1053-60.

\bibitem{Jalili-marandi2009}
Jalili-marandi V, Dinavahi V.
\newblock Large-scale transient stability simulation on graphics processing
  units.
\newblock 2009 IEEE Power\& Energy Society General Meeting  2009:1-6.

\bibitem{Jalili-marandi2010}
Jalili-marandi V, Member S, Dinavahi V, Member S.
\newblock Simd-based large-scale transient stability simulation on the graphics
  processing unit.
\newblock IEEE Trans Power Syst  2010;25(3):1589-99.

\bibitem{Jalili-marandi2012}
Jalili-marandi V, Member S, Zhou Z, Member S.
\newblock Large-scale transient stability simulation of electrical power
  systems on parallel gpus.
\newblock IEEE Trans Parallel Distrib Syst  2012;23(7):1255-66.

\bibitem{Yu2014}
Yu Z, Huang S, Shi L, Chen Y.
\newblock Gpu-based jfng method for power system transient dynamic simulation.
\newblock 2014 International Conference on Power System Technology
  2014;1(Powercon):20-2.

\bibitem{Baijian2012}
Baijian W, Wenxin G, Jiayi H, Fangzong W, Jing Y.
\newblock Gpu based parallel simulation of transient stability using symplectic
  gauss algorithm and preconditioned gmres method.
\newblock In: 2012 Power Engineering and Automation Conference; 2012. p. 1-5.

\bibitem{Liao2016}
Liao X, Wang F.
\newblock Parallel computation of transient stability using symplectic gauss
  method and gpu.
\newblock IET Generation, Transmission \& Distribution  2016;10(15):3727-35.

\bibitem{Gao2014}
Gao W, Chen X.
\newblock Distributed generation placement design and contingency analysis with
  parallel computing technology.
\newblock Journal of Computers  2014;4(April 2009).

\bibitem{Song2017}
Song Y, Chen Y, Huang S, Xu Y, Yu Z, Marti JR.
\newblock Fully gpu-based electromagnetic transient simulation considering
  large-scale control systems for system-level studies.
\newblock IET Generation, Transmission \& Distribution  2017;11(11):2840-51.

\bibitem{Song2018}
Song Y, Member S, Chen Y.
\newblock Efficient gpu-based electromagnetic transient simulation for power
  systems with thread-oriented transformation and automatic code generation.
\newblock IEEE Access  2018;6:25724-36.

\bibitem{Zhou2016}
Zhou G, Zhang X, Lang Y, Bo R, Jia Y, Lin J, et~al.
\newblock Electrical power and energy systems a novel gpu-accelerated strategy
  for contingency screening of static security analysis.
\newblock International Journal of Electrical Power and Energy Systems
  2016;83:33-9.

\bibitem{Chen2017a}
Chen D, Member S, Jiang H, Li Y, Xu D.
\newblock A two-layered parallel static security assessment for large-scale
  grids based on gpu.
\newblock IEEE Trans Smart Grid  2017;8(3):1396-405.

\bibitem{Zhou2017a}
Zhou G, Feng Y, Bo R, Member S, Chien L, Zhang X, et~al.
\newblock Gpu-accelerated batch-acpf solution for n-1 static security analysis.
\newblock IEEE Trans Smart Grid  2017;8(3):1406-16.

\bibitem{Debnath2011}
Debnath JK, Fung WK, Gole AM, Filizadeh S.
\newblock Simulation of large-scale electrical power networks on graphics
  processing units.
\newblock In: 2011 IEEE Electrical Power and Energy Conference; 2011. p.
  199-204.

\bibitem{Debnath2016}
Debnath JK, Gole AM, Fung WK.
\newblock Graphics-processing-unit-based acceleration of electromagnetic
  transients simulation.
\newblock IEEE Trans Power Delivery  2016;31(5):2036-44.

\bibitem{Song2014}
Song Y, Chen Y, Yu Z, Huang S, Chen L.
\newblock A fine-grained parallel emtp algorithm compatible to graphic
  processing units.
\newblock In: 2014 IEEE PES General Meeting | Conference Exposition; 2014. p.
  1-6.

\bibitem{Zhou2014}
Zhou Z, Member S, Dinavahi V, Member S.
\newblock Parallel massive-thread electromagnetic transient simulation on gpu.
\newblock 2015 IEEE Power \& Energy Society General Meeting
  2014;29(3):1045-53.

\bibitem{Zhou2017c}
Zhou Z, Dinavahi V.
\newblock Fine-grained network decomposition for massively parallel
  electromagnetic transient simulation of large power systems.
\newblock IEEE Power Energy Technol Syst J  2017;4(3):51-64.

\bibitem{Wallach1981}
Wallach Y, Handschin E, Bongers C.
\newblock Efficient parallel processing method for power system state
  estimation.
\newblock SAE Preprints  1981;PAS-100(11):4402-6.

\bibitem{Cavin1982}
Cavin RK.
\newblock Multiprocessor static state estimation.
\newblock IEEE Trans Power App Syst  1982;PAS-101(2):302-8.

\bibitem{Aoki1987}
Aoki K.
\newblock A parallel computation algorithm for static state estimation by means
  of matrix inversion lemma.
\newblock IEEE Power Eng Rev  1987;2(3):624-31.

\bibitem{ABUR1990}
Abur A, Tapadiya P.
\newblock Parallel state estimation using multiprocessors.
\newblock Electr Power Syst Res  1990;18(1):67-73.
\newblock Available from:
  \url{https://www.sciencedirect.com/science/article/pii/0378779690900477}.

\bibitem{Lin1992}
Lin SY.
\newblock A distributed state estimator for electric power systems.
\newblock IEEE Trans Power Syst  1992;7(2):551-7.

\bibitem{El-keib1992}
El-Keib AA, Nieplocha J, Singh H, Maratukulam DJ.
\newblock A decomposed state estimation technique suitable for parallel
  processor implementation.
\newblock IEEE Trans Power Syst  1992;7(3):1088-97.

\bibitem{Falcao1995}
Falcao DM, Wu FF, Murphy L.
\newblock Parallel and distributed state estimation.
\newblock IEEE Trans Power Syst  1995;10(2):724-30.

\bibitem{Ebrahimian2000}
Ebrahimian R, Baldick R.
\newblock State estimation distributed processing [for power systems].
\newblock IEEE Trans Power Syst  2000;15(4):1240-6.

\bibitem{Carvalho2000}
Carvalho JB, Barbosa FM.
\newblock Distributed processing in power system state estimation.
\newblock In: 2000 10th Mediterranean Electrotechnical Conference. Information
  Technology and Electrotechnology for the Mediterranean Countries.
  Proceedings. MeleCon 2000 (Cat. No.00CH37099). vol.~3; 2000. p. 1128-31
  vol.3.

\bibitem{Niepolcha2006}
Nieplocha J, Marquez A, Tipparaju V, Chavarria-Miranda D, Guttromson R, Huang
  H.
\newblock Towards efficient power system state estimators on shared memory
  computers.
\newblock In: 2006 IEEE Power Engineering Society General Meeting; 2006. p.~5.

\bibitem{Schneider2009}
Schneider KP, Huang Z, Yang B, Hauer M, Nieplocha Y.
\newblock Dynamic state estimation utilizing high performance computing
  methods.
\newblock In: 2009 IEEE/PES Power Systems Conference and Exposition; 2009. p.
  1-6.

\bibitem{KORRES2013}
Korres GN, Tzavellas A, Galinas E.
\newblock A distributed implementation of multi-area power system state
  estimation on a cluster of computers.
\newblock Electr Power Syst Res  2013;102:20-32.
\newblock Available from:
  \url{https://www.sciencedirect.com/science/article/pii/S0378779613000965}.

\bibitem{Xia2017}
Xia Y, Chen Y, Ren Z, Huang S, Wang M, Lin M.
\newblock State estimation for large-scale power systems based on hybrid
  cpu-gpu platform.
\newblock In: 2017 IEEE Conference on Energy Internet and Energy System
  Integration (EI2); 2017. p. 1-6.

\bibitem{Magana-Lemus2013}
Magaña-Lemus E, Medina-Ríos A, Ramos-Paz A, Montesinos-González VH.
\newblock Periodic steady state determination of power systems using graphics
  processing units.
\newblock In: 2013 10th International Conference on Electrical Engineering,
  Computing Science and Automatic Control (CCE); 2013. p. 274-9.

\bibitem{Magana-Lemus2015}
Magaña-Lemus E, Medina A, Ramos-Paz A.
\newblock Periodic steady state solution of power systems by selective
  transition matrix identification, lu decomposition and graphic processing
  units.
\newblock In: 2015 IEEE Power Energy Society General Meeting; 2015. p. 1-5.

\bibitem{Karimipour2013}
Karimipour H.
\newblock Accelerated parallel wls state estimation for large-scale power
  systems on gpu.
\newblock North American Power Symposium (NAPS)  2013;1(1).

\bibitem{Karimipour2014}
Karimipour H, Dinavahi V.
\newblock On detailed synchronous generator modeling for massively parallel
  dynamic state estimation.
\newblock In: 2014 North American Power Symposium (NAPS); 2014. p. 1-6.

\bibitem{Karimipour2015}
Karimipour H, Dinavahi V.
\newblock Extended kalman filter-based parallel dynamic state estimation.
\newblock IEEE Trans Smart Grid  2015;6(3):1539-49.

\bibitem{Karimipour2017}
Karimipour H, Dinavahi V.
\newblock On false data injection attack against dynamic state estimation on
  smart power grids.
\newblock In: 2017 IEEE International Conference on Smart Energy Grid
  Engineering (SEGE); 2017. p. 388-93.

\bibitem{Rahman2016}
Rahman A, Venayagamoorthy GK.
\newblock Dishonest gauss newton method based power system state estimation on
  a gpu.
\newblock Clemson University Power Systems Conference (PSC)  2016.

\bibitem{Rahman2017}
Rahman MA, Venayagamoorthy GK.
\newblock Convergence of the fast state estimation for power systems.
\newblock SAIEE Africa Research Journal  2017;108(3):117-27.

\bibitem{Kim2011}
Kim MK, Park JK, Nam YW.
\newblock Market-clearing for pricing system security based on voltage
  stability criteria.
\newblock Energy  2011;36(2):1255-64.
\newblock Available from:
  \url{https://www.sciencedirect.com/science/article/pii/S0360544210006456}.

\bibitem{Geng2012}
Geng G, Jiang Q.
\newblock A two-level parallel decomposition approach for transient stability
  constrained optimal power flow.
\newblock IEEE Trans Power Syst  2012;27(4):2063-73.

\bibitem{Jiang2013}
Jiang Q, Zhou B, Zhang M.
\newblock Parallel augment lagrangian relaxation method for transient stability
  constrained unit commitment.
\newblock IEEE Trans Power Syst  2013;28(2):1140-8.

\bibitem{Gong2020}
Gong L, Member S, Fu Y, Member S, Shahidehpour M.
\newblock A parallel solution for the resilient operation of power systems in
  geomagnetic storms.
\newblock IEEE Trans Smart Grid  2020;11(4):3483-95.

\bibitem{D.Vasquez2019}
D~Vasquez A, Sousa T.
\newblock A parallel processing approach to stability analysis considering
  transmission and distribution systems.
\newblock In: 2019 IEEE Milan PowerTech; 2019. p. 1-6.

\bibitem{Qi2019}
Qi S, Li G, Bie Z.
\newblock Hybrid energy flow calculation for electric-thermal coupling system
  based on inexact newton method.
\newblock In: 2019 IEEE Sustainable Power and Energy Conference (iSPEC); 2019.
  p. 2443-9.

\bibitem{Geng2017}
Geng G, Jiang Q, Sun Y.
\newblock Parallel transient stability-constrained optimal power flow using gpu
  as coprocessor.
\newblock IEEE Trans Smart Grid  2017;8(3):1436-45.

\bibitem{Luo2019}
Luo X, Member S, Zhang S, Litvinov E.
\newblock Practical design and implementation of cloud computing for power
  system planning studies.
\newblock IEEE Trans Smart Grid  2019;10(2):2301-11.

\bibitem{Feng2015}
Feng M, Xiaochuan L, Qiang QF Zhang, Litvinov E.
\newblock Cloud computing: an innovative it paradigm to facilitate power system
  operations.
\newblock In: 2015 IEEE Power Energy Society General Meeting; 2015. p. 1-5.

\bibitem{england_2021}
{ISO England}. England I, editor. Working toward a smarter, greener grid.
  Iso-ne.com; 2021.
\newblock [Online; accessed 29-Nov-2021].
\newblock \url{https://bit.ly/3Oh1xhM}.

\bibitem{Morante2006}
Morante Q, Ranaldo N, Vaccaro A, Zimeo E.
\newblock Pervasive grid for large-scale power systems contingency analysis.
\newblock IEEE Trans Ind Inf  2006;2(3):165-75.

\bibitem{Taylor2006}
Taylor GA, Irving MR, Hobson PR, Huang C, Kyberd P, Taylor RJ.
\newblock Distributed monitoring and control of future power systems via grid
  computing.
\newblock In: 2006 IEEE Power Engineering Society General Meeting; 2006. p.~5.

\bibitem{Huang2006}
Huang Q, Qin K, Wang W.
\newblock Development of a grid computing platform for electric power system
  applications.
\newblock IEEE Power Engineering Society General Meeting  2006:1-7.

\bibitem{Huang2008}
Huang Z, Nieplocha J.
\newblock Transforming power grid operations via high performance computing.
\newblock In: 2008 IEEE Power and Energy Society General Meeting - Conversion
  and Delivery of Electrical Energy in the 21st Century; 2008. p. 1-8.

\bibitem{Ali2006}
Ali M, Dong ZY, Li X, Zhang P.
\newblock Rsa-grid: a grid computing based framework for power system
  reliability and security analysis.
\newblock In: 2006 IEEE Power Engineering Society General Meeting; 2006. p.~7.

\bibitem{Al-Khannak2007}
Al-Khannak R, Bitzer B.
\newblock Load balancing for distributed and integrated power systems using
  grid computing.
\newblock In: 2007 International Conference on Clean Electrical Power; 2007. p.
  123-7.

\bibitem{WANG2010}
Wang X, Yan Z, Li L.
\newblock A grid computing based approach for the power system dynamic security
  assessment.
\newblock Computers \& Electrical Engineering  2010;36(3):553-64.
\newblock Available from:
  \url{https://www.sciencedirect.com/science/article/pii/S0045790609001244}.

\bibitem{Rusitschka2010}
Rusitschka S, Eger K, Gerdes C.
\newblock Smart grid data cloud : a model for utilizing cloud computing in the
  smart grid domain.
\newblock First IEEE International Conference on Smart Grid Communications
  2010:483-8.

\bibitem{Mohsenian2010}
Mohsenian-Rad AH, Leon-Garcia A.
\newblock Coordination of cloud computing and smart power grids.
\newblock In: 2010 First IEEE International Conference on Smart Grid
  Communications; 2010. p. 368-72.

\bibitem{Huang2010c}
Huang Q, Zhou M, Zhang Y, Wu Z.
\newblock Exploiting cloud computing for power system analysis.
\newblock In: 2010 International Conference on Power System Technology; 2010.
  p. 1-6.

\bibitem{Markovic2013}
Markovic DS, Zivkovic D, Branovic I, Popovic R, Cvetkovic D.
\newblock Smart power grid and cloud computing.
\newblock Renewable Sustainable Energy Rev  2013;24:566-77.

\bibitem{Hongseok2011}
Hongseok K, Kim YJ, Yang K, Thottan M.
\newblock Cloud-based demand response for smart grid: architecture and
  distributed algorithms.
\newblock In: 2011 IEEE International Conference on Smart Grid Communications
  (SmartGridComm); 2011. p. 398-403.

\bibitem{Bo2014}
Bo ZQ, Wang L, Zhou F, Luo K, Han M, Yin W, et~al.
\newblock Substation cloud computing for secondary auxiliary equipment.
\newblock In: 2014 International Conference on Power System Technology; 2014.
  p. 1853-8.

\bibitem{Yoon2018}
Yoon DH, Kang SK, Kim M, Han Y.
\newblock Exploiting coarse-grained parallelism using cloud computing in
  massive power flow computation.
\newblock Energies  2018;11(9):1-15.

\bibitem{Amazon}
Amazon. Amazon, editor. Amazon web services. Amazon; 2020.
\newblock [Online; accessed 29-Nov-2021].
\newblock \url{https://aws.amazon.com/}.

\bibitem{Cao2017}
Cao Z, Lin J, Wan C, Song Y, Zhang Y, Wang X.
\newblock Optimal cloud computing resource allocation for demand side
  management in smart grid.
\newblock IEEE Trans Smart Grid  2017;8(4):1943-55.

\bibitem{Sheikhi2015}
Sheikhi A, Rayati M, Bahrami S, Ranjbar AM, Sattari S.
\newblock A cloud computing framework on demand side management game in smart
  energy hubs.
\newblock International Journal of Electrical Power and Energy Systems
  2015;64:1007-16.

\bibitem{Rajeev2015}
Rajeev T, Ashok S.
\newblock Dynamic load-shifting program based on a cloud computing framework to
  support the integration of renewable energy sources.
\newblock Appl Energy  2015;146:141-9.

\bibitem{Wang2020}
Wang S, Member S, Wang X, Wu W, Member S.
\newblock Cloud computing and local chip-based dynamic economic dispatch for
  microgrids.
\newblock IEEE Trans Smart Grid  2020;11(5):3774-84.

\bibitem{cloudfuzion2020}
Axceleon. Axceleon, editor. Enfuzion - high performance parallel computing
  software. Axceleon; 2021.
\newblock [Online; accessed 29-Nov-2021].
\newblock \url{http://www.axceleon.com/prod-cloudfuzion/}.

\bibitem{Sarker2018}
Sarker MR, Wang J.
\newblock Security and cloud outsourcing framework for economic dispatch.
\newblock IEEE Trans Smart Grid  2018;9(6):5810-9.

\bibitem{mangasarian_2010}
Mangasarian OL.
\newblock Privacy-preserving linear programming.
\newblock Optim Lett  2010;5(1):165-72.

\bibitem{li_li_deng_2011}
Li W, Li H, Deng C.
\newblock Privacy-preserving horizontally partitioned linear programs with
  inequality constraints.
\newblock Optim Lett  2011;7(1):137-44.

\bibitem{Overlin2018}
Overlin M, Smith C.
\newblock High performance computing techniques with power systems simulations.
\newblock 2018 IEEE High Performance Extreme Computing Conference, HPEC 2018
  2018.

\bibitem{Happ1979}
Happ HH, Pottle C, Wirgau KA.
\newblock Future computer technology for large power system simulation.
\newblock Automatica  1979;15(6):621-9.

\bibitem{Jose1982}
Jose S.
\newblock Effects of special purpose hardware in scientific computation ww with
  emphasis on power system applications.
\newblock IEEE Trans Power App Syst  1982;PAS-101(2):265-70.

\bibitem{Roberge2017}
Roberge V, Tarbouchi M, Okou F.
\newblock Parallel power flow on graphics processing units for concurrent
  evaluation of many networks.
\newblock IEEE Trans Smart Grid  2017;8(4):1639-48.

\bibitem{Pan}
Pan F, Northwest P. PNNL, editor. Hippo : a computation tool for planning
  tomorrow ' s electricity. PNNL; 2017.
\newblock Available from: \url{https://bit.ly/3HFxGNN}.

\bibitem{Palmer2016}
Palmer B, Perkins W, Chen Y, Jin S, Callahan D, Glass K, et~al.
\newblock Gridpacktm: a framework for developing power grid simulations on
  high-performance computing platforms.
\newblock Int J High Perform Comput Appl  2016;30(2):223-40.

\bibitem{Nividia}
Nividia. Nividia, editor. Nividia on demand. Nividia; 2020.
\newblock [Online; accessed 29-Nov-2021].
\newblock \url{https://developer.nvidia.com/cuda-zone}.

\bibitem{MathStories}
Grötschel M.
\newblock Special issue: optimization stories. Selected papers based on the
  presentations at the 21st international symposium on mathematical
  programming, ismp, berlin, germany, august 19–24, 2012.
\newblock Doc Math  2012 Jan.

\bibitem{Gurobi1}
Glockner G. Glockner G, editor. Does gurobi support gpus?. Gurobi; 2021.
\newblock [Online; accessed 29-Nov-2021].
\newblock \url{https://bit.ly/3bbqjBA}.

\bibitem{Jin2016}
Jin S, Chen Y, Diao R, Huang ZH, Perkins W, Palmer B.
\newblock Power grid simulation applications developed using the gridpack™
  high performance computing framework.
\newblock Electr Pow Syst Res  2016;141:22-30.

\bibitem{KOLTSAKLIS2018563}
Koltsaklis NE, Dagoumas AS.
\newblock State-of-the-art generation expansion planning: a review.
\newblock Appl Energy  2018;230:563-89.
\newblock Available from:
  \url{https://www.sciencedirect.com/science/article/pii/S0306261918312583}.

\bibitem{Ng1990}
Ng CP, Jabbour K, Meyer W.
\newblock Loadflow analysis on parallel computers.
\newblock In: Proceedings of the 32nd Midwest Symposium on Circuits and
  Systems,; 1989. p. 10-5 vol.1.

\bibitem{amazon2021}
Services AW. Services AW, editor. Amazon elastic compute cloud : user guide for
  linux instances. Amazon Web Services, Inc.; 2022.
\newblock [Online; accessed 29-Nov-2021].
\newblock \url{https://www.amazonaws.cn/en/ec2/instance-types/}.

\bibitem{AJAGEKAR201976}
Ajagekar A, You F.
\newblock Quantum computing for energy systems optimization: challenges and
  opportunities.
\newblock Energy  2019;179:76-89.
\newblock Available from:
  \url{https://www.sciencedirect.com/science/article/pii/S0360544219308254}.

\bibitem{Tylavsky2003}
Tylavsky DJ, Heydt GT.
\newblock Quantum computing in power system simulation.
\newblock In: 2003 IEEE Power Engineering Society General Meeting (IEEE Cat.
  No.03CH37491). vol.~2; 2003. p. 950-6 Vol. 2.

\bibitem{eskandarpour2020quantum}
Eskandarpour R, Ghosh K, Khodaei A, Zhang L, Paaso A, Bahramirad S.
\newblock Quantum computing solution of dc power flow; 2020.
\newblock ArXiv publication.

\bibitem{Baker20161500SL}
Baker M.
\newblock 1,500 scientists lift the lid on reproducibility.
\newblock Nature  2016;533:452-4.

\bibitem{ReplicationCrisis}
Serra-Garcia M, Gneezy U.
\newblock Nonreplicable publications are cited more than replicable ones.
\newblock Sci Adv  2021;7(21).
\newblock Available from:
  \url{https://ucsdnews.ucsd.edu/pressrelease/a-new-replication-crisis-research-that-is-less-likely-be-true-is-cited-more}.

\bibitem{Cockburn2020ThreatsOA}
Cockburn A, Dragicevic P, Besançon L, Gutwin C.
\newblock Threats of a replication crisis in empirical computer science.
\newblock Commun ACM  2020;63:70-9.

\end{thebibliography}

\newpage
\section{Appendix}


\appendix

\begin{small}
\begin{longtable}{|M{0.05\linewidth} | M{0.05\linewidth} | M{0.25\linewidth} | M{0.25\linewidth} | M{0.15\linewidth}|}

\caption{Literature Survey}
\label{table1}
\scriptsize \\
\toprule
\textbf{Year} &\textbf{Study} &\textbf{Techniques Used} &\textbf{Parallel Machine} &\textbf{Largest Case} \\ \hline
1979 &TSA\footnote{Transient Stability Analysis} &Trapezoidal Integration &CDC 6500 host \& AP-1POB &1723-bus 2764-line 398-unit \\ \hline
1983 &TSA &Trapezoidal Integration \& Fourth-Order Runge-Kutta &VAX-11/780, AP-120B, IBM-3081D \& Cray-1 &103-bus 60-units \\ \hline
1990 &\ac{PF} &Path Graph Factorization \& Fast Decoupled-load Flow &CRAY X/MP-48 &8235-bus \\ \hline
1990 &\ac{PF} &LU decomposition \& Newton-Raphson &Alliant FX/80 &3459-bus 5819-line \\ \hline
1990 &SCED\footnote{Security Constrained Economic Dispatch} &Batch Solution &IAPX 286/287 (16 units) &504-bus 880-line 72 Units 718-contingencies \\ \hline
1991 &\ac{PF} &LU decomposition &IPSC/2 (32 cpus) &IEEE 662 \\ \hline
1991 &TSA &Successive Over Relaxation – Newton &IPSC/2 \& Alliant &662-bus (midwestern U.S) \\ \hline
1991 &TSA &Second Order Runge-Kutta Method &iPSC/2 &256-bus \\ \hline
1991 &TSA &Gauss-Seidel Relaxation \& Newton Method &VAXstation 2000 &662-bus (midwestern U.S.) \\ \hline
1992 &DS\footnote{Dynamic Stability} &Parallel Predictor Corrector \& Newton-Raphson &Cray-2 &50-unit \\ \hline
1992 &DS &Generalized Minimal Residual method \& Newton-Raphson &Cray Y-MP &39-bus \\ \hline
1992 &TSA &Successive Over Relaxation – Newton &Alliant FX/8 (8 cpus) &662-bus (US midwestem) \\ \hline
1992 &TSA &Gauss-Seidel Relaxation \& Newton Method &NCube2 (512 cpus), AP-120B, VAX11/780, Cray-1 \& IBM-3081D &103-bus 60-unit 412-line \\ \hline
1993 &\ac{EMT} &LU Decomposition \& Newton-Raphson &AT386 (8 x Transputers) &1026-bus 2457-line \\ \hline
1993 &TSA &Forward Backward Substitution \& Frequency Domain Relaxation &Cray X/MP-116se \& Cray Y/MP-8/8-64 &1655-bus \\ \hline
1994 &\ac{PF} &Successive Overrelaxation, Gauss-Seidel \& FDLF &NCUBE2 \& iPSC/860 (32 cpus) &2429-bus (Texas) \\ \hline
1994 &TSA &Gauss-Seidel Relaxation \& Newton Method &CRAY Y-MP8/464 (4 cpus) &904-bus (US Network) \\ \hline
1994 &\ac{SCUC} &Dynamic Programming &CRAY Y-MP2/216 &26-unit \\ \hline
1995 &CA\footnote{Contingency Analysis} N-1 &Matrix multiplication and Batch &Apollo DN400 (5 cpus) &2518-bus 665-unit 1983-bus 1651-contingencies \\ \hline
1995 &\ac{ED} &Textured Decomposition \& Han-Powell Method &NCUBE2 (32 cpus) &228-bus \\ \hline
1995 &\ac{PF} &LU factorization \& Newton Raphson \& FDPF &Symmetry 81 (20 cpus) &1106-bus 1530 branch \\ \hline
1995 &TSA &LU factorization, \& Trapazoidal rule \& VDHN &Symmetry Ss1 &2584-bus 6846-line 487-units \\ \hline
1996 &DS &Conjugate Gradient LU factorization \& Newton-Raphson &IPSC/860 (8 cpus) &616-bus 88-unit 995-line (Brazilian southeastern region ) \\ \hline
1996 &\ac{PF} &W-Matrix \& Dependency-Based Substitution Algorithm &CRAY X-MPU216 \& CRAY Y-MPU464 &11670-bus 17288-line \\ \hline
1997 &\ac{OPF} &APP &Sun Sparc-20 &753-bus 1100-line 209-load 12 Ties \\ \hline
1997 &\ac{OPF} &APP &Sun Sparc-20 &753-bus 1100-line 209-load 12 Ties \\ \hline
1997 &\ac{PF} &LU factorization \& conjugate gradient method &Sequent Symmetry (15 cpus) &8235-bus \\ \hline
1997 &TSA &Waveform Relaxation Method &Symmetry Ss1 (20 cpus) &2583-bus 6846-line 487-units \\ \hline
1997 &TSA &Shifted-Picard algorithm, Domain decomposition (cointingency wise) and VDHN &IBM SP2 \& DEC ALPHA AXP-3000/500 (8 workstations) &2583-bus 51-units \\ \hline
1997 &TSA &Factorization Path Tree \& VDHN &IBM-SP2 &616-bus 812-line 92-unit \\ \hline
1997 &\ac{SCUC} &Dynamic Programming &Silicon Graphics 1P22 Network &34-unit \\ \hline
2000 &CA N-1 &Dynamic-load Balancing &IBM RISC 6000 (4 workstations) &1663-bus 1711-contingencies (Brazil South Eastern ) \\ \hline
2000 &\ac{PF} &BBDF \& Factorization Tree Algorithm &Intel 80486 \& 6 transputers &288-bus \\ \hline
2000 &\ac{SSE} &APP \& Gauss – Newton Method &Sun Ultra workstation network &8047-bus 8-region 190-tie \\ \hline
2000 &TSA &Factorization Path Tree &IBM 9076 SP2 (16 cpus) &3021-bus power \\ \hline
2000 &\ac{SCUC} &APP \& Newton Raphson &- &IEEE 118 \\ \hline
2001 &\ac{EMT} &VDHN–Maclaurin Method &AT386 (8 transputers) &IEEE 300 \\ \hline
2002 &\ac{PF} &Newton-GMRES \& Newton Raphson &1.0GHz Pentenium (8 cpus) &30910-bus 39136-line \\ \hline
2004 &CA N-1 &Batch Solution &2.3 GHz Pentium IV &810-bus 1340-line 1300-contingencies \\ \hline
2004 &TSA &Block Bordered Diagonal Form &700MHz Xeon Symmetry computer (12 cpus) &10188-bus 13499-lines 1072 units 3003-load \\ \hline
2005 &DS &hierarchical Block Bordered Diagonal Form (BBDF) \& Newton-Raphson &Intel Xeon (12 cpus) &10188-bus 13499-line 1072-units 3003-load \\ \hline
2005 &\ac{OPF} &Decoupling First-order KKT &166MHz Pentium (29 workstations) &584-bus 118-unit 937-line 11-tielines 5 region (Balkan-5) \\ \hline
2005 &\ac{OPF} &Decoupling First-order KKT &166MHz Pentium (29 workstations) &584-bus 118-unit 937-line 11-tielines 5 region (Balkan-5) \\ \hline
2005 &\ac{PF} &Block Bordered Diagonal Form \& Newton Raphson &Intel 80486 \& 6 transputers &288-bus \\ \hline
2005 &\ac{SCOPF} &Preconditioned GMRES \& Primal-Dual IPM &1.0GHz CPU (16 workstations) &3493-bus 6689-line and 79-contingencies \\ \hline
2005 &\ac{SCOPF} &Preconditioned GMRES \& Primal-Dual IPM &1.0GHz CPU (16 workstations) &3493-bus 6689-line and 79-contingencies \\ \hline
2005 &TSA &Block Bordered Diagonal Form &700MHz Xeon Symmetry computer (12 cpus) &2115 nodes 2614-line 248 genera-tors and 544-load \\ \hline
2006 &\ac{SSE} &Conjugate Gradient \& LU Factorization &SGI Altix3000 (32 cpus) &1177-bus 1770-line \\ \hline
2007 &\ac{PF} &LU factorization of \& Newton-Raphson \& Fast Decoupled Power Flow &APEX20KE , EP20K1500EBC652-1x, FPGA (7 processors) &7917-bus (Northeastern US power grid) \\ \hline
2008 &TSA &Waveform Relaxation Method &IBM PC 1350 ( 6 x 2.4GHz Pentenium) &1923-bus 2280-line171-unit (Northen China) \\ \hline
2009 &CA N-1 &Dynamic-load Balancing &Cray XMT (8 2-Core Intel Xeon cpus) &170000-bus 220648-line 512-contingencies \\ \hline
2009 &DS &LU decomposition /Newton-Raphson \& Fast Poisson Solver \&Fast Fourier Transformation &Intel Xeon CPU host \& NVIDIA Tesla C870 GPU &4,000,000-bus \\ \hline
2009 &\ac{SSE} &Extended Kalman Filter (EKF) \& Newton-Raphson &2.33Ghz 8-Core Xeon E5345 &200-unit \\ \hline
2009 &TSA &trapezoidal rule \& Gauss elimination \& Back substitution &2.5GHz quad-core AMD Phenom host \& Nividia GeForce GTX 280 &1248-bus 1244-line \\ \hline
2009 &TSA &Trapezoidal Integration \& BBDF &IBM PC 1350 ( 6 x 2GHz Pentenium) &1923-bus 2280-line171-unit (Northen China) \\ \hline
2009 &\ac{SCUC} &Monte Carlo &BOINC (100 workstations) &42-unit \\ \hline
2010 &CA N-1 &Dynamic-load Balancing &NWICEB (128 cpus) &14000-bus 17346-contingencies (WECC) \\ \hline
2010 &CA N-1 &Dynamic-load Balancing &500MHz Cray XMT (64 cpus) &170001-bus 220648-line 512-contingencies \\ \hline
2010 &\ac{ED} &Barrier Optimization &- &IEEE 118 \\ \hline
2010 &\ac{PF} &Preconditioned Biconjugate Gradient \& Newton-Raphson &2.83 GHz 2-Core host \& Tesla C870 GPU &IEEE-118 \\ \hline
2010 &TSA &LU Decomposition \& Trapazoidal Integration &2.5GHz quad-core AMD Phenom host \& Nividia GeForce GTX 280 &1248-bus 1244-line \\ \hline
2011 &DS &LU Decomposition \& Distributed Integration &3.2 GHz AMD CPU (64 threads) &16072-bus 19622-line 2361-unit (WECC) \\ \hline
2011 &\ac{EMT} &LU Decomposition \& Trapezoidal Integration &2.13GHz 2-Core Intel \& NVIDIA GTX GeForce 285 &117-bus 21-unit \\ \hline
2012 &\ac{PF} &Forward Backward Substitution \& Newton-Raphson &3.10GHz Intel i3-2100 host \& NVIDIA GeForce GTS 450 &Shandong System \\ \hline
2012 &TSA &Trapazoidal Integration \&LU factorization \&Newton-Raphson iteration &2.5GHz AMDphenom 9850 host \& 4 NVIDIA T10 GPU &9984-bus 2560-units \\ \hline
2012 &TSA &symplectic Gauss Algorithm \&Preconditioned GMRES, &NVIDIA GTX 280 &2383-bus (Polish) \\ \hline
2013 &CA N-1 &Dynamic Master-Slave Scheduling &16 Threads &13030-bus 431-units 5950-load 512-contingencies \\ \hline
2013 &CA N-1 &Dynamic Master-Slave Scheduling &16 Threads &13029-bus 431-unit 5950-load 2000-contingencies \\ \hline
2013 &CA N-1 &Dynamic Master-Slave Scheduling &32 CPUS &13029-bus 431- units 5950-load 4000-contingencies \\ \hline
2013 &\ac{OPF} &ADMM \& Distributed Semi-Definite Programming &3.40 GHz Intel Core i7–2600 CPU &IEEE 37 + 10-bus microgrid \\ \hline
2013 &\ac{OPF} &ADMM \& Distributed Semi-Definite Programming &3.40 GHz Intel Core i7–2600 CPU &IEEE 37 + 10-bus microgrid \\ \hline
2013 &\ac{SSE} &Runge-Kutta Fourth-Order \&Newton-Raphson &2.2GHz Intel Xeon E5606 host \& NVIDIA Tesla C2075 &IEEE-118 \\ \hline
2013 &\ac{SSE} &LU decomposition \&WLS &2.0 GHz 8-Core Intel Xeon E5- 2620 \& NVIDIA Fermi GPU &4992-bus \\ \hline
2013 &\ac{SSE} &Cholesky factorization \&Backward Forward Substitution \& Wegihted Least Square &2.66 GHz Intel Core 2 Duo E8200 PC (11 cpus) &1180-bus (10 x IEEE-118) \\ \hline
2013 &\ac{SCUC} &LR \& Benders &2.4 GHz CPU (30 cpus) &California ISO \\ \hline
2014 &CA N-1 &Trapzoidal Integration \& Dynamic Master-Slave Scheduling and others &256 CPUs &13029-bus 431-unit 5950-load 12488-line 10000-contingencies \\ \hline
2014 &DS &Jacobian-free Newton-GMRES &Xeon-E5-2620 host and NVIDIA K20 &865-bus \\ \hline
2014 &\ac{EMT} &Norton Equivalent Current \&Node Injectivng Current \&Solving Nodal voltage \&Solving Inner Variables \&Solving Control System &Intel Xeon E5645 \& NIVIDIA Tesla C2070 &9-bus RLC circuit \\ \hline
2014 &\ac{EMT} &LU Decomposition \& Trapezoidal Integration &AMD cout host \& NVIDIA GPU &IEEE 39 \\ \hline
2014 &\ac{SSE} &Extended Kalman Filter &3.2 GHz 4-Core AMD PhenomTM II host TeslaTMS2050 GPU &19968-bus 5120 units 30720-measurments \\ \hline
2015 &DS &Schur-complement \& VDHN &48-Core AMD Opteron Interlagos &14653-bus 15994-line 20-unit 1168- wind turbines 19419-load \\ \hline
2015 &DS &Backward Forward Substitution \& Newton-Raphson &Xeon E5603 host and Nvidia Geforce GTX-460 &283-bus \\ \hline
2015 &\ac{PF} &Master–Slave-Splitting &- &IEEE-118 (augmented) \\ \hline
2015 &\ac{SSE} &LU decomposition \&Numerical Differentiation Method &2.2 GHZ Intel Xeon E5606 host \& NVIDIA Tesla C2075 &IEEE-118 \\ \hline
2015 &\ac{SSE} &LU decomposition \& Extended Kalman Filter &2.0 GHz 4-Core Intel Xeon E5-2620 host \& NVIDIA Tesla S2050 &Bus-4992 \& 23152-measurments \\ \hline
2015 &\ac{SCUC} &Convex Relaxation &2.4 GHz Intel Core i5 &IEEE 118 \\ \hline
2015 &\ac{SCUC} &ATC &3.4 GHz CPU &IEEE 4672 \\ \hline
2015 &\ac{SCUC} &ADMM &2.0 GHz and 3 GHz CPUs cluster &IEEE 3375 \\ \hline
2016 &DS &Two-level Schur-complement \& VDHN &2.60 GHz AMD Opteron Interlagos 6238 (44 cpus) &14653-bus 15994-line 23-unit 19419-load 438-pv 730-wind turbine \\ \hline
2016 &\ac{EMT} &Sparsity Techniques \&Matrix Vector Multiplcications &Intel core i7 CPU2600K host \& NVIDIA GeForce GTX 590-GPU &3861-bus 534-line \\ \hline
2016 &\ac{OPF} &Optimality Condition Decomposition \& Newton-Raphson &3.2GHz Intel Core i5 &1416-bus \\ \hline
2016 &\ac{OPF} &Optimality Condition Decomposition \& Newton-Raphson &3.2GHz Intel Core i5 &1416-bus \\ \hline
2016 &PPF\footnote{Probabilistic Power Flow} &Monte Carlo &40-Cores at Royal Institute of Technology HPC facility &1354-bus (PEGASE) \\ \hline
2016 &\ac{PF} &LU decomposition and forward \&back substitution and path tree \&Newton-Raphson &NVIDIA GK110 &23215-bus \\ \hline
2016 &\ac{SSE} &VDHN \& WLS &NVIDIA Tesla K20c GPU &68-bus 81-line 166-measurments \\ \hline
2016 &TSA &Symplectic Gauss method \& Newton method and GMRES &Intel Xeon 8C E5-2650 host \& NVIDIA Tesla K20 &\textbf{2383-bus 327-units} \\ \hline
2016 &\ac{SCUC} &Binary Reduction &3.4 GHz 8-Core i7-3770 &MISO Network \\ \hline
2016 &\ac{SCUC} &APP &3.4 GHz CPU (8 cpus) &1168-bus \\ \hline
2016 &\ac{SCUC} &APP &2.8 GHz CPU &IEEE 118 \\ \hline
2016 &\ac{SCUC} &ADMM &2.30-GHz intel Core i5 CPU &IEEE 118 \\ \hline
2017 &CA N-1 &LU Decomposition \& Newton Raphson &Amazon EC2 v4 instance (256 x Cores Intel Xeon E5-2686 ) &1946-bus 390-unit 2589-line 2589-contingencies (KEPCO2015) \\ \hline
2017 &CA N-1 &LU decomposition &NVIDIA Tesla K40 GPU &9241-bus 2048-contingencies (PEGASE) \\ \hline
2017 &CA N-1 &Batch Solution &3.40 GHz 4-core Intel i7-3770 &15500-bus 2800-unit 20500-line 13922-load 1.44m-contingencies (PJM) \\ \hline
2017 &DS &LU decomposition \& Dishonest Newton Method \& Approximate Minimum Degree &1.2GHz AMD (16 cpus) &(WECC) \\ \hline
2017 &DS &LU decomposition &2.8 GHz Intel i7 (16 cpus) &17000-bus (WECC-Size) \\ \hline
2017 &\ac{EMT} &Layered Directed Acyclic Graph \& Unified Fused Mulitply Add &24-Core 2x Intel Xeon E5-2620 host \& NVIDIA Tesla K20x &IEEE 123 \\ \hline
2017 &\ac{EMT} &Jacobian Domain Decomposition \&Compensation Network Decomposition \&Propagation Delay partitioning \& Vector Multiplication &Intel Xeon E-2620 host \& 2 NVIDIA GP104 GPUs &79872-bus \\ \hline
2017 &\ac{OPF} &ADMM &- &944-bus \\ \hline
2017 &\ac{OPF} &ADMM &- &944-bus \\ \hline
2017 &\ac{PF} &LU decomposition \&Forward Backward Substitution,\&Path Tree \& Newton-Raphson &8-Core Intel Xeon E5-2650 host \& a NVIDIA Telsa K20c &2383-bus \\ \hline
2017 &\ac{PF} &Fast Decouple Power Flow \& Inexact Newton Method &2.27GHz 8-Core Xeon E5607 host \& NVIDIA Tesla M2070 GPU &11624-bus \\ \hline
2017 &\ac{SSE} &WLS \& Jacobian Information Matrix \&Correction \&Vector State computation. &Intel Xeon CPU E3- 1230 V5 \&NVIDA Tesla P100- PCIE-12GB &147841-bus 256784-line 661409-measurmetns \\ \hline
2017 &\ac{SSE} &LU decomposition \&Forward Backward Substitution \& Newton Method &2.2 GHz Intel Xeon host \& NIVIDIA Tesla C2075, GeForce GTX 650 \& GTX 660 &IEEE-118 \\ \hline
2017 &\ac{SSE} &Extended Kalman Filter &2.0 GHz 4-Core Inetl XeonTM E5-2620 host \& 4 Nividia TeslaTMS2050 GPUs &IEEE 118 \\ \hline
2017 &\ac{SCUC} &ADMM \& Surrogate Lagrange Relaxation &2.90GHz Intel i7-4910MQ &10000-unit \\ \hline
2017 &\ac{SCUC} &ADMM &2.8 GHz CPU &IEEE-118 \\ \hline
2018 &\ac{ED} &APP &3.7 GHz CPU &IEEE 118 \\ \hline
2018 &\ac{OPF} &KTT &- &IEEE 118 \\ \hline
2018 &\ac{OPF} &KTT &- &IEEE 118 \\ \hline
2018 &\ac{PF} &Block Bordered Diagonal Form \& Newton–Raphson &2.1 Ghz 8-core Intel Xeon (2cpus) &1354-bus (PEGASE) \\ \hline
2018 &PPF &LU factorization \& Batch &2GHz Intel Xeon E5-2620 host \& 2 NVIDIA Tesla K40 GPUs &9241-bus 10000-contingencies (PEGASE) \\ \hline
2018 &PPF &Fast Decoupled Power Flow \& Layered Directed Acyclic Graph (Batched) &3.4GHz Intel Xeon E3-1230 v5 \& NVIDIA Tesla P100 &3659-bus 4092-unit 20,467-line 1024-load scenarios (PEGASE) \\ \hline
2018 &\ac{SCUC} &ATC &3.2GHz 4-core (4 cpus) &117-bus 3-region \\ \hline
2018 &\ac{SCUC} &ATC &3.1 GHz CPU &IEEE 142 \\ \hline
2019 &\ac{EMT} &LU Decomposition (Crout) &NVIDIA K20x \& P100 &IEEE 14 \\ \hline
2019 &\ac{OPF} &Benders Decomposition &2.4GHz 4-Core CPU &2869-bus (PEGASE) \\ \hline
2019 &\ac{OPF} &Benders Decomposition &2.4GHz 4-Core CPU &2869-bus (PEGASE) \\ \hline
2019 &\ac{SCUC} &APP &Intel Xeon Phi (99 cores) &1168-bus \\ \hline
2019 &\ac{SCUC} &ADMM &3.6 GHz CPU &187-unit \\ \hline
2019 &\ac{UC} &ADMM &2.80 GHz CPU &IEEE 3012 \\ \hline
2020 &\ac{PF} &Parallel Nodal Power Mismatch Vector \& Newton-Raphson &2.60Ghz 4-Core Intel i7 6700HQ host \& NVIDIA GeForce GTX 950M &4068-bus \\ \hline
2020 &SCED &APP &3.7 GHz CPU &IEEE 118 \\ \hline
2020 &\ac{SCOPF} &Schur Complement &2.10 GHz 18-Core Intel Xeon E5-2695 v4 host \& NVIDIA Tesla P100 &13659-bus 20467-contingencies (PEGASE) \\ \hline
2020 &TSA &BBDF (Nested) &2.67 GHz 8 -Core Intel Xeon E7-8837 (8 cpus) &24886-bus \\ \hline
2020 &\ac{UC} &LU Decomposition \& Fast Decoupled Power Flow &2.10 GHz 6 -Cores (2 cpus) &2749-bus \\ \hline
2020 &\ac{SCUC} &ADMM \& Binary Reduction &2.3GHz CPUs (20 cores x 18 nodes) &MISO Network \\ \hline
2020 &\ac{SCUC} &ADMM \& Binary Reduction &2.3GHz CPUs (20 cores x 18 nodes) &MISO Network \\ \hline
2021 &\ac{SCUC} &Multi-Level Formulation &Intel Xeon Cores (100 cores) &IEEE 118 \\ \hline
2021 &\ac{SCUC} &Binary Reduction & &IEEE 118 \\ \hline
2021 &\ac{SCUC} &ATC &Intel i7-8700 k &IEEE 118 \\ \hline
2021 &\ac{SCUC} &ADMM &1.80Ghz Intel i7-8550U &IEEE 13 \\
\hline

\end{longtable}

\end{small}

\end{document}